\documentclass[12pt]{article}
\usepackage[dvips]{graphicx}
\usepackage{epsfig}
\usepackage{latexsym,amsmath,amsfonts,amssymb}
\usepackage[latin1]{inputenc}
\usepackage[american]{babel}
\usepackage[dvips]{graphicx}
\usepackage{bbm}
\def\pl{Phys. Lett.}

\def\im{Invent. Math.}

\newcommand{\be}{\begin{equation}}
\newcommand{\ee}{\end{equation}}
\newcommand{\beq}{\begin{equation}}
\newcommand{\eeq}{\end{equation}}
\newcommand{\bea}{\begin{eqnarray}}
\newcommand{\eea}{\end{eqnarray}}

\newcommand{\ba}{\begin{eqnarray}}
\newcommand{\ea}{\end{eqnarray}}
\begin{document}
\baselineskip=15.5pt
\pagestyle{plain}
\setcounter{page}{1}


\def\del{{\partial}}
\def\vev#1{\left\langle #1 \right\rangle}
\def\cn{{\cal N}}
\def\co{{\cal O}}
\def\IC{{\mathbb C}}
\def\IR{{\mathbb R}}
\def\IZ{{\mathbb Z}}
\def\RP{{\bf RP}}
\def\CP{{\bf CP}}
\def\Poincare{{Poincar\'e }}
\def\tr{{\rm tr}}
\def\tp{{\tilde \Phi}}

\def\TL{\hfil$\displaystyle{##}$}
\def\TR{$\displaystyle{{}##}$\hfil}
\def\TC{\hfil$\displaystyle{##}$\hfil}
\def\TT{\hbox{##}}
\def\HLINE{\noalign{\vskip1\jot}\hline\noalign{\vskip1\jot}}
\def\seqalign#1#2{\vcenter{\openup1\jot
   \halign{\strut #1\cr #2 \cr}}}
\def\lbldef#1#2{\expandafter\gdef\csname #1\endcsname {#2}}
\def\eqn#1#2{\lbldef{#1}{(\ref{#1})}%
\begin{equation} #2 \label{#1} \end{equation}}
\def\eqalign#1{\vcenter{\openup1\jot
     \halign{\strut\span\TL & \span\TR\cr #1 \cr
    }}}
\def\eno#1{(\ref{#1})}
\def\href#1#2{#2}
\def\half{{1 \over 2}}

\def\ads{{\it AdS}}
\def\adsp{{\it AdS}$_{p+2}$}
\def\cft{{\it CFT}}

\newcommand{\ber}{\begin{eqnarray}}
\newcommand{\eer}{\end{eqnarray}}

\newcommand{\beqar}{\begin{eqnarray}}
\newcommand{\cN}{{\cal N}}
\newcommand{\cO}{{\cal O}}
\newcommand{\cA}{{\cal A}}
\newcommand{\cT}{{\cal T}}
\newcommand{\cF}{{\cal F}}
\newcommand{\cC}{{\cal C}}
\newcommand{\cR}{{\cal R}}
\newcommand{\cW}{{\cal W}}
\newcommand{\eeqar}{\end{eqnarray}}
\newcommand{\tht}{\thteta}
\newcommand{\lm}{\lambda}\newcommand{\Lm}{\Lambda}
\newcommand{\eps}{\epsilon}


\newcommand{\nonu}{\nonumber}
\newcommand{\oh}{\displaystyle{\frac{1}{2}}}
\newcommand{\dsl}
   {\kern.06em\hbox{\raise.15ex\hbox{$/$}\kern-.56em\hbox{$\partial$}}}
\newcommand{\id}{i\!\!\not\!\partial}
\newcommand{\as}{\not\!\! A}
\newcommand{\ps}{\not\! p}
\newcommand{\ks}{\not\! k}
\newcommand{\D}{{\cal{D}}}
\newcommand{\dv}{d^2x}
\newcommand{\Z}{{\cal Z}}
\newcommand{\N}{{\cal N}}
\newcommand{\Dsl}{\not\!\! D}
\newcommand{\Bsl}{\not\!\! B}
\newcommand{\Psl}{\not\!\! P}
\newcommand{\eeqarr}{\end{eqnarray}}
\newcommand{\ZZ}{{\rm \kern 0.275em Z \kern -0.92em Z}\;}


\def\del{{\delta^{\hbox{\sevenrm B}}}} \def\ex{{\hbox{\rm e}}}
\def\azb{A_{\bar z}} \def\az{A_z} \def\bzb{B_{\bar z}} \def\bz{B_z}
\def\czb{C_{\bar z}} \def\cz{C_z} \def\dzb{D_{\bar z}} \def\dz{D_z}
\def\im{{\hbox{\rm Im}}} \def\mod{{\hbox{\rm mod}}} \def\tr{{\hbox{\rm Tr}}}
\def\ch{{\hbox{\rm ch}}} \def\imp{{\hbox{\sevenrm Im}}}
\def\trp{{\hbox{\sevenrm Tr}}} \def\vol{{\hbox{\rm Vol}}}
\def\rl{\Lambda_{\hbox{\sevenrm R}}} \def\wl{\Lambda_{\hbox{\sevenrm W}}}
\def\fc{{\cal F}_{k+\cox}} \def\vev{vacuum expectation value}
\def\nodiv{\mid{\hbox{\hskip-7.8pt/}}}
\def\ie{{\em i.e.}}
\def\ie{\hbox{\it i.e.}}

\def\CC{{\mathchoice
{\rm C\mkern-8mu\vrule height1.45ex depth-.05ex
width.05em\mkern9mu\kern-.05em}
{\rm C\mkern-8mu\vrule height1.45ex depth-.05ex
width.05em\mkern9mu\kern-.05em}
{\rm C\mkern-8mu\vrule height1ex depth-.07ex
width.035em\mkern9mu\kern-.035em}
{\rm C\mkern-8mu\vrule height.65ex depth-.1ex
width.025em\mkern8mu\kern-.025em}}}

\def\RR{{\rm I\kern-1.6pt {\rm R}}}
\def\NN{{\rm I\!N}}
\def\ZZ{{\rm Z}\kern-3.8pt {\rm Z} \kern2pt}
\def\IB{\relax{\rm I\kern-.18em B}}
\def\ID{\relax{\rm I\kern-.18em D}}
\def\II{\relax{\rm I\kern-.18em I}}
\def\IP{\relax{\rm I\kern-.18em P}}
\newcommand{\CS}{{\scriptstyle {\rm CS}}}
\newcommand{\CSs}{{\scriptscriptstyle {\rm CS}}}
\newcommand{\rc}{\nonumber\\}
\newcommand{\bear}{\begin{eqnarray}}
\newcommand{\eear}{\end{eqnarray}}
\newcommand{\W}{{\cal W}}
\newcommand{\F}{{\cal F}}
\newcommand{\x}{{\cal O}}
\newcommand{\LL}{{\cal L}}

\def\mani{{\cal M}}
\def\calo{{\cal O}}
\def\calb{{\cal B}}
\def\calw{{\cal W}}
\def\calz{{\cal Z}}
\def\cald{{\cal D}}
\def\calc{{\cal C}}
\def\to{\rightarrow}
\def\ele{{\hbox{\sevenrm L}}}
\def\ere{{\hbox{\sevenrm R}}}
\def\zb{{\bar z}}
\def\wb{{\bar w}}
\def\nodiv{\mid{\hbox{\hskip-7.8pt/}}}
\def\menos{\hbox{\hskip-2.9pt}}
\def\dr{\dot R_}
\def\drr{\dot r_}
\def\ds{\dot s_}
\def\da{\dot A_}
\def\dga{\dot \gamma_}
\def\ga{\gamma_}
\def\dal{\dot\alpha_}
\def\al{\alpha_}
\def\cl{{closed}}
\def\cls{{closing}}
\def\vev{vacuum expectation value}
\def\tr{{\rm Tr}}
\def\to{\rightarrow}
\def\too{\longrightarrow}


\def\a{\alpha}
\def\b{\beta}
\def\c{\gamma}
\def\d{\delta}
\def\e{\epsilon}           
\def\f{\phi}               
\def\vf{\varphi}  \def\tvf{\tilde{\varphi}}
\def\vp{\varphi}
\def\g{\gamma}
\def\h{\eta}
\def\j{\psi}
\def\k{\kappa}                    
\def\l{\lambda}
\def\m{\mu}
\def\n{\nu}
\def\o{\omega}  \def\w{\omega}
\def\q{\theta}  \def\th{\theta}                  
\def\r{\rho}                                     
\def\s{\sigma}                                   
\def\t{\tau}
\def\u{\upsilon}
\def\x{\xi}
\def\z{\zeta}
\def\pt{\tilde{\varphi}}
\def\tt{\tilde{\theta}}
\def\lab{\label}
\def\6{\partial}
\def\wg{\wedge}
\def\atanh{{\rm arctanh}}
\def\bpsi{\bar{\psi}}
\def\bt{\bar{\theta}}
\def\bvf{\bar{\varphi}}

%

\newfont{\namefont}{cmr10}
\newfont{\addfont}{cmti7 scaled 1440}
\newfont{\boldmathfont}{cmbx10}
\newfont{\headfontb}{cmbx10 scaled 1728}
\newcommand{\re}{\,\mathbb{R}\mbox{e}\,}
\newcommand{\hyph}[1]{$#1$\nobreakdash-\hspace{0pt}}
\providecommand{\abs}[1]{\lvert#1\rvert}
\newcommand{\Nugual}[1]{$\mathcal{N}= #1 $}
\newcommand{\sub}[2]{#1_\text{#2}}
\newcommand{\partfrac}[2]{\frac{\partial #1}{\partial #2}}
\newcommand{\bsp}[1]{\begin{equation} \begin{split} #1 \end{split} \end{equation}}
\newcommand{\calF}{\mathcal{F}}
\newcommand{\calO}{\mathcal{O}}
\newcommand{\calM}{\mathcal{M}}
\newcommand{\calV}{\mathcal{V}}
\newcommand{\bbZ}{\mathbb{Z}}
\newcommand{\bbC}{\mathbb{C}}

\numberwithin{equation}{section}

\newcommand{\Tr}{\mbox{Tr}}    


%
\renewcommand{\theequation}{{\rm\thesection.\arabic{equation}}}
\begin{titlepage}
\vspace{0.1in}

\begin{center}
\Large \bf On the gravity dual of Chern-Simons-matter theories with unquenched flavor
\end{center}
\vskip 0.2truein
\begin{center}
Eduardo Conde  \footnote{eduardo@fpaxp1.usc.es}
 and
Alfonso V. Ramallo\footnote{alfonso@fpaxp1.usc.es}\\
\vspace{0.2in}
\it{
Departamento de  F\'\i sica de Part\'\i  culas, Universidade
de Santiago de
Compostela\\and\\Instituto Galego de F\'\i sica de Altas
Enerx\'\i as (IGFAE)\\E-15782, Santiago de Compostela, Spain
}

\vspace{0.2in}
\end{center}
\vspace{0.2in}
\centerline{{\bf Abstract}}
We find solutions of type IIA supergravity which are dual to three-dimensional Chern-Simons-matter theories with unquenched fields in the fundamental representation of the gauge group (flavors). In the holographic dual the addition of flavor is performed by means of D6-branes that are extended along the Minkowski gauge theory directions and are delocalized in the internal space in such a way that the system is ${\cal N}=1$ supersymmetric and the flavor group is abelian.  For massless flavors the corresponding geometry has the form of a product space $AdS_4\times {\cal M}_6$, where ${\cal M}_6$ is a six-dimensional compact manifold whose metric is obtained by squashing the Fubini-Study metric of  ${\mathbb C}{\mathbb P}^3$ with suitable constant factors which depend on the number of flavors. We compute the effect of dynamical quarks in several observables and, in some cases, we compare our results with the ones corresponding to the ${\cal N}=3$ supergravity solutions generated by localized flavor branes. We also show how to generalize our results to include massive flavors.

\smallskip
\end{titlepage}
\setcounter{footnote}{0}

\tableofcontents

\section{Introduction}

The recent results on the $AdS_4/CFT_3$ correspondence constitute a rich framework in which fundamental questions about the holographic correspondence \cite{jm} can be posed and, in some cases, highly non-trivial answers can be obtained. The crucial breakthrough in this subject took place with the results of Bagger and Lambert \cite{BL} and Gustavsson \cite{Gustavsson:2007vu}, who proposed that the low energy theory on multiple M2-branes is given by a new class of maximally supersymmetric Chern-Simons-matter theories. Inspired by these results, Aharony et al. (ABJM) \cite{Aharony:2008ug} constructed an ${\cal N}=6$ supersymmetric Chern-Simons-matter theory which is now believed to describe the dynamics of multiple M2-branes at a ${\mathbb C}^4/{\mathbb Z}_k$ singularity. 

The ABJM theory is a $U(N)\times U(N)$ Chern-Simons gauge theory with levels $(k,-k)$ and bifundamental matter fields. In the large $N$ limit this theory admits a supergravity description in M-theory in terms of the $AdS_4\times 
{\mathbb S}^7/{\mathbb Z}_k$ geometry. If we represent ${\mathbb S}^7$ as a $U(1)$ Hopf bundle over ${\mathbb C}{\mathbb P}^3$, the ${\mathbb Z}_k$ orbifold acts by quotienting the ${\mathbb S}^1$ fiber. When the Chern-Simons level $k$ is large the size of the fiber is small and the system is better described in terms of type IIA supergravity by performing a dimensional reduction  to ten dimensions along the Hopf fiber of ${\mathbb S}^7/{\mathbb Z}_k$. In this ten-dimensional description the geometry is of  the form $AdS_4\times {\mathbb C}{\mathbb P}^3$ with fluxes and preserves 24 supersymmetries. 

The precise knowledge of the field theory dual to a system of multiple  M2-branes on 
${\mathbb C}^4/{\mathbb Z}_k$ has allowed to test some of the non-trivial predictions of the AdS/CFT correspondence. In particular, in \cite{Drukker:2010nc} it was checked by means of  a purely field theoretic calculation, using matrix model techniques  and localization, that the number of low energy degrees of freedom of $N$ coincident M2-branes scales as $N^{{3\over 2}}$ for large $N$, as predicted by the gravity dual (see also \cite{Herzog:2010hf}). Moreover, the ABJM model has been generalized in several directions. By adding fractional M2-branes, the gravity dual of $U(M)\times U(N)$ Chern-Simons-matter theories with $M\not=N$ was constructed in \cite{Aharony:2008gk} . If the sum of Chern-Simons levels for the two gauge groups is non-zero, the corresponding gravity dual can be found in massive type IIA supergravity by considering solutions in which the Romans mass parameter is non-vanishing \cite{Gaiotto:2009mv} (see also \cite{Fujita:2009kw}).  The ABJM construction has been extended  to include different quivers and gauge groups with several amount of supersymmetry in refs. \cite{Imamura:2008nn}-\cite{Ooguri:2008dk}. 

In this paper we will study the generalization of the ABJM model which is obtained by adding flavors, \ie\ fields transforming in the fundamental representations $(N,1)$ and $(1,N)$ of the $U(N)\times U(N)$ gauge group. The holographic dual of such a system was proposed in refs. \cite{Hohenegger:2009as,Gaiotto:2009tk}. In the type IIA description the addition of massless flavor is achieved by considering D6-branes that fill the $AdS_4$ space and wrap an ${\mathbb R}{\mathbb P}^3$ submanifold inside
the ${\mathbb C}{\mathbb P}^3$, while preserving ${\cal N}=3$ supersymmetry
(see also \cite{Fujita:2009xz}-\cite{Benini:2011cm} for different setups  with D6-branes in Chern-Simons-matter theories). When the number of flavors is small one can study the system in the quenched approximation, in which the D6-branes are considered as  probes in the 
$AdS_4\times {\mathbb C}{\mathbb P}^3$  background. This quenched approach  has been adopted in refs. \cite{Hikida:2009tp}-\cite{Ammon:2009wc}, where different observables of the Chern-Simons-matter theory with flavor have been analyzed.

In the present article we will study the holographic dual of the ABJM theory with unquenched flavor. For a system of localized and coincident D6-branes, the corresponding
 gravity dual in M-theory is a purely geometric background which, in the near horizon limit, is a space of the type $AdS_4\times {\cal M}_7$, where ${\cal M}_7$ is a tri-Sasakian seven-dimensional manifold whose cone is an eight-dimensional hyperk\"ahler manifold \cite{Gauntlett:1997pk}. Notice that the  backreacted  metric always has an Anti-de-Sitter factor. This is related to the fact that the dual Chern-Simons theory remains conformally invariant after the addition of flavor (see ref. \cite{Bianchi} for a perturbative calculation of the beta functions and a study of the corresponding fixed points).  However, the tri-Sasakian metric of ${\cal M}_7$ has, in general, a complicated structure which makes difficult to use it for many purposes. For this reason, in this paper we study the backreaction induced by a smeared continuous distribution of a large number $N_f$ 
 of flavor branes. This approach was initiated in \cite{Bigazzi:2005md} in the context of non-critical strings and  has been successfully applied to study unquenched flavor in several other setups (see \cite{Nunez:2010sf} for an extensive review and references to the original works). 

In order to obtain the gravity dual of a field theory with unquenched flavor one has to solve the equations of motion of supergravity with brane sources. These sources typically modify the Bianchi identities of the forms and, as they contribute to the energy-momentum tensor of the system, they also modify the Einstein equations. If the flavor branes are localized the sources contain Dirac $\delta$-functions and, as a consequence, solving the equations of motion is, in general, a difficult task. On the contrary, if the sources are delocalized there are no $\delta$-function terms in the equations of motion and finding explicit analytic solutions is much simpler. 
 Notice that when the branes are smeared, they are not coincident anymore and, therefore, the flavor symmetry  for $N_f$ flavors is $U(1)^{N_f}$ rather than $U(N_f)$. Moreover, the solutions with smeared unquenched flavor are generically less supersymmetric than the ones with localized flavor, due to the fact that we are superposing branes with different orientations in the internal space. Indeed, in our flavored ABJM case the solutions will be ${\cal N}=1$ SUSY instead of being ${\cal N}=3$.

The backreaction of the flavor branes induces a deformation of the unquenched solution which, in particular, results in a suitable squashing of the metric. In order to determine precisely this flavor deformation one has to write the metric in a way such that it can be squashed without breaking all supersymmetry. We will argue below that for the ABJM case in the type IIA description the convenient way of writing the  ${\mathbb C}{\mathbb P}^3$ metric is as ${\mathbb S}^2$-bundle over ${\mathbb S}^4$. 
After representing the ${\mathbb C}{\mathbb P}^3$ metric in this way, the flavor deformation just amounts to squashing the ${\mathbb S}^2$ fiber with respect to the 
${\mathbb S}^4$  base, as well as to changing the radii of both $AdS_4$ and
 ${\mathbb C}{\mathbb P}^3$ factors of the metric (similar squashed deformations of 
  ${\mathbb C}{\mathbb P}^3$ were considered recently in \cite{Tomasiello:2007eq,Aldazabal:2007sn,Ooguri:2008dk}). In our solutions,  the squashing factors are constants which have a precise dependence on the number of flavors $N_f$ and  encode the effects due to loops of fundamentals. Indeed, we will be able to determine the effects due to dynamical flavors in several observables such as, among others,  the free energy on the sphere, the quark-antiquark potential energy or the dimension of meson operators. 

The organization of the rest of this paper is the following. In section \ref{ABJM-unflavored} we review the ABJM $AdS_4\times{ \mathbb C}{\mathbb P}^3$ solution and we rewrite it in terms of the ${\mathbb S}^2$ fibration over ${\mathbb S}^4$. In section \ref{deformed ABJM} we study the deformations of the ABJM unflavored background with ${\cal N}=1$ supersymmetry which preserve the  ${\mathbb S}^2$-bundle structure. The corresponding supergravity solutions are obtained by solving a system of first-order BPS equations which we partially integrate in general.  In particular we find two Anti-de-Sitter solutions which correspond to the original unflavored ABJM model and to the squashed ${\cal N}=1$  model of reference \cite{Ooguri:2008dk}. In section \ref{deformed ABJM} we also find running solutions which approach the ABJM background in the IR.

In section \ref{flavor-quenched} we study the supersymmetric embeddings of flavor D6-brane probes in the deformed geometries. The backreaction for branes corresponding to massless quarks is analyzed in section \ref{unquenched-massless}, where an ansatz for the background deformed by the massless fundamental fields is proposed and a set of first-order BPS equations is obtained. Section \ref{flavored-AdS} is devoted to studying the solutions of the flavored BPS system that are Anti-de-Sitter. In section \ref{Flavor-effects} we analyze the effect of the unquenched flavor in several observables. In section \ref{unquenched-massive} we study the backreaction  of unquenched massive flavors. Finally, in section \ref{Conclusions} we summarize our results and discuss some directions for future work. The article is completed with several appendices, where some detailed calculations  not included in the main text are performed and some other aspects of our work are explored.

\section{The ABJM solution}
\label{ABJM-unflavored}
The metric of the ABJM background is given by:
\beq
ds^2\,=\,L^2\,ds^2_{AdS_4}\,+\,4\,L^2\,ds^2_{{\mathbb C}{\mathbb P}^3}\,\,,
\label{ABJM-metric}
\eeq
where $ds^2_{AdS_4}$ and $ds^2_{{\mathbb C}{\mathbb P}^3}$ are respectively
 the  $AdS_4$  and ${\mathbb C}{\mathbb P}^3$ metrics. The former,  in Poincare coordinates,  is given by:
\beq
ds^2_{AdS_4}\,=\,r^2\,d x_{1,2}^2\,+\,{dr^2\over r^2}\,\,,
\label{AdS4metric}
\eeq
with $d x_{1,2}^2$ being the Minkowski metric in 2+1 dimensions. This solution depends on two integers $N$ and $k$ which  represent, in the gauge theory dual, the rank of the gauge groups and the Chern-Simons level, respectively. In string units, 
the $AdS_4$  radius $L$ can be written in terms of $N$ and $k$ as:
\beq
L^4\,=\,2\pi^2\,{N\over k}\,\,.
\label{ABJM-AdSradius}
\eeq
Moreover, for this background the dilaton is constant and given by:
\beq
e^{\phi}\,=\,{2L\over k}\,\,=\,\,2\sqrt{\pi}\,\Big(\,{2N\over k^5}\,\Big)^{{1\over 4}}\,\,.
\label{ABJMdilaton}
\eeq
This solution of type IIA supergravity is also endowed with a RR two-form $F_2$ and a RR four-form $F_4$ whose expression can be written as:
\beq
F_2\,=\,2k\,J\,\,,\qquad\qquad
F_4\,=\,{3\over 2}\,k\,L^2\,\Omega_{AdS_4}\,=\,{3\pi\over \sqrt{2}}\,\,
\big(\,kN\,\big)^{{1\over 2}}\,\Omega_{AdS_4}\,\,,
\label{F2-F4-ABJM}
\eeq
with $J$ being the K\"ahler form of ${\mathbb C}{\mathbb P}^3$ and $\Omega_{AdS_4}$ is the volume element of the $AdS_4$ metric (\ref{AdS4metric}).  This solution is a good  gravity dual of the Chern-Simons-matter theory when the $AdS$ radius is large in string units and the string coupling $e^{\phi}$  is small. By looking at eqs. (\ref{ABJM-AdSradius}) and (\ref{ABJMdilaton}) this amounts to the condition $N^{{1\over 5}}<<k<< N$.

The 
${\mathbb C}{\mathbb P}^3$  metric in (\ref{ABJM-metric}) is the canonical Fubini-Study metric. In the context of the ABJM solution the ${\mathbb C}{\mathbb P}^3$ space is usually represented as foliated by the $T^{1,1}\sim{\mathbb S}^2\times {\mathbb S}^3$ manifold. Here we will 
 write the ${\mathbb C}{\mathbb P}^3$  metric   in a form which is more convenient for our purposes. We will regard  ${\mathbb C}{\mathbb P}^3$ as an ${\mathbb S}^2$-bundle over ${\mathbb S}^4$, with the fibration constructed by using the self-dual $SU(2)$ instanton on the four-sphere. Explicitly, 
$ds^2_{{\mathbb C}{\mathbb P}^3}$ will be written as:
\beq
ds^2_{{\mathbb C}{\mathbb P}^3}\,=\,{1\over 4}\,\,\Big[\,
ds^2_{{\mathbb S}^4}\,+\,\big(d x^i\,+\, \epsilon^{ijk}\,A^j\,x^k\,\big)^2\,\Big]\,\,,
\label{CP3=S4-S2}
\eeq
where $ds^2_{{\mathbb S}^4}$ is the standard metric for the unit four-sphere, $x^i$ ($i=1,2,3$) are cartesian coordinates that parametrize  the unit two-sphere ($\sum_i (x^i)^2\,=\,1$) and $A^i$ are the components of the non-abelian one-form connection corresponding to the $SU(2)$ instanton. Mathematically, the representation (\ref{CP3=S4-S2}) is obtained when  ${\mathbb C}{\mathbb P}^3$ is constructed as the twistor space of the four-sphere. We shall now introduce  a specific system of  coordinates  to represent the metric (\ref{CP3=S4-S2}). First of all,  let $\omega^i$ ($i=1,2,3$) be the $SU(2)$ left-invariant one-forms which satisfy $d\omega^i={1\over2}\,\epsilon_{ijk}\,\omega^j\wedge\omega^k$ (for an explicit representation of the $\omega^i$'s in terms of angular coordinates, see (\ref{w123})). Together with a new coordinate $\xi$, the $\omega^i$'s can be used to parameterize the metric of  a four-sphere ${\mathbb S}^4$ as:
\beq
ds^2_{{\mathbb S}^4}\,=\,
{4\over(1+\xi^2)^2}
\left[d\xi^2+{\xi^2\over4}\left((\omega^1)^2+(\omega^2)^2+(\omega^3)^2
\right)\right]\,\,,
\label{S4metric}
\eeq
where $0\le \xi<+\infty$ is a non-compact coordinate. The $SU(2)$ instanton one-forms $A^i$ can be written in these coordinates as:
\beq
A^{i}\,=\,-{\xi^2\over 1+\xi^2}\,\,\omega^i\,\,. 
\label{A-instanton}
\eeq
Let us next parametrize the $x^i$ coordinates of the ${\mathbb S}^2$ by means of two angles $\theta$ and $\varphi$ ($0\le\theta<\pi$, $0\le\varphi<2\pi$), namely:
\beq
x^1\,=\,\sin\theta\,\cos\varphi\,\,,\qquad\qquad
x^2\,=\,\sin\theta\,\sin\varphi\,\,,\qquad\qquad
x^3\,=\,\cos\theta\,\,.
\eeq
Then, one can easily prove that:
\beq
\big(d x^i\,+\, \epsilon^{ijk}\,A^j\,x^k\,\big)^2\,=\,(E^1)^2\,+\,(E^2)^2\,\,,
\eeq
where  $E^1$ and $E^2$ are the following one-forms:
\bear
&&E^1=d\theta+{\xi^2\over1+\xi^2}\left(\sin\varphi\,\omega^1-\cos\varphi\,\omega^2\right)\,,
\rc\rc
&&E^2=\sin\theta\left(d\varphi-{\xi^2\over1+\xi^2}\,\omega^3\right)+{\xi^2\over1+\xi^2}\,
\cos\theta\left(\cos\varphi\,\omega^1+\sin\varphi\,\omega^2\right)\,.
\label{Es}
\eear
Therefore,  the canonical Fubini-Study metric of $ {\mathbb C}{\mathbb P}^3$ can be written in terms of the one-forms defined above as:
\beq
ds^2_{{\mathbb C}\mathbb{P}^3}\,=\,{1\over 4}\,\Big[\,ds^2_{{\mathbb S}^4}\,+\,
(E^1)^2\,+\,(E^2)^2\,\Big]\,\,.
\label{CP3-metric}
\eeq
As a check, one can verify that the volume of ${\mathbb C}\mathbb{P}^3$ obtained from the above metric is $\pi^3/ 6$. We shall  now consider a rotated version of the forms $\omega^i$ by the  two angles $\theta$ and $\varphi$. Accordingly, we define three new one-forms  $S^i$ $(i=1,2,3)$ as:
\bear
&&
S^1=\sin\varphi\,\omega^1-\cos\varphi\,\omega^2\,,\rc\rc
&&
S^2=\sin\theta\,\omega^3-\cos\theta\left(\cos\varphi\,\omega^1+
\sin\varphi\,\omega^2\right)\,,\rc\rc
&&
S^3=-\cos\theta\,\omega^3-\sin\theta\left(\cos\varphi\,\omega^1+
\sin\varphi\,\omega^2\right)\,.
\label{rotomega}
\eear 
In terms of the forms defined in (\ref{rotomega})
 the line element  of the four-sphere is just obtained by substituting $\omega^i\to S^i$ in (\ref{S4metric}). Let us next define the one-forms ${\cal S}^{\xi}$   and ${\cal S}^{i}$ as:
\beq
{\cal S}^{\xi}\,=\,{2\over 1+\xi^2}\,d\xi\,\,,\qquad\qquad
{\cal S}^{i}\,=\,{\xi\over 1+\xi^2}\,S^i \,\,,\qquad(i=1,2,3)\,\,,
\label{calS}
\eeq
in terms of which the metric of the four-sphere is just 
$ds^2_{{\mathbb S}^4}=({\cal S}^{\xi})^2+\sum_i({\cal S}^{i})^2$.  The RR two-form $F_2$ can be written in terms of the one-forms defined in
(\ref{Es}) and (\ref{calS}) as:
\beq
F_2\,=\,{k\over 2}\,\Big(\,E^1\wedge E^2\,-\,\big(
{\cal S}^{\xi}\wedge {\cal S}^{3}\,+\,{\cal S}^1\wedge {\cal S}^{2}\big)\,\Big)\,\,.
\label{F2-ansatz}
\eeq
Notice that $F_2$ is a closed two-form due to the relation:
\bear
&&d\big(E^1\wedge E^2\big)\,=\,d\big(
{\cal S}^{\xi}\wedge {\cal S}^{3}\,+\,{\cal S}^1\wedge {\cal S}^{2}\big)\,=\,\rc\rc
&&=\,E^1\wedge ({\cal S}^{\xi}\wedge {\cal S}^{2}\,-\,{\cal S}^1\wedge {\cal S}^{3}\big)\,+\,
E^2\wedge ({\cal S}^{\xi}\wedge {\cal S}^{1}\,+\,{\cal S}^2\wedge {\cal S}^{3}\big)\,\,.
\label{dE1E2}
\eear
Notice that there is a non-trivial ${\mathbb C}{\mathbb P}^1={\mathbb S}^2$ in the geometry, which in our coordinates is parametrized by the angles $\theta$ and $\varphi$ at a fixed point in the base ${\mathbb S}^4$. As one can readily verify by a simple calculation from (\ref{F2-ansatz}),  the flux of the RR two-form $F_2$ through this ${\mathbb C}{\mathbb P}^1$ is given by: 
\beq
{1\over 2\pi}\,\,\int_{{\mathbb C}{\mathbb P}^1}\,F_2\,=\,k\,\,.
\label{F2-flux}
\eeq
Eq. (\ref{F2-flux}) is essential to understand the meaning of $k$ as the Chern-Simons level of  the gauge theory. Indeed, let us consider a fractional D2-brane, \ie\ a D4-brane wrapping a ${\mathbb C}{\mathbb P}^1$ two-cycle and extended along the Minkowski directions. For such a brane there is a coupling  to the worldvolume gauge field $A$ of the type  
$\int_{{\rm Min}_{1,2}}AdA \int_{{\mathbb C}{\mathbb P}^1}\,F_2$ which, taking into account  (\ref{F2-flux}), clearly induces a Chern-Simons coupling for the gauge field $A$ with level $k$.

Some basic facts of the geometry of the ABJM solution in our coordinates are compiled in appendix \ref{ABJMgeometry}. In particular, it is shown how the uplifted solution in M-theory corresponds to the space $AdS_4\times {\mathbb S}^7/{\mathbb Z_k}$, where the $ {\mathbb S}^7$ is realized as an $ {\mathbb S}^3$-bundle over 
$ {\mathbb S}^4$. Also, the non-trivial cycles of ${\mathbb C}{\mathbb P}^3$ are displayed.

\section{Deforming the ABJM background}
\label{deformed ABJM}

We will now analyze the generalization of the ABJM background obtained by performing a certain deformation of the metric which preserves some amount of supersymmetry. Specifically, we shall  consider a ten-dimensional string frame metric of the form:
\beq
ds^2_{10}\,=\,h^{-{1\over 2}}\,\,dx^2_{1,2}\,+\,h^{{1\over 2}}\,\,ds^2_{7}\,\,,
\label{metric-ansatz}
\eeq
where $h$ is a warp factor and $ds^2_{7}$ is a seven-dimensional metric containing an  
${\mathbb S}^2$ fibered over an  ${\mathbb S}^4$ in the same way as in ${\mathbb C}\mathbb{P}^3$, namely:
\beq
ds^2_{7}\,=\,dr^2+
e^{2f}\,ds^2_{{\mathbb S}^4}\,+\, e^{2g}\Big[\,(E^1)^2\,+\,(E^2)^2\Big]\,\,,
\label{7metric-ansatz}
\eeq
with $h$, $f$ and $g$ being functions of the radial variable $r$. Notice that $f$ and $g$ determine the sizes of the ${\mathbb S}^4$ and ${\mathbb S}^2$ of the internal ${\mathbb C}\mathbb{P}^3$ manifold. If $f\not=g$ we will say that the ${\mathbb C}\mathbb{P}^3$ is squashed. We will verify below that this squashing is compatible with supersymmetry when the functions of the ansatz satisfy certain first-order BPS equations. 

The type IIA supergravity solutions we are looking for are endowed with a RR two-form $F_2$, for which we will adopt the same ansatz as in (\ref{F2-ansatz}). 
 In addition, as it is always the case for the solutions associated to D2-branes, there is a RR four-form $F_4$ given by:
\beq
F_4\,=\,K(r)\,\,d^3x\,\wedge dr\,\,,
\label{F4-ansatz}
\eeq
where $K(r)$ is a function of the radial coordinate $r$. Moreover, we will assume that the dilaton $\phi$ depends only on $r$. 

Notice that the Bianchi identities $dF_2=dF_4=0$ are automatically satisfied. Moreover,  the Hodge dual of $F_4$ is equal to:
\beq
{}^*\,F_4= -K\,h^2\,e^{4f+2g}\,{\rm Vol} ({\mathbb S}^4)\,\wedge E^1\wedge E^2\,\,,
\eeq
and, thus,  the equation of motion of the four-form $F_4$ (namely $d{}^*\,F_4\,=\,0$), leads to:
 \beq
 K\,h^2\,e^{4f+2g}\,=\,constant\,\equiv\,\beta\,\,,
 \label{EOM-F4}
 \eeq
 where the constant $\beta$ should be determined from a quantization condition. 
 Thus, it follows that $K(r)$ can be written in terms of the other functions of the ansatz, namely:
 \beq
 K\,=\,\beta\,h^{-2}\,e^{-4f-2g}\,\,.
 \label{K-beta}
 \eeq
 Notice that $F_4$ is subjected to the following flux quantization condition:
\beq
{1\over 2 \kappa_{10}^2}\,\,\int_{M_6}\,{}^*\,F_4\,=\,\pm N\,T_{D_2}\,\,,
\eeq 
where $M_6$ is the six-dimensional angular manifold  enclosing the D2-brane. Using our ansatz this quantization condition is converted into:
\beq
{1\over 2 \kappa_{10}^2\,T_{D_2}}\,\,\int_{M_6}\,{}^*\,F_4\,=\,-
{1\over (2\pi)^5}\,
\int_{M_6} Kh^2 e^{4f+2g}\,{\rm Vol} ({\mathbb S}^4)\,\wedge {\rm Vol} ({\mathbb S}^2)\,=\,
-\,{\beta \over 3\pi^2}\,\,,
\eeq
 where ${\rm Vol} ({\mathbb S}^n)$ denotes the volume form of a unit $n$-sphere and
 we are using string units. Therefore, the coefficient $\beta$  should be related to the number of D2-branes as:
\beq
\beta\,=\,3\pi^2\,N\,\,.
\label{beta}
\eeq
and the function $K$ is related to the other functions in the ansatz as:
\beq
K\,=\,3\pi^2\,N\,h^{-2}\,e^{-4f-2g}\,\,.
 \label{K-N}
 \eeq

We will determine the functions entering our ansatz by requiring that our background preserves (at least) two supersymmetries. As shown in detail in appendix \ref{SUSY}, this requirement is fulfilled if the dilaton $\phi$, the warp factor $h$ and the functions $f$ and $g$ satisfy the following system  of first-order BPS equations:
\bear
&&\phi'\,=\,-{3k\over 8}\,e^{\phi}\,h^{-{1\over 4}}\,\big(\,
e^{-2g}-2e^{-2f}\,\big)\,-\,{e^{\phi}\over 4}\,K\,h^{{3\over 4}}\,\,,\rc\rc
&&h'\,=\,{k\over 2}\,e^{\phi}\, h^{{3\over 4}}\, \big(\,e^{-2g}\,-\,2e^{-2f}\,\big)\,-\,
e^{\phi}\,K\,h^{{7\over 4}}\,\,,\rc\rc
&&f'\,=\,{k\over 4} \,h^{-{1\over 4}}\,e^{\phi}\,\big[\,e^{-2f}\,-\,e^{-2g}\,\big]
\,+\,e^{-2f+g}\,\,,\rc\rc
&&g'\,=\,{k\over 2}\,e^{\phi}\,h^{-{1\over 4}}\, e^{-2f}\,+\,
e^{-g}\,-\,e^{-2f+g}\,\,.
\label{BPS-full-r}
\eear
In the first two equations of the system (\ref{BPS-full-r}) the function $K$ should be understood as given by (\ref{K-N}).  This BPS system is obtained after imposing the following projections on the Killing spinors:
\bear
&&\Gamma_{47}\,\epsilon\,=\,\Gamma_{56}\,\epsilon\,=\,\Gamma_{89}\,\epsilon\,\,,\rc\rc
&&\Gamma_{012}\,\epsilon\,=\,-\epsilon\,\,,\rc\rc
&&\Gamma_{3458}\,\epsilon\,=\,-\epsilon\,\,,
\label{full-projections}
\eear
where the $\Gamma_{a_1\cdots a_n}$ denote antisymmetrized products of flat Dirac matrices in the basis of one-forms written in (\ref{ten-dim-frame}).

The BPS system (\ref{BPS-full-r}) can be rewritten in a compact fashion in terms of two calibration forms. In order to recast (\ref{BPS-full-r}) in this way, 
let us next define the calibration seven-form ${\cal K}$ as:
\beq
{\cal K}\,=\,{1\over 7!}\,\,{\cal K}_{a_0\,\cdots a_6}\,\,e^{a_0\,\cdots a_6}\,\,,
\eeq
where the components ${\cal K}_{a_0\,\cdots a_6}$ are the fermionic bilinears:
\beq
{\cal K}_{a_0\,\cdots a_6}\,=\,e^{{\phi\over 3}}\,\,h^{{1\over 4}}\,\epsilon^{{\dagger}}\,\Gamma_{a_0\,\cdots a_6}\,\epsilon\,\,,
\label{cal-K-def}
\eeq
with $\epsilon$ being a Killing spinor of the background and the prefactor in (\ref{cal-K-def}) is included to account for the proper normalization of $\epsilon$ (see (\ref{radial-normalization-epsilon})).  By using the projections satisfied by $\epsilon$, one can verify that ${\cal K}$ is given by:
\beq
{\cal K}\,=\,-e^{012}\,\wedge\big(\,
e^{3458}\,-\,e^{3469}\,+\, e^{3579}\,+\,e^{3678}\,+\,e^{4567}\,+\,e^{4789}\,+\,e^{5689}
\,\big)\,\,.
\label{cal-K-explicit}
\eeq
In a background generated by D2-branes, it is natural to have also a calibration three-form. Accordingly, we also define the three-form  ${\cal \tilde K}$, as:
\beq
{\cal \tilde K}\,=\,{1\over 3!}\,{\cal \tilde K}_{a_0\,a_1\,a_2}\,e^{a_0\,a_1\,a_2}\,\,,
\qquad\qquad\qquad
{\cal \tilde K}_{a_0\,a_1\,a_2}\,=\,
e^{{\phi\over 3}}\,\,h^{{1\over 4}}\,\epsilon^{{\dagger}}\,\Gamma_{a_0\, a_1\,a_2}\,\epsilon\,\,.
\eeq
Using again the projections satisfied by the spinor $\epsilon$, one can show that:
\beq
{\cal \tilde K}\,=\,e^{012}\,\,.
\label{cal-tildeK-explicit}
\eeq
One can now verify  that the BPS equations  (\ref{BPS-full-r}) can be  compactly recast as:
\beq
{}^*\,F_2\,=\,-d\,\Big(\,e^{-\phi}\,{\cal K}\,\Big)\,\,,\qquad\qquad
d\,\Big(\,e^{-\phi}\,h^{-{1\over 2}}\,{}^*{\cal K}\,\Big)\,=\,0\,\,,\qquad\qquad
F_4\,=\,-d\big(\,e^{-\phi}\,{\cal \tilde K}\,\big)\,\,.
\label{calibration-conds}
\eeq

\subsection{Partial integration}
\label{partial-integration}

Let us  now carry out  some simple manipulations of the BPS system (\ref{BPS-full-r}), which will allow us to perform a partial integration. First of all, let us define the function $\Lambda$ as follows:
\beq
e^{\Lambda}\equiv e^{\phi}\,h^{-{1\over 4}}\,\,.
\label{Lambda-def}
\eeq
Clearly, from this definition one has:
\beq
\Lambda'\,=\,\phi'\,-\,{h'\over 4h}\,\,.
\label{Lambda-prime}
\eeq
Moreover, 
it is easy to prove that $\Lambda$, $f$ and $g$ close the following system of first-order differential equations:
\bear
&&\Lambda'\,=\,k\,e^{\Lambda-2f}\,-\,{k\over 2}\,e^{\Lambda-2g}\,\,,\rc\rc
&&f'\,=\,{k\over 4}\,\,e^{\Lambda-2f}\,-\,{k\over 4}\,e^{\Lambda-2g}\,+\,e^{-2f+g}\,\,,\rc\rc
&&g'\,=\,{k\over 2}\,\,e^{\Lambda-2f}\,+\,e^{-g}\,-\,e^{-2f+g}\,\,.
\label{3-system}
\eear
Notice that the function $K$ has disappeared from the system (\ref{3-system}) and that $\phi$ and $h$ only appear through the combination $\Lambda$. Actually, by combining the first two equations in (\ref{BPS-full-r}) one proves that $K$ can be written as:
\beq
K\,=\,{d\over dr}\,\,\Big(\,e^{-\phi}\,\,h^{-{3\over 4}}\,\Big)\,\,.
\label{K-phi-h}
\eeq

The warp factor $h$ and the dilaton $\phi$ can be obtained from the solution of the system (\ref{3-system}). Indeed, by using again the first two equations in (\ref{BPS-full-r}) together with  (\ref{K-beta}) , one arrives at:
\beq
h'\,+\,{4\over 3}\,h\,\phi'\,=\,-4\pi^2N\,e^{\Lambda-4f-2g}\,\,,
\label{h-beta}
\eeq
where we have used the definition (\ref{Lambda-def}). Eliminating $\phi'$ between (\ref{Lambda-prime}) and (\ref{h-beta}), we get:
\beq
h'\,+\,h\,\Lambda'\,=\,-3\pi^2N\,\,e^{\Lambda-4f-2g}\,\,.
\label{h-dif-eq}
\eeq
Eq. (\ref{h-dif-eq}) is a first-order differential equation for $h$ that can be solved by the method of variation of constants. The result is:
\beq
h(r)\,=\,\,e^{-\Lambda(r)}\,\,\Big[\,\alpha\,-\,3\pi^2N\,
\int^r\,\,
e^{2\Lambda(z)-4f(z)-2g(z)}\,\,dz\,\,\Big]\,\,,
\label{warp-integral}
\eeq
where $\alpha$ is a constant of integration that should be adjusted appropriately. We have not been able to integrate the BPS system (\ref{3-system}) in general. However, we have found some important particular solutions which we will discuss in the next two subsections and in appendix \ref{squashed-unflavored}. In some of these solutions the supersymmetry is enhanced with respect to the two supersymmetries preserved by the generic solution of (\ref{3-system}). 

\subsection{Anti-de-Sitter solutions}
\label{AdS-unflavored}

We will be mostly interested in backgrounds with the Anti-de-Sitter geometry and in their corresponding deformations. In order to find these solutions systematically, 
let us now introduce a new radial variable $\tau$, related to $r$ as follows:
\beq
e^f\,{d\over dr}\,=\,{d\over d\tau}\,\,.
\label{r-tau}
\eeq
If the dot denotes derivative with respect to $\tau$, the system of equations (\ref{3-system}) reduces to:
\bear
&&\dot\Lambda\,=\,k\,e^{\Lambda-f}\,-\,{k\over 2}\,e^{\Lambda+f-2g}\,\,,\rc\rc
&&\dot f\,=\,{k\over 4}\,\,e^{\Lambda-f}\,-\,{k\over 4}\,e^{\Lambda+f-2g}\,+\,e^{-f+g}\,\,,
\rc\rc
&&\dot g\,=\,{k\over 2}\,\,e^{\Lambda-f}\,+\,e^{f-g}\,-\,e^{-f+g}\,\,.
\label{3-system-tau}
\eear
Let us next define the following combination of functions:
\beq
\Sigma\,\equiv \Lambda-f\,\,,\qquad\qquad
\Delta\,\equiv f-g\,\,.
\label{Sigma-Delta-def}
\eeq
Notice that $\Delta$ measures the squashing of the ${\mathbb S}^4$ and 
 ${\mathbb S}^2$ internal directions. Actually, 
the right-hand-side of the equations in (\ref{3-system-tau}) depends only on $\Sigma$ and $\Delta$ and it is straightforward  to  find the following system of two equations involving $\Sigma$ and $\Delta$:
\bear
&&\dot\Sigma\,=\,{k\over 4} e^{\Sigma}\,\Big(\,3-e^{2\Delta}\,\Big)
\,-\,e^{-\Delta}\,\,,
\rc\rc
&&\dot \Delta\,=\,-{k\over 4}\,e^{\Sigma}\,
\Big(\,1+e^{2\Delta}\,\Big)\,-\,e^{\Delta}\,+2\,e^{-\Delta}\,\,.
\label{Sigma-Delta-sys}
\eear
One can take $\Sigma$, $\Delta$ and (say) $g$ as independent functions. In fact,  $g$ can be obtained by simple integration  once $\Sigma$ and $\Delta$ are known, due to the equation:
\beq
\dot g\,=\,{k\over 2}\,e^{\Sigma}\,+\,e^{\Delta}\,-\,e^{-\Delta}\,\,.
\label{g-Sigma-Delta}
\eeq

We have thus reduced the full BPS system to a set of two coupled differential equations for the functions $\Sigma$ and $\Delta$. 

We will now obtain some particular solutions of (\ref{g-Sigma-Delta})
in which the squashing factor $\Delta$ is constant, as expected for an $AdS$ background.  It follows from the second equation in (\ref{Sigma-Delta-sys}) that, in this case, $\Sigma$ must be also constant.  Actually, we can eliminate $\Sigma$ from the equations $\dot\Sigma=\dot\Delta=0$ and get a simple algebraic equation for $\Delta$. In order to express this equation in simple terms, let us define the quantity $q$ as:
\beq
q\equiv e^{2\Delta}\,=\,e^{2f-2g}
\,\,.
\label{q-def}
\eeq
Then,  $\dot\Sigma=\dot\Delta=0$ implies the following quadratic equation for $q$:
\beq
q^2-6q+5=0\,\,,
\label{q-eq}
\eeq
which has two solutions, namely:
\beq
q=1\,\,,\qquad q=5\,\,.
\label{q-sol}
\eeq
Notice that $q=1$ corresponds to the ${\cal N}=6$ ABJM background, while $q=5$ should correspond to a background of reduced SUSY of the form $AdS_4\times \overline{ {\mathbb C}{\mathbb P}^3}$,  with $\overline{ {\mathbb C}{\mathbb P}^3}$ being a squashed version of ${\mathbb C}{\mathbb P}^3$. This 
$AdS_4\times \overline{ {\mathbb C}{\mathbb P}^3}$ background was proposed in ref. \cite{Ooguri:2008dk} to be the gravity  dual of an ${\cal N}=1$ superconformal Chern-Simons-matter gauge theory with $Sp(2)\times U(1)\cong SO(5)\times U(1)$ global symmetry. We will describe this solution and some generalizations in appendix \ref{squashed-unflavored}. In the remaining of this subsection we will concentrate in showing how the ABJM background reviewed in section \ref{ABJM-unflavored} is obtained in this formalism from the  $q=1$ solution. First of all, we notice that when $q=e^{2\Delta}=1$, the system (\ref{Sigma-Delta-sys}) leads to the following solution for $\Sigma$:
\beq
e^{\Sigma}\,=\,{2\over k}\,\,.
\eeq
Using these values of $\Sigma$ and $\Delta$ in (\ref{g-Sigma-Delta}) one readily gets that $f=g=\tau$ and, by using (\ref{r-tau}) one shows that the radial variables $\tau$ and $r$ are related as $r=e^{\tau}$. Therefore, in the original variable $r$, one has:
\beq
e^{f}=e^{g}=r\,\,.
\label{fg-ABJM}
\eeq
Taking into account that  the function $\Lambda$ defined in (\ref{Lambda-def}) is $\Lambda=\Sigma+\Delta+g$,  we get that $e^{\Lambda}=2r/k$ and we can obtain $h$, $\phi$ and $K$ from eqs. (\ref{warp-integral}), (\ref{Lambda-def}) and (\ref{K-phi-h}) respectively. The dilaton obtained in this way is constant and is  just the one written in (\ref{ABJMdilaton}), while $h$ and $K$ are given by:
\beq
h={2\pi^2 N\over k}\,\,{1\over r^4}\,\,,\qquad\qquad
K\,=\,{3k^2\over 4\pi^2 N}\,\,r^2\,\,.
\label{hK-ABJM}
\eeq
By rescaling the Minkowski coordinates as $x^{\mu}\to\bar\lambda x^{\mu}$ with $\bar\lambda=\pi\,\sqrt{{N\over 2k}}$, one can verify that, indeed, the metric and RR four-form for this solution coincide with the ones in (\ref{ABJM-metric}) and (\ref{F2-F4-ABJM}).

\subsection{Running solutions}
\label{running-solutions}

We will now solve the BPS system (\ref{3-system})  in a power series expansion in  the radial coordinate $r$. The aim is to find new solutions that approach the $AdS_4\times {\mathbb C} {\mathbb P}^3$ background in the IR limit $r\to 0$. We begin by rewriting the system (\ref{3-system}) in a more convenient form. Let us define the new function $F$ as:
\beq
F\,=\,{k\over 2}\,\,e^{\Lambda}\,\,.
\eeq
Then, one can recast (\ref{3-system}) as:
\bear
&&F'\,=\,F^2\,\,\Big[\,2\,e^{-2f}\,-\,e^{-2g}\,\Big]\,\,,\rc\rc
&&\big(e^{f}\big)'\,=\,{F\over 2}\,\,\Big[\,e^{-f}\,-\,e^{f-2g}\,\Big]\,+\,e^{g-f}\,\,,\rc\rc
&&\big(e^{g}\big)'\,=\,F\,e^{g-2f}\,+\,1\,-\,e^{2g-2f}\,\,.
\label{F-f-g-system}
\eear
The ${\cal N}=6$ ABJM solution (without squashing ) can be simply written as $F= e^{f}=e^{g}=r$.  We will now solve (\ref{F-f-g-system}) in a series expansion  in powers of $r$ around this solution. We will look for a solution in which $F$, $e^{f}$ and $e^{g}$ take the form:
\beq
F\,=\,r\big[1+a_1\,r\,+a_2\,r^2\,+\,\cdots\big],\,\,\,\,
e^{f}\,=\,r\big[1+b_1\,r\,+b_2\,r^2\,+\,\cdots\big],\,\,\,\,
e^{g}\,=\,r\big[1+c_1\,r\,+c_2\,r^2\,+\,\cdots\big].\,\,\,\,\qquad
\eeq
By substituting this ansatz on  the system (\ref{F-f-g-system}) and solving for the different powers of $r$ up to third order, one can find the following solution:
\bear
&&F\,=\,r\big[1+3c\,r\,+8c^2\,r^2\,+\,20 c^3 r^3\,+\,\cdots\,],\rc\rc
&&e^{f}\,=\,r\big[\,1\,+\,{c\over 2}\,r\,+\,{7\over 8}\,c^2\,r^2\,+\,{25\over 16}\,c^3\,r^3\,
\,+\,\cdots\,\big]\,\,,\rc\rc
&&e^{g}\,=\,r\big[\,1\,+\,c\,r\,+\,2c^2\, r^2\,+\,4 c^3\,r^3\,+\,\cdots
\,\big]\,\,,
\label{F-f-g-expanded}
\eear
where $c$ is an arbitrary constant (if $c=0$ we come back to the $AdS_4\times {\mathbb C}{\mathbb P}^3$ solution). Plugging the expansions (\ref{F-f-g-expanded}) into the right-hand side of (\ref{warp-integral}) and adjusting the integration constants in such a way that $h$ vanishes at $r=\infty$, one gets the following expression of the warp factor $h$:
\beq
h\,=\,{2\pi^2 N\over k}\,\,\,\Big[\,{1\over r^4}\,+\,{c^2\over r^2}\,-\,
{23 c^3\,+\,12 c^3\,\log (r)\over r}\,+\,\cdots\,\Big]\,\,.
\eeq
Similarly, the dilaton runs as:
\beq
e^{\phi}\,=\,\ 2\sqrt{\pi}\,\Big(\,{2N\over k^5}\,\Big)^{{1\over 4}}\,\,\Big[\,
1\,+\,3c\,r\,+\,{33 c^2\, r^2\over 4}\,+\,
3\,c^3\,\Big(\,5-\,\log (r)\,\Big)\,r^3\,+\,\cdots\,\Big]\,\,,
\eeq
whereas the function $K$ is given by:
\beq
K\,=\,{3k^2\over 4\pi^2 N}\,r^2\,\Big(\,1\,-\,4cr\,+\,2c^3\,\big(29+12\,\log(r)\,\big)\,r^3
\,+\,\cdots\,\Big)\,\,.
\eeq
Notice that, when the constant $c$ is non-vanishing, the geometry is not Anti-de-Sitter and the internal space is squashed by an $r$-dependent function. 

\section{SUSY embeddings of flavor D6-branes}
\label{flavor-quenched}

In this section we will study the addition of flavor D6-branes to a background of the type studied in section \ref{deformed ABJM}.  We will analyze certain configurations in which the D6-branes preserve some amount of supersymmetry of the background. In the present section we will work in the probe approximation, corresponding to having quenched quarks on the field theory side, in which the background supergravity solution is not affected by the presence of the flavor D6-branes. The effect of the backreaction will be considered in detail in sections \ref{unquenched-massless} and \ref{unquenched-massive}. 

 Generically, we will consider configurations corresponding to massless quarks which extend along the three Minkowski directions $x^{\mu}$, the radial coordinate $r$ and that wrap a three-dimensional cycle of ${\mathbb C}{\mathbb P}^3$. On general grounds \cite{Hohenegger:2009as,Gaiotto:2009tk} it is expected that this three-cycle of ${\mathbb C}{\mathbb P}^3$ is a special lagrangian cycle which can be identified with ${\mathbb R} {\mathbb  P}^3$. Let us show how this ${\mathbb R}{\mathbb P}^3$ arises in our coordinates. With this purpose let us parametrize the $SU(2)$ left invariant one-forms $\omega_i$ of the four-sphere metric (\ref{S4metric})  in terms of three angles $\hat\theta$, $\hat\varphi$ and $\hat \psi$, namely:
\bear
\omega^1 & = & \cos\hat\psi\,d\,\hat\theta+\sin\hat\psi\,\sin\hat\theta\,d\hat\varphi\,\,, \rc
\omega^2 & = & \sin\hat\psi\,d\,\hat\theta-\cos\hat\psi\,\sin\hat\theta\,d\hat\varphi\,\,, \rc
\omega^3 & = & d\hat\psi+\cos\hat\theta \,d\hat\varphi\,\,,
\label{w123}
\eear
with $0\le \hat\theta\le \pi$, $0\le\hat\varphi<2\pi$, $0\le\hat\psi \le 4\pi$.  In order to write down a coordinate description of the D6-brane configuration, let us choose the following set of worldvolume coordinates:
\beq
\zeta^{\alpha}\,=\,(x^{\mu}, r, \xi, \hat \psi, \varphi)\,\,.
\eeq
In these coordinates our embedding is defined by the conditions:
\beq
\hat\theta\,\,,\hat\varphi\,\,=\,{\rm constant}\,\,,\qquad\qquad \theta={\pi\over 2}\,\,.
\label{RP3-cycle}
\eeq
Notice that $\xi$ and $\hat\psi$ vary inside the ${\mathbb S}^4$, whereas  $\varphi$ varies inside the ${\mathbb S}^2$.  Actually,  $\xi$ and $\hat\psi$ parametrize an ${\mathbb S}^2\subset {\mathbb S}^4$, while $\theta={\pi\over 2}$ is a maximal ${\mathbb S}^1\subset {\mathbb S}^2$.  The induced worldvolume metric  is:
\beq
ds_7^2\,=\,h^{-{1\over 2}}\,dx_{1,2}^2\,+\,h^{{1\over 2}}\, dr^2\,+\,ds_3^2\,\,,
\label{wvmetric-massless}
\eeq
where the metric  $ds_3^2$   of the three-cycle is given by:
\beq
ds_3^2\,=\,4\,h^{{1\over 2}}\,e^{2g}\,\,\Big[\,q\,\Big(\,
{d\xi^2\over (1+\xi^2)^2}\,+\,
{\xi^2\over 4(1+\xi^2)^2}\,(\,d\hat\psi\,)^2\,\Big)\,+\,{1\over 4}\,
\big(\,d\varphi\,-\,{\xi^2\over 1+\xi^2}\,d\hat\psi\,\big)^2\,\,\Big]\,\,,
\eeq
with $q$ being the squashing factor defined in (\ref{q-def}). 
Let us verify that this three-dimensional metric corresponds to  (a squashed) ${\mathbb R}{\mathbb P}^3=S^3/{\mathbb Z}_2$. We first perform the  following change of variable from $\xi$ to a new angular variable $\alpha$, defined as:
\beq
\xi\,=\,\tan\Big({\alpha\over 2}\Big)\,\,,\qquad\qquad
0\le\alpha <\pi\,\,.
\label{alpha-angle}
\eeq
In terms of $\alpha$  the metric $ds_3^2$  becomes:
\beq
ds_{3}^2\,=\,h^{{1\over 2}}\,e^{2g}\,\,\Big[\, q\,(d\alpha)^2\,+\,q\,\sin^2\alpha \, \Big({d\hat\psi\over 2}\Big)^2\,+\,\big(\,d\varphi\,+\,{1\over 2}\,\big(\,\cos\alpha\,-\,1\,\big)\,d\hat\psi\,\Big)^2\,\Bigg]\,\,.
\label{induced-alpha}
\eeq
Let us  next define new angles  $\beta$ and $\psi$ as:
\beq
\beta\,=\,{\hat\psi\over 2}\,\,,\qquad\qquad
\psi\,=\,\varphi\,-\,{\hat\psi\over 2}\,\,.
\label{RP3-angles}
\eeq
Then, the metric (\ref{induced-alpha}) becomes:
\beq
ds_{3}^2\,=\,h^{{1\over 2}}\,e^{2g}\,\,\Big[\, q\,(d\alpha)^2\,+\,q\,\sin^2\alpha \,(d\beta)^2\,+\, \big(\,d\psi\,+\,\cos\alpha\,d\beta\,\big)^2\,\Big]\,\,.
\label{wv-metric-RP3}
\eeq
It is clear from (\ref{RP3-angles}) that  $0\le \beta < 2\pi$. Moreover, by comparing the volume obtained with the metric (\ref{induced-alpha}) with the one obtained with (\ref{wv-metric-RP3}), one  concludes that $0\le \psi<2\pi$ and that
our three-cycle is indeed a squashed   ${\mathbb R}{\mathbb P}^3$ manifold inside ${\mathbb C}{\mathbb P}^3$.

Let us now verify that the cycle (\ref{RP3-cycle}) preserves some amount of supersymmetry.  First of all, let us note that the embedding (\ref{RP3-cycle}) can be characterized by the three differential conditions:
\beq
{\cal S}^1\,=\,{\cal S}^3\,=\,E^1\,=\,0\,\,,
\label{diff-conditions-cycle}
\eeq
which are integrable due to Frobenius theorem  since $d{\cal S}^1\,=\,d{\cal S}^3\,=\,dE^1\,=\,0$ when (\ref{diff-conditions-cycle}) holds and $\theta=\pi/2$. From this result one can verify that the cycle is calibrated by the form ${\cal K}$ for a metric given by the general form (\ref{metric-ansatz}) and (\ref{7metric-ansatz}). Indeed, the only non-zero component of the pullback of ${\cal K}$ (denoted by $\hat {\cal K}$) is the one containing $e^{3469}$ in (\ref{cal-K-explicit}) and one has:
\beq
\hat {\cal K}\,=\,{2\xi\over (1+\xi^2)^2}\,\,h^{{1\over 4}}\,e^{2f+g}\,
d^3\,x \wedge dr\wedge d\xi\wedge d\psi\wedge d\varphi\,\,,
\eeq
which one can easily show that coincides with the volume form derived from the worldvolume metric (\ref{wvmetric-massless}).  In appendix \ref{Kappa} we have confiormed, by making use of kappa symmetry, that these embeddings preserve the supersymmetry of the background. Actually, in that appendix we generalize the embeddings (\ref{RP3-cycle}) to the case in which $\theta$ is not constant.  If $\theta=\theta(r)$, the supersymmetric configurations are those where the following first-order BPS equation is satisfied:
\beq
e^g\,{d\theta\over dr}\,=\,\cot\theta\,\,.
\label{BPS-theta}
\eeq
Notice that, indeed, the solutions of (\ref{BPS-theta})  with constant $\theta$ must necessarily  have $\theta=\pi/2$. Moreover, one can easily show by analyzing (\ref{BPS-theta}) that, when $\theta$ is not constant, the radial coordinate reaches a minimal value $r_*$ in the corresponding brane embedding. Therefore, one can interpret these D6-brane configurations with varying $\theta$  as flavor branes that add massive flavors to the Chern-Simons-matter theory. The corresponding quark mass is related to the minimal distance $r_*$.  Notice also that one can generate a whole continuous  family of embeddings equivalent to the ones studied so far by acting with the different isometries of the metric. In the next sections we will construct supergravity solutions that incorporate the deformation of the geometry due to the presence of such a continuous set of branes.

\section{Backreacted massless flavor}
\label{unquenched-massless}

Let us now study the backreaction of the flavor branes on the backgrounds of the ABJM type. With this purpose in mind we first analyze the modification of the Bianchi identity introduced by D6-brane sources. This modification is determined by the WZ term of the D6-branes, which for a collection of $N_f$ of them is given by:
\beq
S_{WZ}=\,T_{D_6}\,\,\sum_{i=1}^{N_f}\,\,
\int_{{\cal M}_7^{(i)}}\,\,\hat C_7\,\,,
\eeq
where the hat denotes the pullback to the D6-brane worldvolume.
Let us rewrite this expression in terms of a charge distribution three-form $\Omega$ as:
\beq
S_{WZ}=\,T_{D_6}\,\,\int_{{\cal M}_{10}}\,\,
C_7\wedge \Omega\,\,,
\label{WZ-smeared}
\eeq
with $\Omega$ being only non-vanishing at the location of the D6-branes. 
The term (\ref{WZ-smeared})  induces a violation of the Bianchi identity of $F_2$. Indeed, let us write the supergravity plus branes  action in terms of the $RR$ eight-form $F_8$ and its seven-form potential $C_7$. This action contains a contribution of the form:
\beq
-{1\over 2\kappa_{10}^2}\,\,{1\over 2}\,\,\int_{{\cal M}_{10}}\,F_{8}\wedge {}^*F_{8}\,+\,
T_{D_6}\,\int_{{\cal M}_{10}}\,C_7\wedge \Omega\,\,.
\label{C7-action}
\eeq
It is manifest from (\ref{C7-action}) that the D6-branes act as  a source for the RR seven-form potential $C_7$. Actually,  the equation of motion of $C_7$  derived from (\ref{C7-action}) gives rise to the following Maxwell equation for $F_8$:
\beq
{1\over 2\kappa_{10}^2}\,d{}^*F_{8}\,=\,T_{D_6}\,\,\Omega\,\,,
\eeq
which, as $F_2=^*F_{8}$,  is equivalent, as claimed,  to the violation of the Bianchi identity of $F_2$, namely:
\beq
dF_2\,=\,2\pi\,\,\Omega\,\,,
\label{dF2}
\eeq
where we have used the fact that, in our units, we have $ 2\kappa_{10}^2\,T_{D_6}\,=\,2\pi$. 

Following the general procedure reviewed in \cite{Nunez:2010sf}, we will consider a large number $N_f$ of flavor branes and we will substitute the discrete set of branes by a continuous distribution characterized by the three-form $\Omega$. We will assume that the flavor branes are delocalized in such a way that there are no Dirac $\delta$-functions in the expression of $\Omega$. In this way we will be able to find solutions of the different field equations of supergravity with sources. Actually, it is easy to find an expression of $\Omega$ which preserves both the two supersymmetries and the form of the metric of the deformed unflavored solutions. Notice, however, that the ansatz for $F_2$ must be modified in order to satisfy (\ref{dF2}). In fact, by looking at (\ref{dE1E2}) it is easy to find the appropriate modification of our ansatz (\ref{F2-ansatz}). Indeed, the two-form $F_2$ written in (\ref{F2-ansatz}) is closed because there is a precise balance between the term $E^1\wedge E^2$ (along the fibered ${\mathbb S}^2$ ) and the two other terms with components along the ${\mathbb S}^4$. Clearly, to get a non-closed two-form $F_2$ without distorting much the 
 ${\mathbb S}^2$-${\mathbb S}^4$ structure of our ansatz, one should squash the two type of terms in (\ref{F2-ansatz}) by means of some squashing factor $\eta$. Accordingly, we will adopt the following ansatz for $F_2$ in this flavored case:
\beq
F_2\,=\,{k\over 2}\,\,\Big[\,\,
E^1\wedge E^2\,-\,\eta\,
\big({\cal S}^{\xi}\wedge {\cal S}^{3}\,+\,{\cal S}^1\wedge {\cal S}^{2}\big)
\,\,\Big]\,\,.
\label{F2-flavored}
\eeq
Notice that eq. (\ref{F2-flux}) is still satisfied and, therefore, the constant $k$ continues to be the Chern-Simons level of the gauge theory.  In this section we will consider the case of massless flavors, which corresponds to taking $\eta$ constant (see section \ref{unquenched-massive} for the case of massive flavors). The precise relation between $\eta$ and the number of flavors is obtained below. The violation of 
the Bianchi identity for $F_2$  can be verified by computing the exterior derivative of the two-form written in  (\ref{F2-flavored}) which, in turn, yields the expression of the smearing form $\Omega$, namely:
\beq
\Omega\,=\,{k\over 4\pi}\,\,
\big(\,1-\eta\,\big)\,\Big[\,
E^1\wedge ({\cal S}^{\xi}\wedge {\cal S}^{2}\,-\,{\cal S}^1\wedge {\cal S}^{3}\big)\,+\,
E^2\wedge ({\cal S}^{\xi}\wedge {\cal S}^{1}\,+\,{\cal S}^2\wedge {\cal S}^{3}\big)\,
\Big]\,\,. 
\label{Omega}
\eeq
Let us argue that this is the correct smearing form for a collection of flavor branes, embedded as in section \ref{flavor-quenched}, giving rise to massless flavors . Notice that the smearing form should contain the volume element of the space transverse to the brane worldvolume. In the case of the set of embeddings (\ref{RP3-cycle}) this space should be spanned by the three one-forms ${\cal S}^1$, ${\cal S}^3$ and $E^1$ (see eq. (\ref{diff-conditions-cycle})). Thus, one expects to have a term of the type $E^1\wedge {\cal S}^1\wedge {\cal S}^3$  in $\Omega$, which is indeed contained in our ansatz (\ref{Omega}). It is also easy to find embeddings with $E^2={\cal S}^2={\cal S}^3=0$, which contribute to the $E^2\wedge{\cal S}^2\wedge{\cal S}^3$ component of $\Omega$. Presumably, there are other embeddings generating the other components of the charge-density three-form written in (\ref{Omega}).

In order to relate the squashing coefficient $\eta$ to the number of flavors $N_f$, let us compare the smeared DBI action with the DBI action of a single brane. The former is given by:
\beq
S_{DBI}^{smeared}\,=\,-T_{D_6}\,\int_{{\cal M}_{10}}\,e^{-\phi}\,
{\cal K}\wedge \Omega\,\,,
\eeq
where we have taken into account that ${\cal K}$ is a calibration form for the D6-brane worldvolume. By using the explicit expression of ${\cal K}$ (eq. (\ref{cal-K-explicit})) and our ansatz for $\Omega$, we get:
\beq
{\cal K}\wedge \Omega\,=\,{k(1-\eta)\over \pi}\,h^{{1\over 4}}\,e^{2f+g}\,
d^3 x\wedge dr\wedge {\rm Vol}({\mathbb S}^4)\wedge {\rm Vol}({\mathbb S}^2)\,\,,
\eeq
with ${\rm Vol}({\mathbb S}^4)$ being the volume form of the metric (\ref{S4metric}) and
${\rm Vol}({\mathbb S}^2)=\sin\theta\,d\theta\wedge d\varphi$. Integrating over ${\mathbb S}^4$ and ${\mathbb S}^2$ gives a factor $32\pi^3/3$ and we can write the smeared DBI action  as:
\beq
S_{DBI}^{smeared}\,=\,\int \,d^3x\,dr\,{\cal L}_{DBI}^{smeared}\,\,,
\label{smeared-action}
\eeq
where the DBI lagrangian density of the smeared set of flavor branes is:
\beq
{\cal L}_{DBI}^{smeared}\,=\,
-{32\pi^2\,k\,(1-\eta)\,\over 3}\,\,T_{D_6}\,e^{-\phi}\,
h^{{1\over 4}}\,e^{2f+g}\,\,.
\label{Lsmeared}
\eeq
Let us now compare the action (\ref{smeared-action})  for the whole set of smeared branes with the one corresponding to a single representative brane, which we will choose to be the one written in (\ref{RP3-cycle}). In terms of the angular coordinates 
defined in (\ref{alpha-angle}) and (\ref{RP3-angles}) the DBI action for the embedding can be written as:
\beq
S_{DBI}^{single}\,=\,-T_{D_6}\,\int\,d^3x\,dr\,d\alpha\,d\beta\,d\psi\,e^{-\phi}\,
\sqrt{-\det g_7}\,
\equiv\,\int \,d^3x\,dr\,{\cal L}_{DBI}^{single}\,\,,
\eeq
where $g_7$ is the induced metric written in (\ref{wvmetric-massless}) and (\ref{wv-metric-RP3}). By integrating over the angular variables one easily gets the effective lagrangian density for the representative embedding, namely:
\beq
{\cal L}_{DBI}^{single}\,=\,-8\pi^2\,T_{D_6}\,e^{-\phi}\,
h^{{1\over 4}}\,e^{2f+g}\,\,.
\label{Lsingle}
\eeq
Since all the embeddings of the family of D6-branes are related by isometries, they are equivalent and they should give the same action. Thus  we should have:
\beq
{\cal L}_{DBI}^{smeared}\,=\,N_f\,{\cal L}_{DBI}^{single}\,\,.
\label{micro-macro}
\eeq
It is now straightforward to use the lagrangian densities (\ref{Lsmeared}) and (\ref{Lsingle}) in (\ref{micro-macro}) to find the precise relation between the squashing factor $\eta$ and the number of flavors $N_f$. One gets:
\beq
\eta\,=\,1\,+\,{3N_f\over 4k}\,\,.
\label{etaNf-k}
\eeq
Notice that $\eta$ depends linearly on the deformation parameter $\epsilon$, defined as:
\beq
\epsilon\,\equiv\,{N_f\over k}\,\,.
\label{epsilon-def}
\eeq
Indeed, it follows from (\ref{etaNf-k}) that $\eta=1+3\epsilon/4$. Interestingly, 
$\epsilon$ can be rewritten in terms of gauge theory quantities  as:
\beq
\epsilon\,=\,{N_f\over N}\,\lambda\,\,,
\eeq
where $\lambda\,=\,N/k$ is the 't Hooft coupling of the Chern-Simons-matter theory. As we will show below, the deformations of the metric, dilaton and RR four-form will also depend on $\epsilon$, similarly to what happens in other flavored backgrounds such as the D3-D7 system (see \cite{Nunez:2010sf} for a review and further references).

\subsection{Flavored BPS system} 
\label{AdS-flavored}
We are now in the position of addressing the central problem of this paper, namely finding the backgrounds dual to Chern-Simons-matter theories with flavors. We will adopt an ansatz for the metric as the one written in (\ref{metric-ansatz}) and (\ref{7metric-ansatz}), in which the line element is parametrized by a warp factor $h$ and two squashing functions $f$ and $g$.  Moreover, the RR four-form $F_4$ will depend on the function $K$ as in (\ref{F4-ansatz}), while $F_2$ will be given by (\ref{F2-flavored}). By imposing on the Killing spinors the projection conditions (\ref{full-projections}), one can find a system of first-order BPS equations for the different functions of the ansatz. Actually, these  BPS equations imposed by supersymmetry are readily obtained from (\ref{BPS-full-r}) by performing the substitution:
\beq
k\,e^{-2f}\,\,\to\,\,k\,\eta\,e^{-2f}\,\,.
\eeq
In this way, one gets:
\bear
&&\phi'\,=\,-{3k\over 8}\,e^{\phi}\,h^{-{1\over 4}}\,\big(\,
e^{-2g}-2\eta\,e^{-2f}\,\big)\,-\,{e^{\phi}\over 4}\,K\,h^{{3\over 4}}\,\,,\rc\rc
&&h'\,=\,{k\over 2}\,e^{\phi}\, h^{{3\over 4}}\, \big(\,e^{-2g}\,-\,2\eta\,e^{-2f}\,\big)\,-\,
e^{\phi}\,K\,h^{{7\over 4}}\,\,,\rc\rc
&&f'\,=\,{k\over 4} \,h^{-{1\over 4}}\,e^{\phi}\,\big[\,\eta\,e^{-2f}\,-\,e^{-2g}\,\big]
\,+\,e^{-2f+g}\,\,,\rc\rc
&&g'\,=\,{k\over 2}\,e^{\phi}\,h^{-{1\over 4}}\, \eta\,e^{-2f}\,+\,
e^{-g}\,-\,e^{-2f+g}\,\,.
\label{BPS-flavored}
\eear
The fulfillment of the system (\ref{BPS-flavored}) guarantees the preservation of two supercharges, which corresponds to ${\cal N}=1$ supersymmetry in three dimensions. Moreover, one can show that the BPS system can be rewritten as in (\ref{calibration-conds}), in terms of the calibration forms ${\cal K}$ and $\tilde {\cal K}$, which are written in (\ref{cal-K-explicit}) and (\ref{cal-tildeK-explicit}) in the frame basis. Furthermore, in appendix \ref{eoms} we show that (\ref{BPS-flavored}) implies the fulfillment of the equations of motion of ten-dimensional type IIA supergravity  with sources corresponding  to delocalized branes. 

The system (\ref{BPS-flavored}) is very similar to the unflavored one in (\ref{BPS-full-r}).  Therefore, we can follow the same procedure as in section \ref{deformed ABJM} to perform a partial integration of the system of differential equations and to find some of its particular solutions. First of all, we define the function $\Lambda$ as in (\ref{Lambda-def}), in terms of which the system (\ref{BPS-flavored}) becomes:
\bear
&&\Lambda'\,=\,k\,\eta\,e^{\Lambda-2f}\,-\,
{k\over 2}\,e^{\Lambda-2g}\,\,,\rc\rc
&&f'\,=\,{k\,\eta\over 4}\,e^{\Lambda-2f}\,-\,
{k\over 4}\,e^{\Lambda-2g}\,+\,e^{-2f+g}\,\,,\rc\rc
&&g'\,=\,{k\,\eta\over 2}\,e^{\Lambda-2f}\,+\,e^{-g}\,-\,e^{-2f+g}\,\,.
\label{3-system-flavored}
\eear
As in (\ref{3-system}), the function $K$ does not appear anymore in the system (\ref{3-system-flavored}) and can be  determined in terms of the functions appearing in (\ref{3-system-flavored}). Indeed, one can immediately show that $K$ can be obtained from $\phi$ and $h$ by means of the expression written in (\ref{K-phi-h}). Moreover, eqs. (\ref{h-beta}) and (\ref{h-dif-eq}) also hold in this unflavored case and one can integrate the warp factor in terms of $\Lambda$ as in (\ref{warp-integral}). 

The system (\ref{3-system-flavored}) can be further reduced by introducing the new radial variable $\tau$ defined in (\ref{r-tau}) and by defining the functions $\Sigma$ and $\Lambda$ as in (\ref{Sigma-Delta-def}). The 
resulting system of  equations is:
\bear
&&\dot\Sigma\,=\,{k\over 4} e^{\Sigma}\,\Big(\,3
\eta-\,e^{2\Delta}\,\Big)
\,-\,e^{-\Delta}\,\,,
\rc\rc
&&\dot \Delta\,=\,-{k\over 4}\,e^{\Sigma}\,
\Big(\,\eta+\,e^{2\Delta}\,\Big)\,-\,e^{\Delta}\,+2\,e^{-\Delta}\,\,.
\label{Sigma-Delta-sys-flav}
\eear
In the next section we will concentrate on studying a particular solution of the system 
(\ref{Sigma-Delta-sys-flav}) which leads to the Anti-de-Sitter geometry in this flavored case.

\section{Flavored Anti-de-Sitter solutions}
\label{flavored-AdS}

In close analogy with the study carried out in section \ref{AdS-unflavored}, let us consider solutions of the reduced  system (\ref{Sigma-Delta-sys-flav}) in which the functions $\Sigma$ and $\Delta$ are constant. Notice that, according to the definition in (\ref{q-def}), constant $\Delta$ implies that the squashing parameter $q=e^{2\Delta}$ is also constant. Actually, by imposing $\dot \Sigma=\dot\Delta=0$ in (\ref{Sigma-Delta-sys-flav}) one can straightforwardly prove that $q$ must satisfy the following quadratic equation:
\beq
q^2\,-\,3(1+\eta)\,q\,+5\eta\,=\,0\,\,,
\label{q-eq-flav}
\eeq
which reduces to (\ref{q-eq}) when $\eta=1$. The relation  (\ref{q-eq-flav}) can be regarded as the relation between the deformation $\eta$  of the RR two-form  and the internal deformation of the  ${\mathbb C}{\mathbb P}^3$  metric, parametrized by the squashing factor $q$. Both parameters are related to the number of flavors or, to be more precise, to the deformation parameter $\epsilon$ defined in (\ref{epsilon-def}). Actually, by solving  (\ref{q-eq-flav})  as a quadratic equation in $q$, one gets:
\beq
q\,=\,{3(1+\eta)\pm \sqrt{9\eta^2-2\eta+9}\over 2}\,\,.
\label{q-sol-flav}
\eeq
By using (\ref{etaNf-k}) one can obtain the squashing factor $q$ in terms of $N_f$ and $k$, namely:
\beq
q\,=\,3+{9\over 8}\,\,{N_f\over k}\,\pm\,2\sqrt{1+{3\over 4}\,{N_f\over k}\,+\,
\Big({3\over 4}\Big)^4\,\Big({N_f\over k}\Big)^2}\,\,.
\label{q-Nf}
\eeq
The two signs in (\ref{q-sol-flav})  give rise to the two possible branches. 
The minus sign in (\ref{q-sol-flav}) corresponds to the flavored ABJM model (it reduces to $q=1$ when $\eta=1$), while the plus sign corresponds to the squashed model with $SO(5)$ global symmetry which is discussed in appendix \ref{squashed-unflavored}. 
 Notice that  the discriminant in (\ref{q-sol-flav}) is never negative and, therefore, the parameter $\eta$ can be arbitrary. Actually, when $\eta\to\infty$ one has the following behavior in the two branches:
\beq
\lim_{\eta\to\infty}\,q\,=\,
\begin{cases}5/3\,\,,&{\rm for\, the\,} $-$ \,{\rm  branch}\,\,,\cr\cr
\infty\,\,,&{\rm for\, the}\, $+$\,{\rm  branch}\,\,.
\end{cases}
\label{q-largeNf}
\eeq
Similarly, one can compute the squashing factor for the case in which $N_f$ is small. At second  order in $N_f/k$, one gets:
\beq
q\,\approx\,
\begin{cases}1+{3\over 8}\,{N_f\over k}-{45\over 256}\,\big({N_f\over k}\big)^2\,\,,&
\qquad{\rm for\, the\,} $-$ \,{\rm  branch}\,\,,\cr\cr
5+{15\over 8}\,{N_f\over k}\,+\,{45\over 256}\,\big({N_f\over k}\big)^2\,\,,&
\qquad{\rm for\, the}\, $+$\,{\rm  branch}\,\,.
\end{cases}
\label{q-smallNf}
\eeq
It is interesting to point out that, in the two branches in (\ref{q-Nf}), the squashing factor $q$ takes values in ranges that are disjoint. In the flavored ABJM case $1\le q\le 5/3$, whereas $q\ge 5$ in the other branch. In what follows we will restrict ourselves to the ABJM model with flavor and thus $q$ should be understood as the right-hand-side of (\ref{q-Nf}) with the minus sign. Let us write the complete supergravity solution in this case. First of all,   it follows from the system (\ref{Sigma-Delta-sys-flav}) that
$\Sigma$ is given by:
\beq
{k\over 2}\,e^{\Sigma}\,=\,{2\over\sqrt q}\,\,{2-q\over \eta+ q}\,\,.
\label{sigma-flavored}
\eeq
Moreover,  the equation for $g$ in (\ref{3-system-flavored}) can be rewritten as:
\beq
{d\over dr}\,\,\big(\,e^{g}\,\big)\,=\,{1\over \sqrt{q}}\,\,
\Big[\,{k\eta \over 2}\,e^{\Sigma}\,+\,\sqrt{q}\,-\,{1\over \sqrt{q}}\,\Big]\,\,.
\label{eg-derivative}
\eeq
Since the right-hand side of (\ref{eg-derivative})  is constant, it follows that $e^{g}$ is a linear function of $r$. Actually, by making use of (\ref{sigma-flavored}) one can prove that $g$ and $f$ can be written as:
\beq
e^{g}\,=\,{r\over b}\,\,,\,\qquad\qquad
e^{f}\,=\,{\sqrt{q}\over b}\,\,r\,\,,
\label{f-g-flavored}
\eeq
with the coefficient $b$ being given in terms of the squashing parameters $\eta$ and $q$ by:
\beq
b\,=\,{q(\eta+q)\over 2(q+\eta q-\eta)}\,\,.
\label{b-squashing}
\eeq
Let us next compute the  warp factor $h$ by using the general expression (\ref{warp-integral}). A glance at the right-hand side of (\ref{warp-integral}) reveals that we have to compute the function $\Lambda$ first. However, it follows  from the definition (\ref{Sigma-Delta-def})  that  $e^{\Lambda}=e^{\Sigma}\,e^{f}$ and, therefore, we can obtain $e^{\Lambda}$ from (\ref{sigma-flavored}) and (\ref{f-g-flavored}). One gets:
\beq
e^{\Lambda}\,=\,{8\over k}\,\,{2-q\over q\, (\eta+ q)^2}\,\,
\big[\,q+\eta q-\eta\,]\,r\,\,.
\eeq
Taking into account this result,  we can write the warp factor $h$ as:
\beq
h\,=\,{\beta \over 24 \,k}\,\,
{q^3\,(\eta+\ q)^4\,(2-q)\over 
\big[\,q+\eta q-\eta\,]^5}\,\,
{1\over r^4}\,\,,
\label{warp-flavored}
\eeq
where we have adjusted the integration constant in (\ref{warp-integral}) by requiring that $h$ vanishes at $r\to\infty$. Notice that, as $h\sim r^{-4}$, this solution does indeed lead to an Anti-de-Sitter metric.  Actually, the $AdS_4$ radius $L$ is related to the warp factor $h$ as:
\beq
L^4\,=\,r^4\,\,h\,\,.
\label{RAdS-h}
\eeq
From (\ref{warp-flavored}) we can extract the expression of $L$,  which can be written as:
\beq
L^4\,=\,2\pi^2\,\,{N\over  k }\,\,
{(2-q)\,b^4\over 
q(q+\eta q\,-\,\eta)}\,\,.
\label{radii}
\eeq
Using these results we can represent the ten-dimensional metric for this solution as:
\beq
ds^2\,=\,L^2\,\,ds^2_{AdS_4}\,+\,
{L^2\over b^2}\,\Big[\,q\,ds^2_{{\mathbb S}^4}\,+\,
(E^1)^2\,+\,(E^2)^2\,\Big]\,\,.
\label{metric-AdS-flavored}
\eeq
In order to write the $AdS_4$ part of the metric as in (\ref{AdS4metric}) we have to rescale the Minkowski coordinates as $x^{\mu}\to L^2\, x^{\mu}$, where $L$ is the same as in (\ref{radii}).  
From (\ref{metric-AdS-flavored}) the interpretation of the parameter $b$ is rather clear: it represents the relative squashing, due to the flavor,  of the ${\mathbb C}{\mathbb P}^3$ part of the metric with respect to the $AdS_4$ part. It is also interesting to
 rewrite the metric  (\ref{metric-AdS-flavored}) in terms of the variables used in (\ref{S4metric}). One has:
\beq
ds^2\,=\,L^2\,\,ds^2_{AdS_4}\,+\,ds^2_{6}\,\,,
\label{flavored-metric-total}
\eeq
where $ds^2_{6}$ is the metric of the deformed  internal six-dimensional manifold,  written in terms of the $SU(2)$ instanton on the four-sphere, as:
\beq
ds^2_{6}\,=\,{L^2\over b^2}\,\,\Big[\,
q\,ds^2_{{\mathbb S}^4}\,+\,\big(d x^i\,+\, \epsilon^{ijk}\,A^j\,x^k\,\big)^2\,\Big]\,\,.
\label{internal-metric-flavored}
\eeq
Let us now obtain the remaining non-vanishing fields for this solution. First of all, the
constant dilaton can be found by using (\ref{Lambda-def}). One gets:
\beq
e^{\phi}\,=\,{4\sqrt{\pi}\over \eta+q}\,\,
{(2-q)^{{5\over 4}}\over \big[q(q+\eta q\,-\,\eta)\big]^{{1\over 4}}}\,\,
\Bigg({2N\over k^5}\Bigg)^{{1\over 4}}\,\,.
\label{dilaton-flavored}
\eeq
This expression can be rewritten in a more compact form as:
\beq
e^{-\phi}\,=\,{b\over 4}\,{\eta+q\over 2-q}\,{k\over L}\,\,.
\label{dilaton-flavored-squashings}
\eeq
Finally, let us write the four-form $F_4$ for these solutions. After rescaling the Minkowski coordinates as before, 
we can write $F_4$ as proportional to the volume element $\Omega_{AdS_4}$ of $AdS_4$, namely:
\beq
F_4\,=\,{L^6\,K\over r^2}\,\,
\,\Omega_{AdS_4}\,\,.
\eeq
The function $K$ was written in (\ref{K-beta}) in terms of $h$, $f$ and $g$. Taking into account that $e^{2f}=q\, e^{2g}$ and the relation (\ref{RAdS-h}), one can rewrite  $F_4$ as:
\beq
F_4\,=\,{\beta\over q^2}\,\,{r^6\,e^{-6g}\over L^2}\,\,\,\Omega_{AdS_4}
\,\,.
\eeq
Using the expressions of $g$ and $L$ in (\ref{f-g-flavored}) and (\ref{radii}) and the one for $\beta$ written in (\ref{beta}), we arrive at:
\beq
F_4\,=\,{3\pi\over 16\sqrt{2}}\,\,{q^{{5\over 2}}\,\,(\eta+q)^{4}\over 
(q+\eta q\,-\,\eta)^{{7\over 2}}\,\,\sqrt{2-q}}\,\,
\sqrt{k\,N}\,\,\,\Omega_{AdS_4}\,\,.
\label{F4-flavored}
\eeq
We can rewrite this result more compactly as:
\beq
F_4\,=\,{3k\over 4}\,\,\,{(\eta+q)b\over 2-q}\,\,L^2\,\,\Omega_{AdS_4}\,\,.
\eeq
Eqs. (\ref{flavored-metric-total}) and (\ref{internal-metric-flavored}) are the flavored generalization of the ABJM metric written in (\ref{ABJM-metric}) and (\ref{CP3=S4-S2}). 
Notice that the radius $L$ is not the same in both cases (compare (\ref{ABJM-AdSradius}) and (\ref{radii})) and, in addition, the flavored metric is deformed by the parameters $b$ and $q$. The RR two-form $F_2$ for the flavored solution was written in (\ref{F2-flavored}) (it was our starting point) and, together with the $F_4$ written in (\ref{F4-flavored}), generalize (\ref{F2-F4-ABJM}). Finally, the constant dilaton also gets corrected by the effect of the matter fields, as one can see by comparing eqs. (\ref{dilaton-flavored}) and (\ref{ABJMdilaton}).  

Let us finish this section by discussing the regime of validity of our supergravity dual. On general grounds we must require that the curvature of the space is small in string units (or, equivalently, that the curvature radius is large) and that the string coupling $e^{\phi}$  is small (otherwise we should describe the system in eleven-dimensional supergravity). Thus, the two conditions that make our type IIA supergravity approximation valid are:
\beq
L>>1\,\,,
\qquad\qquad\qquad
e^{\phi}<<1\,\,.
\label{regimen-general}
\eeq
Let us analyze the two conditions in (\ref{regimen-general}) in two different regimes of the  deformation parameter $\epsilon=N_f/k$. If $\epsilon$ is of the order one or less, the squashing parameters are also of this same order and they do not modify the order of magnitude of $L$ and $e^{\phi}$. Therefore, (\ref{regimen-general}) leads to the same conditions as in the unflavored case, namely:
\beq
N^{{1\over 5}}<<k<<N\,\,,
\label{small-epsilon-condition}
\eeq
with $N_f$ being, at most,  of the order of the Chern-Simons level $k$.  Let us consider next the opposite limit, namely when $N_f>>k$. In this case, as $q$ remains finite (see (\ref{q-largeNf})) and $\eta$ is large,  it follows from (\ref{radii}) that:
\beq
L^4\approx 2\pi^2 \,{N\over k}\,\,
{(2-q)\,b^4\over q(q-1)}\,\,{1\over \eta}\,\,.
\eeq
Moreover, since $\eta\sim{N_f\over k}$ in this limit, one has:
\beq
L^4\sim {N\over N_f}\,\,.
\label{L-largeNf}
\eeq
Similarly, for $N_f>>k$ the dilaton behaves as:
\beq
e^{\phi}\approx 4\sqrt{\pi}\,
\Big({2N\over k^5}\Big)^{{1\over 4}}\,\,
{(2-q)^{{5\over 4}}\over q^{{1\over 4}}
(q-1)^{{1\over 4}}}\,\,\eta^{-{5\over 4}}
\sim \Big({N\over N_f^5}\Big)^{{1\over 4}}\,\,.
\label{dilaton-largeNf}
\eeq
Thus, the conditions (\ref{regimen-general}) for $N_f>>k$ are equivalent to:
\beq
N^{{1\over 5}}<< N_f<<N\,\,.
\label{large-epsilon-condition}
\eeq
Notice that conditions similar to (\ref{small-epsilon-condition}) and (\ref{large-epsilon-condition}) were found in ref. \cite{Gaiotto:2009tk} for general tri-Sasakian manifolds. 

Let us now discuss the regime of validity of the DBI+WZ action used to describe the flavor branes \footnote{We are grateful to A. Cotrone for discussions on what follows. }. In principle the DBI action is considered to be valid when $g_s\,N_f$ is small \cite{Callan:1988wz}. Indeed, $g_s\,N_f$ is the effective coupling for the process in which an open string ends on the $N_f$ branes.  However, as argued in refs. \cite{Bigazzi:2008zt,HoyosBadajoz:2008fw}, when the flavor branes are smeared the situation is more subtle and the effective coupling $g_s\,N_f$ is further suppressed due to the fact that the branes are separated a large distance in $\sqrt{\alpha'}$ units and only a small fraction of the $N_f$ branes will be available for an open string process. In general, if $R$ denotes the typical radius of an internal dimension of the geometry in $\sqrt{\alpha'}$ units, the number of flavor branes involved in a typical process will be of order $N_f/R^d$, where $d$ is the codimension of the flavor  branes in the internal space. Thus, we should require that  $g_s\,N_f/R^d$  be small.  In our case $d=3$ and $R$ is just the radius $L$. Therefore, we should require that $e^{\phi}\,N_f\,/L^3$ be small.  When $\epsilon=N_f/k$ is small this condition is satisfied if (\ref{small-epsilon-condition}) holds since  $e^{\phi}\,N_f\,/L^3\sim\epsilon \sqrt{k / N}$ in this case. In the  opposite  large $\epsilon$  regime, $L$ and $e^{\phi}$ behave as in (\ref{L-largeNf}) and (\ref{dilaton-largeNf}) respectively and one has:
\beq
{e^{\phi} N_f\over L^3}\sim \sqrt{N_f\over N}\,\,.
\eeq
Thus, we should require that $N_f<<N$, which is the same condition obtained by imposing that the curvature is small in string units. Notice also that the typical separation scale between two D6-branes is:
\beq
D\sim {L\over N_f^{{1\over 3}}}\,\,.
\eeq
For consistency we should require that the distribution of branes is dense enough to be described by a continuos charge density. This condition amounts to require that 
$D<<L$, which is clearly satisfied if $N_f$ is large. On the other hand, in accordance with our discussion above, we should also require that $D$ must be greater than 1 (in $\sqrt{\alpha'}$ units) which, for large $N_f$,  leads to the condition $N_f<< N^{{3\over 7}}$. Notice that this requirement is more restrictive than the one written in (\ref{large-epsilon-condition}).

\section{Some flavor effects}
\label{Flavor-effects}

Having a simple supergravity solution with no evident pathologies for the unquenched flavor gives us a great opportunity to explore the effects of dynamical matter in several observables. Moreover, since our solution is an Anti-de-Sitter background, we have at our disposal several techniques and holographic prescriptions to evaluate the observables of the unquenched theory in a neat form. Furthermore, some of these observables can also be computed for the localized flavor solutions, which gives us a unique chance to compare with our results and to explore the effects of the smearing technique. In the next subsections we will analyze these flavor effects for some of these observables.  We will show that,  although the two setups (localized and smeared) have different amount of supersymmetry and flavor group, the results are very similar and, in the case of some observable such as the free energy, they are amazingly close. This is an indication of the fact that the ${\cal N}=3\to{\cal N}=1$ breaking introduced by the smearing is, in fact, rather mild.

\subsection{Free energy on the sphere}
Let us consider the euclidean version of the conformal field theory formulated in a three-sphere. The corresponding free energy is given by:
\beq
F({\mathbb S}^3)\,=\,-\log |\,Z_{{\mathbb S}^3}\,|\,\,,
\eeq
where $Z_{{\mathbb S}^3}$ is the euclidean path integral. The holographic calculation of this quantity  in $AdS_4$ gives \cite{Emparan:1999pm}:
\beq
F({\mathbb S}^3)\,=\,{\pi L^2\over 2 G_N}\,\,,
\eeq
where $L$ is the $AdS_4$ radius and $G_N$ is the effective four-dimensional Newton constant. In our case,  $G_N$ is related to the ten-dimensional Newton constant $G_{10}$ by means of the equation:
\beq
{1\over G_N}\,=\,{1\over G_{10}}\,e^{-2\phi}\,\,{\rm Vol}({\cal M}_6)\,\,,
\eeq
where ${\cal M}_6$ is the internal manifold and ${\rm Vol}({\cal M}_6)$ its volume. For our flavored Anti-de-Sitter solutions this volume can be readily computed from the metric of 
${\cal M}_6$ written in (\ref{internal-metric-flavored}) and is equal to:
\beq
{\rm  Vol}({\cal M}_6)\,=\,{32\pi^3\over 3}\,\,{q^2 L^6\over b^6}\,\,.
\eeq
Taking into account that, in our units, the ten-dimensional Newton constant $G_{10}$ is given by $G_{10}\,=\,8\pi^6$, and using the value of the dilaton for our solution (eq. (\ref{dilaton-flavored-squashings})), we get:
\beq
F({\mathbb S}^3)\,=\,{k^2\over 24 \pi^2}\,\,{q^2 (\eta+q)^2\over (2-q)^2\,b^4}\,\,L^6\,\,.
\eeq
Using the value of the AdS radius $L$ for our geometry written in (\ref{radii}), we get that $ F({\mathbb S}^3)$ can be represented as:
\beq
F({\mathbb S}^3)\,=\,{\pi\sqrt{2}\over 3}\,\,k^{{1\over 2}}\,\,N^{{3\over 2}}\,
\xi\Big({N_f\over k}\Big)\,\,,
\label{F-flavored}
\eeq
where $\xi\Big({N_f\over k}\Big)$ is given by:
\beq
\xi\Big({N_f\over k}\Big)\equiv{1\over 16}\,\,
{q^{{5\over 2}}\,(\eta+q)^4\over (2-q)^{{1\over 2}}\,\,(q+\eta q\,-\,\eta)^{{7\over 2}}}\,\,.
\label{xi-smeared}
\eeq
Notice that $\xi=1$ for the unflavored case and we recover the ABJM result, namely:
\beq
F_{ABJM}({\mathbb S}^3)\,=\,{\pi\sqrt{2}\over 3}\,\,k^{{1\over 2}}\,\,N^{{3\over 2}}\,=\,
{\pi\sqrt{2}\over 3}\,{N^2\over \sqrt{\lambda}}\,\,,
\eeq
where, in the last step, we have written the result in terms of the 't Hooft coupling $\lambda=N/k$. For small values of $N_f/k$ we can expand $\xi$ as:
\beq
\xi\,=\,1\,+\,{3\over 4}\,{N_f\over k}\,-\,{9\over 64}\,\Big(\,{N_f\over k}\,\Big)^2\,+\,
{\cal O}\Big(\Big(\,{N_f\over k}\,\Big)^3\,\Big)\,\,.
\label{xi-smeared-flavored}
\eeq
Thus, the free energy of the flavored theory can be expanded in powers of $N_f/k$ as:
\beq
F({\mathbb S}^3)\,=\,{\pi\sqrt{2}\over 3}\,{N^2\over \sqrt{\lambda}}\,+\,
{\pi\sqrt{2}\over 4}\,N_f\,N\,\sqrt{\lambda}\,-\,{3\pi\sqrt{2}\over 64}\,\,
N_f^2\,\lambda^{{3\over 2}}\,+\,\cdots\,\,.
\eeq
If, on the contrary,  $N_f/k$  is large, one can verify that, at leading order, $\xi$ behaves as:
\beq
\xi\sim {225\over 256}\,\sqrt{{5\over 2}}\,\sqrt{{N_f\over k}}\,\,\approx
1.389\,\sqrt{{N_f\over k}}\,\,,
\label{xi-large-epsilon}
\eeq
and, therefore, the free energy in this large $N_f$ case becomes:
\beq
F({\mathbb S}^3)\sim {75\sqrt{5}\,\pi\over 256}\,\,\,
N_f^{{1\over 2}}\,\,N^{{3\over 2}}\,\,.
\eeq

\begin{figure}[ht]
\begin{center}
\includegraphics[width=0.7\textwidth]{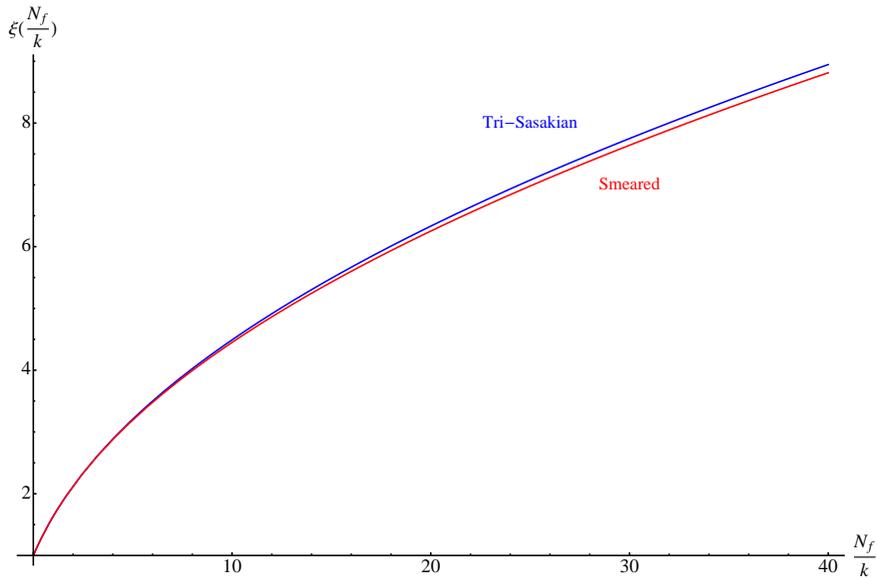}
\end{center}
\caption[xi]{ Comparison of the flavor correction factor $\xi$ obtained with our smeared setup (lower red curve) and the one corresponding to the tri-Sasakian geometry (upper blue curve). \label{xi}} 
\end{figure}

Let us next compare our results with the ones obtained with the ${\cal N}=3$ tri-Sasakian geometry that corresponds to a localized stack of flavor D6-branes. In M-theory these geometries are obtained as the base $X_7({\bf t})$  of a hyperk\"ahler cone ${\cal M}_8 ({\bf t})$, labelled by three natural numbers ${\bf t}=(t_1, t_2, t_3)$. These cones are obtained as  hyperk\"ahler quotients of the form  ${\mathbb H}^3///U(1)$, where the $U(1)$ action is characterized by the three charges ${\bf t}$. The dual  to the ${\cal N}=3$ Chern-Simons-matter theories with $N_f$ fundamentals has charges ${\bf t}=(N_f, N_f, k)$ (see \cite{Gaiotto:2009tk}). The volume of $X_7({\bf t})$ has been computed in \cite{Lee:2006ys} by using localization techniques. From this result one can obtain the corresponding free energy $F({\mathbb S}^3)$ 
\cite{Gaiotto:2009tk}, which matches the matrix model  field theory calculation \cite{Santamaria:2010dm} . One gets an expression like (\ref{F-flavored}) with a flavor correction factor $\xi$  simply given by:
\beq
\xi^{3-S}\,=\,{1+{N_f\over k}\over \sqrt{1+{N_f\over 2k}}}\,\,.
\label{xi-3S}
\eeq
Let us expand $\xi^{3-S}$  in powers of $N_f/k$, namely:
\beq
\xi^{3-S}\,=\,1\,+\,{3\over 4}\,{N_f\over k}\,-\,{5\over 32}\,\Big(\,{N_f\over k}\,\Big)^2\,+\,
{\cal O}\Big(\Big(\,{N_f\over k}\,\Big)^3\,\Big)\,\,.
\eeq
This result is indeed very similar to our result (\ref{xi-smeared-flavored}). For large  $N_f/k$ one has:
\beq
\xi^{3-S}\sim \sqrt{2}\,\sqrt{{N_f\over k}}\,\,,
\eeq
which  is again amazingly close to the value we have found in (\ref{xi-large-epsilon}). 
Actually, one can plot together the two functions $\xi$ written in (\ref{xi-smeared}) and (\ref{xi-3S}) and check that the two curves are, indeed, almost identical (see figure \ref{xi}). 

\subsection{Wilson loops and quark-antiquark potentials}

Most of the results derived for Wilson loops in ${\cal N}=4$ super-Yang-Mills in  four dimensions can be adapted to our flavored setup. To illustrate this fact let us consider the calculation of the quark-antiquark static potential. Due to conformal invariance, the quark-antiquark potential must be Coulombic, namely of the form:
\beq
V_{q\bar q}\,=\,-{Q\over d}\,\,,
\label{q-barq-pot}
\eeq
where $d$ is the distance between the quark and the antiquark and the coefficient $Q$ measures the strength of the Coulombic potential. The holographic calculation of $Q$ is just the same as the one performed in \cite{Wilson} and yields the result:
\beq
Q\,=\,{4\pi^2 L^2\over 
\big[\,\Gamma\big(\,{1\over 4}\,\big)\,\big]^4}\,\,.
\eeq
Using the value of the AdS radius $L$ (see (\ref{radii})), one can write $Q$ as:
\beq
Q\,=\,{4\pi^3 \sqrt{2\lambda}\over 
\big[\,\Gamma\big(\,{1\over 4}\,\big)\,\big]^4}\,
\sigma\,\,,
\eeq
where $\sigma$ parametrizes the screening effects due to dynamical quarks, and is given by:
\beq
\sigma\,=\,\sqrt{
{2-q\over q(q+\eta q-\eta)}}\,\,\,\,b^2\,=\,
{1\over 4}\,\,
{q^{{3\over 2}}\,\,(\eta+q)^2\,(2-q)^{{1\over 2}}
\over
(q+\eta q-\eta)^{{5\over 2}}}\,\,.
\label{screening-sigma}
\eeq
For small $N_f/k$, one can expand $\sigma$ as:
\beq
\sigma\,=\,1\,-\,{3\over 8}\,{N_f\over k}\,+\,{9\over 64}\,\,\Big({N_f\over k}\Big)^2\,+\,\cdots\,\,.
\label{sigma-small-Nf}
\eeq
Clearly, the fact that the first correction is negative means that
the screening makes the Coulombic attraction  between the quark and the antiquark smaller, as expected on physical grounds.  In the opposite limit, when $N_f/k$ is large,  $\sigma$ is small. Actually, one gets:
\beq
\sigma\to
{1\over 4}\,{q^{{3\over 2}}\,(2-q)^{{1\over 2}}\over
(q-1)^{{5\over 2}}}\,\,\eta^{-{1\over 2}}\,=\,
\sqrt{{125\over 128}}\,\,\sqrt{{k\over N_f}}\,\,,
\qquad\qquad
{\rm for\, \,\,}\,  \,\,{N_f\over k}>> 1\,\,.
\label{sigma-largeNf}
\eeq
Similarly, the calculation of the circular Wilson loop can be done by applying the same techniques as in the $AdS_5\times S^5$  background (see \cite{Drukker:2008zx}). The result depends on the $AdS$ radius $L$, namely:
\beq
<\,W\,>\,\sim\,e^{L^2}\,\,.
\eeq
Using our value of $L$, we get:
\beq
<\,W\,>\,\sim\,e^{\pi\sqrt{2\lambda}\,\,\sigma}\,\,,
\eeq
and, as before,  the screening effects are encoded in $\sigma$\,\,. It is interesting to compare again our results with the ones found in refs. \cite{Gaiotto:2009tk,Santamaria:2010dm} from the tri-Sasakian geometry. The corresponding screening factor $\sigma$ is given by:
\beq
\sigma^{3-S}\,=\,{1\over \sqrt{1+{N_f\over 2k}}}\,
\approx\,
\begin{cases}
1\,\,-\,{1\over 4}\,{N_f\over k} \,\,,\qquad
&{\rm for\,\,\,} N_f << k\,\,,\cr\cr
\sqrt{2}\,\,\sqrt{{k\over N_f }}\,\,,&{\rm for\, \,\,}\,  N_f>> k\,\,.
\end{cases}
\label{sigma-3-S}
\eeq
By comparing the right-hand side of (\ref{sigma-3-S}) with (\ref{sigma-small-Nf}) and (\ref{sigma-largeNf}) we conclude that our result is qualitatively similar to the one of \cite{Gaiotto:2009tk,Santamaria:2010dm}, although our smeared setup gives rise to a larger screening effect.

\subsection{Dimensions of scalar meson operators}

Let us analyze how the dimensions of the meson operators (bilinears in the fundamental fields) change when the effect of dynamical unquenched matter is taken into account. With this purpose, let us consider a D6-brane probe in the flavored background which fluctuates  around the static BPS configuration (\ref{RP3-cycle}). The induced metric on the D6-brane worldvolume for this static configuration is given by:
\beq
{\cal G}_{\alpha\beta}\,d\zeta^{\alpha}\,d\zeta^{\beta}\,=\,L^2\,
ds^2_{AdS_4}\,+\,
{L^2\over b^2}\,\,\Big[\, q\,(d\alpha)^2\,+\,q\,\sin^2\alpha \,(d\beta)^2\,+\,
\big(\,d\psi\,+\,\cos\alpha\,d\beta\,\big)^2\,\Big]\,\,.
\label{induced-D6-meson}
\eeq
For simplicity we will concentrate on the case in which only the angle $\theta$ varies with respect to the value written in (\ref{RP3-cycle}) and we will consider a perturbed configuration in which  the angle $\theta$ is given by:
\beq
\theta(r)\,=\,{\pi\over 2}\,+\,\lambda(r)\,\,,
\eeq
with $\lambda(r)$ being a small fluctuation of the transverse scalars of the BPS embedding (\ref{RP3-cycle}) (it should not be confused with the 't Hooft coupling). In order to study the equation of motion for  $\lambda$ let us compute the worldvolume induced metric $g_7$ at second order in $\lambda$. We represent $g_7$ as:
\beq
g_7\,=\,{\cal G}\,+\,g'\,\,,
\eeq
where ${\cal G}$ is the metric (\ref{induced-D6-meson}).
At second order in $\lambda$ the metric perturbation  $g'$ is the following:
\beq
g'_{\alpha\beta}\,\, d\zeta^{\alpha}\,d\zeta^{\beta}\,=\,
{L^2\over b^2}\,\,\Big[\,\lambda^2\,\big(\,d\psi\,+\,\cos\alpha\,d\beta\,\big)^2\,+\,
(\lambda')^2\,(dr)^2\,\Big]\,\,,
\eeq
with $\lambda'\,=\,d\lambda/dr$. The DBI lagrangian density  for the flavor D6-brane is just:
\beq
{\cal L}_{DBI}\,=\,-T_{D_6}\,e^{-\phi}\,\sqrt{-\det g_7}\,\,.
\label{DBI-D6}
\eeq
By plugging the results written above for $g_7$, one gets the following second-order result:
\beq
{\cal L}_{DBI}\,=\,-T_{D_6}\,e^{-\phi}\,\sqrt{-\det {{\cal G}}}\,\,\Big[\,
{r^2\over 2 b^2}\,\,(\lambda')^2\,-\,{1\over 2}\,\lambda^2\,\Big]\,\,.
\eeq
Similarly, the WZ term of the lagrangian is:
\beq
{S}_{WZ}\,=\,T_{D_6}\,\int_{{\cal M}_7}\,e^{-\phi}\,\,\hat{\cal K}\,\,,
\eeq
with ${\cal K}$ being the calibration form and $\hat{\cal K}$ its
pullback to the D6-brane worldvolume ${\cal M}_7$. One can easily show that, at second order, one has:
\beq
\hat{\cal K}\,=\,\Big(\,1\,-\,\lambda^2\,-\,{r\over b}\,\lambda\,\lambda'\,\Big)\,
{\rm Vol}\big(\,{\cal M}_7\,\big)\,\,,
\eeq
where ${\rm Vol}\big(\,{\cal M}_7\,\big)$ is the volume form of the metric ${\cal G}$.  After integrating by parts, we can write the langrangian density for the WZ part of the action as:
\beq
{\cal L}_{WZ}\,=\,-T_{D_6}\,e^{-\phi}\,\sqrt{-\det {{\cal G}}}\,\,\Big[\,
\Big(1\,-\,{3\over 2b}\,\Big)\,\lambda^2\,\Big]\,\,.
\eeq
The total lagrangian density is thus:
\beq
{\cal L}\,=\,-T_{D_6}\,e^{-\phi}\,\sqrt{-\det {{\cal G}}}\,\,\Big[\,
{r^2\over 2 b^2}\,\,(\lambda')^2\,+\,{1\over 2}\,
\Big(1\,-\,{3\over b}\,\Big)\,\lambda^2\,\Big]\,\,,
\eeq
and the corresponding equation of motion for $\lambda$ is:
\beq
{1\over r^2}\,\partial_r\,\big[\,r^4\,\,\partial_r\lambda\,\big]\,+\,b(3-b)\,\lambda\,=\,0\,\,.
\label{fluct-lambda}
\eeq
Let us assume that the fluctuation $\lambda(r)$ in $AdS_4$ behaves for large $r$ as:
\beq
\lambda\,\sim\,c_1\,r^{-2a_1}\,+\,c_2\,r^{-2a_2}\,\,,\qquad\qquad a_2>a_1\,\,,
\eeq
where $a_1$ ($a_2$) corresponds to the non-normalizable (normalizable) mode. The associated conformal dimension of the dual operator $\bar\psi\,\psi$  is:
\beq
\Delta\,=\,{\rm dim} ( \bar\psi\,\psi)\,=\, {3\over 2}\,+\,a_2-a_1\,\,.
\label{Delta-general}
\eeq
In order to find the values of the exponents $a_1$ and $a_2$ in our case, let us assume that there is a solution of the fluctuation equation (\ref{fluct-lambda}) in the form:
\beq
\lambda\sim r^{\alpha}\,\,.
\eeq
It is now  straightforward to show that the exponent $\alpha$ can take the values
$\alpha=-b, b-3$. It follows that $a_1$ and $a_2$ are given by:
\beq
a_1\,=\,{b\over 2}\,\,,\qquad\qquad
a_2\,=\,{3-b\over 2}\,\,.
\eeq
Thus, it follows from (\ref{Delta-general}) that the dimension $\Delta$ is just given by:
\beq
{\rm dim} (\bar\psi\psi)\,=\,3-b\,\,.
\eeq
Let us expand this result in the number of flavors. As the parameter $b$ written in (\ref{b-squashing}) is given by the following power  series in $N_f/ k$:
\beq
b\,=\,1\,+{3\over 16}\,{N_f\over k}\,-\,{63\over 512}\,\,\Big(\,{N_f\over k}\,\Big)^2\,+\,
\cdots\,\,,
\eeq
the dimension of the dual operator is:
\beq
{\rm dim} (\bar\psi\psi)\,=\,2\,-\,{3\over 16}\,{N_f\over k}\,+\,{63\over 512}\,\,\Big(\,{N_f\over k}\,\Big)^2\,+\,\cdots\,\,.
\eeq
This equation shows how the canonical dimension ${\rm dim} (\bar\psi\psi)=2$ is corrected by the addition of dynamical quarks in the regime in which $N_f/k$ is not large. Let us consider next the opposite limit in which $N_f/k$ is large. In this case one can easily verify that:
\beq
 b\to {5\over 4}\,\,,\qquad\qquad\qquad
\Big(\,{N_f\over k}\to\infty\,\Big)\,\,,
\eeq
and therefore
\beq
{\rm dim} (\bar\psi\psi)\to {7\over 4}\,\,,\qquad\qquad
\Big(\,{N_f\over k}\to\infty\,\Big)\,\,.
\eeq

\subsection{Dimension of high spin operators}

A high spin operator can be holographically realized as a rotating string \cite{Gubser:2002tv}. The anomalous dimension $\Delta$ of such operator can be computed in the large $\lambda$ limit. Indeed, the calculation is just the same as in \cite{Gubser:2002tv} and the result for the difference between the dimension $\Delta$ and the spin $S$ can be written as:
\beq
\Delta-S\,=\,f(\lambda,\epsilon)\,\,\log S\,\,,
\eeq
with $f(\lambda,\epsilon)$ being the so-called cusp anomalous dimension which depends on the 't Hooft coupling $\lambda$ and on the flavor deformation parameter $\epsilon$ (the unflavored result was obtained in \cite{Aharony:2008ug}). From the calculations in \cite{Gubser:2002tv}, we have:
\beq
f(\lambda,\epsilon)\,=\,{L^2\over \pi}\,\,.
\eeq
In terms of the gauge theory parameters the cusp anomalous dimension can be written as:
\beq
f(\lambda,\epsilon)\,=\,\sqrt{2\lambda}\,\,\,\sigma\,\,,
\eeq
where $\sigma$ is the screening factor of the quark-antiquark potential  defined in
(\ref{screening-sigma}), which also encodes the effects of the unquenched quarks on the anomalous dimensions of high spin operators.

\subsection{Particle-like branes}

As our final example let us analyze how quark loops effects change the dimensions of the operators dual to  some particle-like brane configurations. In general, the conformal dimension of the operator dual to an object of mass $m$ in the $AdS_4$ space of radius $L$ is given by:
\beq
\Delta\,=\,m\,L\,\,.
\label{Delta-mR}
\eeq
First of all, let us consider the case of D0-branes which, according to \cite{Aharony:2008ug}, are dual to di-monopole operators with charges $(1,1)$ under the two gauge groups. These operators are equivalent to Wilson line operators carrying $k$ fundamental indices of one group and $k$ anti-fundamental indices of the other group.  From the value of the dilaton in (\ref{dilaton-flavored}) we immediately obtain the mass of a D0-brane 
($m_{D0}=1/g_s$):
\beq
m_{D_0}\,=\,{\eta+q\over 4\sqrt{\pi}}\,\,\,\,
{q^{{1\over 4}}\,\,(q+\eta q-\eta)^{{1\over 4}}\over (2-q)^{{5\over 4}}}\,\,\,
\Big(\,{k^5\over 2 N}\,\Big)^{{1\over 4}}\,\,.
\eeq
The conformal dimension of the gauge dual is just obtained by applying (\ref{Delta-mR}). We get:
\beq
\Delta_{D_0}\,=\,{1\over 8}\,\,{q \,(\eta+q)^2\over 
(q+\eta q-\eta)\,\, (2-q)}\,\,k\,\,.
\label{Delta-D0}
\eeq
As a check, we notice that (\ref{Delta-D0}) reduces to $k/2$ for $\eta=q=1$, which is the value expected for an operator which is the product of $k$ bi-fundamentals of dimension $1/2$. 

In reference \cite{Gaiotto:2009tk} it was argued from the field theory side that the dimension of the di-monopole operators of an ${\cal N}=3$ Chern-Simons-matter theory is corrected by the fundamentals as $\Delta_{D0}\to\Delta_{D0}+{N_f\over 2}$. In order to compare this result with our expression (\ref{Delta-D0}), let us evaluate 
 $\Delta_{D0}$ for $N_f$ small and large with respect to the Chern-Simons-Level $k$. One gets:
\beq
\Delta_{D_0}\,\approx\,
\begin{cases}
{k\over 2}\,\,+\,{9N_f\over 16} \qquad
&{\rm for\,\,\,} N_f << k\,\,,\cr\cr
{45\over 64}\,N_f &{\rm for\, \,\,}\,  N_f>> k\,\,.
\end{cases}
\label{Delta-limits}
\eeq
Notice that (\ref{Delta-limits}) is not very different from the result obtained in \cite{Gaiotto:2009tk}, specially for small $N_f$ (although both results refer to theories with different amount of SUSY and flavor group).

We now consider di-baryons in the flavored geometry. They should correspond to D4-branes wrapped on (deformed)  ${\mathbb C} {\mathbb P}^2$, which we will take to be given by the same angular embedding as in the unflavored case in (\ref{CP2-embedding}), namely it will be defined by $\varphi=\theta=\pi/2$. In terms of the angle $\chi$ defined in (\ref{chi-CP2}), the induced metric in the four-cycle is given by the following deformation of the Fubini-Study metric:
\beq
{L^2\over b^2}\,\,\Big[\,(d\chi)^2\,+\,\cos^2{\chi\over 2}\,
\Big(\big(\,1+\,(q-1)\,\sin^2{\chi\over 2}\,\big)\,\,
\big(\,(\omega^1)^2\,+\,(\omega^3)^2\,\big)\,+\,
q\,\sin^2{\chi\over 2}\,(\omega^2)^2\,\Big)\,\Big]\,\,.
\eeq
The volume $V_4$ of this cycle can be immediately obtained by integration, namely:
\beq
V_4\,=\,{8\pi^2\,\over 3}\,\,{(2+q)\,L^4\over b^4}\,\,,
\eeq
and the mass of the wrapped D4-brane is:
\beq
m_{D_4}\,=\,{e^{-\phi}\over (2\pi)^4}\,\,V_4\,=\,{L^3\over 24\pi^2\,b}\,\,
{(2+q)(\eta+q)\over 2-q}\,\,k\,\,.
\eeq
The corresponding conformal dimension $\Delta_{D4}= m_{D_4}\,L$ is just:
\beq
\Delta_{D_4}\,=\,{1\over 96 }\,
{q^2\,(\eta+q)^4\,(2+\eta)\over
(q+\eta q-\eta)^4}\,\,N\,\,.
\label{Delta-D4-flavored}
\eeq
As a check of the formula (\ref{Delta-D4-flavored}) one can verify that its right-hand side reduces to the unflavored value  $N/2$ when $N_f=0$. Actually, 
the  dimension $\Delta_{D4}$ does not vary much with the number of flavors. For small
$N_f/k$ one can expand
\beq
\Delta_{D_4}\,=\
\Big(\,1+\,{1\over 8}\,{N_f\over k}\,-\,{33\over 256}\,\Big({N_f\over k}\Big)^2\,
+\,\cdots\,\Big)\,\,{N\over 2}\,\,.
\eeq
Moreover, for large $N_f/k$ the dimension approaches the following constant asymptotic limit:
\beq
\Delta_{D_4}\to {125\over  256}\,\,N\,\,.
\eeq
which is very close to the unflavored value $N/2$. Actually, it was argued in \cite{Lee:2006ys,Fujita:2009xz} from the analysis of the tri-Sasakian geometry dual to ${\cal N}=3$ theories  that $\Delta_{D_4}$ should not be changed when fundamentals are added. Again, we see that the results obtained with our smeared geometry are not very different from the ones found with the localized ${\cal N}=3$  backgrounds.

\section{Backreaction with massive flavors}
\label{unquenched-massive}

Let us write an ansatz for the backreacted background in the case that the quarks introduced by the flavor D6-branes are massive. According to what happens in other setups \cite{Nunez:2010sf} analyzed with the smearing technique, we will modify the ansatz of $F_2$ by substituting $N_f$ by $N_f\,p(r)$, where $p(r)$ is a function of the radial coordinate to be determined. This new function should satisfy the following conditions:
\bear
&&p(r)=0\qquad\qquad {\rm if}\,\,\,\,r<r_*\,\,,\rc\rc
&&\lim_{r\to\infty}\,\,p(r)\,=\,1\,\,,
\label{conditions-p}
\eear
where $r_*$ is related to the mass of the quarks.  Therefore, the new ansatz for $F_2$ is:
\beq
F_2\,=\,{k\over 2}\,\,E^1\wedge E^2\,-\,{1\over 2}\,
\Big(k+{3N_f\over 4}\,\,p(r)\Big)\,
\Big({\cal S}^{\xi}\wedge {\cal S}^{3}\,+\,{\cal S}^1\wedge {\cal S}^{2}\Big)\,\,.
\eeq
Now, the smearing form $\Omega=dF_2/2\pi$ takes the form:
\bear
&&\Omega\,=\,-{3N_f\over 16\pi}\,\,p(r)\,\,
\Big[\,
E^1\wedge ({\cal S}^{\xi}\wedge {\cal S}^{2}\,-\,{\cal S}^1\wedge {\cal S}^{3}\big)\,+\,
E^2\wedge ({\cal S}^{\xi}\wedge {\cal S}^{1}\,+\,{\cal S}^2\wedge {\cal S}^{3}\big)\,
\Big]\,\,-\rc\rc
&&\qquad\qquad\qquad\qquad
-{3N_f\over 16\pi}\,\,p'(r)\,\,dr\wedge 
\Big({\cal S}^{\xi}\wedge {\cal S}^{3}\,+\,{\cal S}^1\wedge {\cal S}^{2}\Big)\,\,,
\label{massive-Omega}
\eear
and has new components (the last line in (\ref{massive-Omega})) which were not present in the massless case. Notice, however, that  the BPS equations for this massive case can be simply obtained by changing:
\beq
k\eta\to k+{3 N_f\over 4}\,p(r)\,\,,
\eeq
 in the system for massless matter (\ref{3-system-flavored}) . 

Let us see how one can obtain the function $p(r)$ by comparing the smeared WZ action of the D6-brane with the one corresponding to a single massive embedding, which was studied in appendix \ref{Kappa}. First of all we notice that, with our new expression (\ref{massive-Omega}) for the smearing form $\Omega$, we get:
\beq
{\cal K}\wedge \Omega\,=\,{3N_f\over 4 \pi}\,\,\Big[\,p(r)\,+\,{1\over 2}\,e^{g}\,p'(r)\,\Big]\,
h^{{1\over 4}}\,e^{2f+g}\,
d^3 x\wedge dr\wedge {\rm Vol}({\mathbb S}^4)\wedge {\rm Vol}({\mathbb S}^2)\,\,.
\eeq
By integrating over the ${\mathbb S}^4$ and ${\mathbb S}^2$ (which amounts to multiplying by $32\pi^3/3$) and using this result in (\ref{WZ-smeared}) (with $C_7$ as in (\ref{C7-explicit})), one gets the effective WZ lagrangian density in the $x^{\mu}$ and $r$ variables, namely:
\beq
{\cal L}_{WZ}\,=\,8\pi^2\,N_f\,T_{D6}\, e^{-\phi}\,h^{{1\over 4}}\,e^{2f+g}\,
\Big[\,p(r)\,+\,{1\over 2}\,e^{g}\,p'(r)\,\Big]\,\,.
\label{smeared-lag-massive}
\eeq
By comparing with the lagrangian (\ref{WZ-lagrangian-local}) multiplied by $N_f$, we get that one can identify $p(r)$ with:
\beq
p(r)\,=\,\big(\sin\theta(r)\big)^2\,\Theta(r-r_*)\,\,.
\eeq
By using the expression (\ref{general-integral-kappa}) of $\theta(r)$, one gets:
\beq
p(r)\,=\,\Big[\,1\,-\,\exp\big[\,-2\int_{r_*}^{r}\,e^{-g(z)}\,dz\,\big]\,\Big]\,
\Theta(r-r_*)\,\,,
\label{p(r)}
\eeq
where we used the fact that $\theta=0$ when $r=r_*$. 
One can also get this same result by comparing the smeared and localized  DBI actions. Notice that, when the kappa symmetry equation (\ref{kappa-ode}) for the embedding holds, the DBI lagrangian density (\ref{DBI-loc}) for a massive embedding reduces to:
\beq
{\cal L}_{DBI}(BPS)\,=\,-8\pi^2\,T_{D6}\, e^{-\phi}\,h^{{1\over 4}}\,e^{2f+g}\,\,.
\eeq
The smeared DBI lagrangian is just the same as in (\ref{smeared-lag-massive})  with opposite sign. Then, it follows that $p(r)$ must satisfy the following differential equation:
\beq
e^{g}\,p'(r)\,+\,2p(r)\,=\,2\,\,.
\label{ODE-p}
\eeq
It is straightforward to verify that (\ref{p(r)}) is the most general solution of the ODE (\ref{ODE-p}) satisfying  the required conditions (\ref{conditions-p}). Moreover, by inspecting (\ref{ODE-p}) one realizes that, in this massive case, it is quite convenient to introduce a new radial variable $\rho$ such that:
\beq
{d\rho\over d r}\,=\,e^{-g}\,\,.
\eeq
Denoting by a dot the derivative with respect to $\rho$, one finds that the first-order differential equation (\ref{ODE-p}) becomes:
\beq
\dot p\,+\,2 p\,=\,2\,\,,
\qquad\qquad (\rho>\rho_*)\,\,,
\label{ode-p-rho}
\eeq
where $\rho_*$ is the value of the $\rho$ coordinate that corresponds to the minimal value $r_*$ of $r$. Eq. (\ref{ode-p-rho}) can be solved as:
\beq
p(\rho)\,=\,\Big[\,1\,-\,e^{2(\rho_*-\rho)}\,\Big]\,\,\Theta(\rho-\rho_*)\,\,,
\eeq
where we have required the continuity of $p(\rho)$ at $\rho=\rho_*$. Therefore, in the $\rho$ variable the function $p(\rho)$ is known and, in order to determine the background, one has to integrate the system:
\bear
&&\dot\Lambda\,=\,\Big(k+{3N_f\over 4}\,p(\rho)\,\Big)\,e^{\Lambda-2f+g}\,-\,
{k\over 2}\,e^{\Lambda-g}\,\,,\rc\rc
&& \dot f\,=\,{1\over 4}\,\Big(k+{3N_f\over 4}\,p(\rho)\,\Big)\,e^{\Lambda-2f+g}\,-\,
{k\over 4}\,e^{\Lambda-g}\,+\,e^{-2f+2g}\,\,,\rc\rc
&&\dot g\,=\,{1\over 2}\,\Big(k+{3N_f\over 4}\,p(\rho)\,\Big)\,e^{\Lambda-2f+g}\,+\,1\,-\,e^{-2f+2g}\,\,.
\label{3-system-massive}
\eear
We will not attempt here to solve the system (\ref{3-system-massive}). Let us just mention that, when the mass of the quarks is non-zero, there should be solutions of 
(\ref{3-system-massive})  that interpolate between two $AdS$ spaces, namely, the original unflavored ABJM geometry in the IR and our flavored $AdS$ space in the deep UV. We leave the verification of this fact and the detailed analysis of the system 
(\ref{3-system-massive}) for a future work.

\section{Conclusions}
\label{Conclusions}

Let us recapitulate our main results. We have studied the holographic dual of unquenched flavor in the ABJM Chern-Simons-matter theory in the type IIA description. We have found a geometry of the type $AdS_4\times {\cal M}_6$, where $ {\cal M}_6$ is a compact six-dimensional manifold whose metric we have explicitly determined. In order to get this result we have considered a setup in which the transverse internal space is filled with a continuous set of D6-branes which act as sources for the violation of the Bianchi identity of the RR two-form $F_2$ and deform the original unflavored $AdS_4\times {\mathbb C}\,{\mathbb P}^3$ geometry. To describe this deformation we have written the ${\mathbb C}\,{\mathbb P}^3$ manifold  as an ${\mathbb S}^2$-bundle over  ${\mathbb S}^4$ and we have argued that the flavor induces a squashing of the ${\mathbb S}^4$ base with respect to the ${\mathbb S}^2$ fiber and, at the same time,  changes  the size of both the Anti-de-Sitter and  the internal space. These squashing factors induced by the backreaction of the flavor branes are constant and depend non-linearly on the deformation parameter $\epsilon$ defined in (\ref{epsilon-def}). 

Contrary to what happens to other flavored backgrounds  obtained with the smearing method (see, for example, those of refs. \cite{CNP,conifold}) our supergravity solution has a good UV. Actually, since the metric has an Anti-de-Sitter factor, the holographic methods are firmly  established 
 and it is possible to apply a whole battery of techniques to perform a clean analysis of the flavor screening effects in different observables. This good behavior of the flavored solutions is a reflection of the fact that the D6-branes lift to pure geometry in eleven dimensions and, as a consequence, it is possible to have a nice geometrical description of matter in these systems.  We have carried out this study for some of these observables in section \ref{Flavor-effects} and, in some cases, we have compared our results with the ones obtained from the localized tri-Sasakian geometry, in order to determine how these quantities  change with the smearing of the flavor branes. Although a more detailed understanding of the field theory dual to our background would be desirable, we showed that the flavor effects on several quantities match the ones expected for a background in which dynamical quarks have been incorporated.

Our results can be generalized in several directions which we now briefly sketch. First of all, it is clear that the running solutions such as the ones  of section \ref{running-solutions} and appendix \ref{squashed-unflavored} deserve further analysis. In particular, it would be very interesting to determine if there are non-trivial flows that connect different fixed points (see \cite{Ahn:2008ua}). Another point that would be worth to investigate is the integration of the BPS system (\ref{3-system-massive}) for massive flavors. In this case one expects to find solutions which interpolate between the $AdS_4\times {\mathbb C}{\mathbb P}^3$ unflavored geometry in the IR and our massless flavored Anti-de-Sitter solutions in the UV.

Our analysis of the flavor effects in section \ref{Flavor-effects} is clearly incomplete. One could also study Wilson loops in representations different from the fundamental such as, for example, Wilson loops in the antisymmetric representations. Moreover, one could also compute the spectra of meson masses (and not just the dimension of the meson operators). In the unflavored case this analysis was performed in \cite{Jensen:2010vx}. The expectation is that the screening effects would reduce the binding energy and, as a consequence, the masses would increase with respect to the unflavored values (similar results for the theories on the conifold were obtained in \cite{Bigazzi:2009gu}). 

Another possible direction for future research could be trying to generalize the flavored supergravity solutions found here. One of these generalizations could be constructing a black hole solution with flavor (with finite temperature and, eventually, finite chemical potential) along the lines of \cite{D3-D7} and then studying its corresponding thermodynamical and hydrodynamical properties. Another possibility could be adding fractional branes to our solution, in such a way that the ranks of the two gauge groups is different \cite{Aharony:2008gk}, which would amount to introduce a NSNS $B$ field. Notice that, according to the analysis of \cite{Aharony:2009fc}, even the unflavored ABJM solution should have a non-vanishing flat $B$ field due to the so-called Freed-Witten anomaly. It would be interesting to explore how this issue changes when the dynamical quarks are added to the background. Another topic of interest would by trying  to find a flavored solution of type IIA supergravity with non-zero Romans mass which, according to \cite{Gaiotto:2009mv}, would correspond to a theory in which the sum of the two Chern-Simons levels are non-vanishing. Finally, one could also try to apply the same methodology to add flavor to backgrounds with reduced supersymmetry, whose internal space is not ${\mathbb C}{\mathbb P}^3$. 

We are already working along some of these lines and we hope to report on these topics elsewhere.

\section*{Acknowledgments}

We are grateful to F. Benini, F. Bigazzi, A. Cotrone, J. Gaillard, N. Jokela, 
C. N\'u\~nez,  A. Paredes, D. Rodriguez-Gomez, S. Cremonesi, J. Tarrio and D. Zoakos for very useful discussions. 
This  work  was funded in part by MICINN   under grant
FPA2008-01838,  by the Spanish Consolider-Ingenio 2010 Programme CPAN (CSD2007-00042), by Xunta de Galicia (Conseller\'\i a de Educaci\'on and grant INCITE09 206 121 PR) and by FEDER. AVR thanks the Galileo Galilei Institute for Theoretical Physics for the hospitality and the INFN for partial support during the completion of this work. E.C. is supported by a Spanish FPU fellowship, and thanks the FRont Of Galician-speaking Scientists for unconditional support.

\vskip 1cm
\renewcommand{\theequation}{\rm{A}.\arabic{equation}}
\setcounter{equation}{0}
\appendix

\section{ABJM geometry}
\label{ABJMgeometry}

Let us study the uplift of the ABJM metric (\ref{ABJM-metric}) to eleven-dimensional supergravity. The corresponding uplifting formula for the metric is:
\beq
ds_{11}^2\,=\,e^{-{2\phi\over 3}}\,\,ds^2_{10}\,+\,e^{{4\phi\over 3}}\,
\big(dx_{11}-A_1\big)^2\,\,,
\label{uplifted-metric}
\eeq
where we take the eleven-dimensional coordinate $x_{11}$ to take values in the range $x_{11}\in [0,2\pi)$ and $A_1$ is the one-form potential for the type IIA field strength $F_2$:
\beq
F_2\,=\,d A_1\,\,.
\eeq
For the ABJM solution of section \ref{ABJM-unflavored}, the actual value of $A_1$ is:
\beq
A_1\,=\,-{k\over 2}\,\Big(\,\cos\theta\,d\varphi\,+\,\xi\,{\cal S}^3\,\Big)\,\,,
\eeq
where ${\cal S}^3$ has been defined in (\ref{calS}). Let us next define a new angular variable $\psi$ as:
\beq
\psi\,\equiv \, {2\over k}\,x_{11}\,\,,\qquad\qquad \psi\in \big[0, {4\pi\over k}\big)\,\,,
\label{psi-x11}
\eeq
as well as a new one-form $E^3$ as:
\beq
E^3\equiv d\psi\,+\,\cos\theta\,d\varphi\,+\,\xi\,{\cal S}^3\,\,.
\label{E3}
\eeq
Then, the uplifted metric  (\ref{uplifted-metric}) can be written as the one corresponding to the product space $AdS_4\times {\mathbb S}^7/ {\mathbb Z}_k$, namely :
\beq
ds_{11}^2\,=\,{R^2\over 4}\,\,ds_{AdS_4}^2\,+\,R^2\,
ds^2_{{\mathbb S}^7/ {\mathbb Z}_k}\,\,,
\eeq
where $R$ is given by
\beq
R^6\,=\,2^5\,\pi^2\,N\,k\,\,,
\eeq
and the ${\mathbb S}^7/ {\mathbb Z}_k$ metric is:
\beq
ds^2_{{\mathbb S}^7/ {\mathbb Z}_k}\,=\,{1\over 4}\,\,
\Big[\,ds^2_{{\mathbb S}^4}\,+\,(E^1)^2\,+\,(E^2)^2\,+\,(E^3)^2
\,\Big]\,\,,
\label{S7/Zk}
\eeq
with $ds^2_{{\mathbb S}^4}$ being the metric of the four-sphere written in (\ref{S4metric}). As a check one can verify that the eight-dimensional cone with metric 
$dr^2+r^2\,ds^2_{{\mathbb S}^7/ {\mathbb Z}_k}$ is locally flat. Moreover, the metric 
(\ref{S7/Zk}) can be written as an ${\mathbb S}^3$-bundle over ${\mathbb S}^4$. Indeed, let  $\tilde\omega^i$ $(i=1,2,3$)  be a second set of left-invariant one-forms, defined in terms of the angles $\theta$, $\varphi$ and $\psi$ as:
\bear
&&\tilde\omega^1\,=\,-\sin\varphi\,d\theta\,+\,\cos\varphi\,\sin\theta\,d\psi\,\,,\rc
&&\tilde\omega^2\,=\,\cos\varphi\,d\theta\,+\,\sin\varphi\,\sin\theta\,d\psi\,\,,\rc\
&&\tilde\omega^3\,=\,d\varphi\,+\,\cos\theta\,d\psi\,\,.
\label{tilde-omega-i}
\eear
These forms satisfy $d\tilde\omega^i={1\over 2}\,\epsilon^{ijk}\tilde\omega^j\wedge\tilde\omega^k$. In terms of the $\tilde\omega^i$ the metric (\ref{S7/Zk}) takes the form:
\beq
ds^2_{{\mathbb S}^7/ {\mathbb Z}_k}\,=\,{1\over 4}\,\,
\Big[\,ds^2_{{\mathbb S}^4}\,+\,\sum_i\,\big(\,\tilde\omega^i+A^i\,\big)^2\,\Big]\,\,,
\eeq
where the one-forms $A^i$ are  the components of the $SU(2)$ instanton connection written in (\ref{A-instanton}). 

Let us come back to the ${\mathbb C} {\mathbb P}^3$  metric,  written in (\ref{CP3-metric}) as an ${\mathbb S}^2$-bundle over ${\mathbb S}^4$. As mentioned at the end of section \ref{ABJM-unflavored}, there is a non-trivial ${\mathbb C} {\mathbb P}^1\subset {\mathbb C} {\mathbb P}^3$, which is spanned by the angles $\theta$ and $\varphi$ for a fixed point of the ${\mathbb S}^4$. Let us now check that, in our coordinates, the embedding corresponding to a ${\mathbb C} {\mathbb P}^2$  is obtained by taking the ${\mathbb S}^2$ angles $\theta$ and $\varphi$ to be constant and given by:
\beq
\varphi=\theta={\pi\over 2}\,\,.
\label{CP2-embedding}
\eeq
In this case the pullbacks of $E^1$ and $E^2$ are given by:
\beq
\hat E^1\,=\,{\xi^2\over 1+\xi^2}\,\omega^1\,\,,\qquad\qquad
\hat E^2\,=\,-{\xi^2\over 1+\xi^2}\,\omega^3\,\,.
\eeq
Then, the pullback of the   ${\mathbb C}{\mathbb P}^3$ metric (\ref{CP3-metric}) is just: 
\beq
{d\xi^2\over (1+\xi^2)^2}\,+\,{\xi^2\over 4(1+\xi^2)}\,\Big[\,
(\omega^1)^2\,+\,(\omega^3)^2\,\Big]\,+\,
{\xi^2\over 4(1+\xi^2)^2}\,(\omega^2)^2\,\,.
\eeq
Introducing the new angle $\chi$ as:
\beq
\xi\,=\,\cot\Big( {\chi\over 2}\Big)\,\,,\qquad\qquad\qquad\qquad
0\le\chi\le \pi\,\,,
\label{chi-CP2}
\eeq
we can rewrite the above metric as:
\beq
{1\over 4}\,\,\Big[\,
\big(d\chi\big)^2\,+\,
\cos^2{\chi\over 2}\,\Big[\,(\omega^1)^2\,+\,(\omega^3)^2\,+\,
\sin^2{\chi\over 2}\,(\omega^2)^2\,\Big]\,\Big]
\,\,,
\eeq
which is, indeed, the Fubini-Study metric of  ${\mathbb C} {\mathbb P}^2$. Notice that  the total volume for this metric is just $\pi^2/2$.  As an application of this result, 
let us consider now a D4-brane wrapped on the ${\mathbb C} {\mathbb P}^2$. Its mass is given by:
\beq
m_{D4}\,=\,T_{D4}\,V_{{\mathbb C} {\mathbb P}^2}\,\,,
\eeq
where $T_{D4}$  and $V_{{\mathbb C} {\mathbb P}^2}$ are:
\beq
T_{D4}\,=\,{1\over (2\pi)^4\,e^{\phi}}\,\,,\qquad\qquad
V_{{\mathbb C} {\mathbb P}^2}\,=\,{\pi^2\over 2}\,\,L^4\,\,,
\eeq
where the dilaton has been written in (\ref{ABJMdilaton}) and $L$ is the $AdS_4$ radius for the unflavored ABJM solution displayed in (\ref{ABJM-AdSradius}). Using these values 
of $g_s$ and $L$, we get:
\beq
m_{D4}\,=\,{1\over 2\sqrt{\pi}}\,\,
\Big(\,{kN^3\over 2}\,\Big)^{{1\over 4}}\,\,.
\eeq
The resulting conformal dimension $\Delta_{D4}$ is just $m_{D4}\,L$, namely:
\beq
\Delta_{D4}\,=\,{N\over 2}\,\,,
\eeq
which is just the expected result since these branes are dual to dibaryon operators which are products of $N$ bifundamental fields, each of them  of dimension  $1/2$.

\vskip 1cm
\renewcommand{\theequation}{\rm{B}.\arabic{equation}}
\setcounter{equation}{0}

\section{SUSY analysis}
\label{SUSY}

In this appendix we will find the system of BPS first-order equations satisfied by our supergravity solutions, as well as the corresponding Killing spinors.  To find the supersymmetric configurations, we will use the SUSY variations of the dilatino $\lambda$ and the gravitino $\psi_{\mu}$  of the type IIA SUGRA in string frame (which we take from \cite{Martucci:2005rb}):
\bear
&&\delta\,\lambda\,=\,\Big[\,{1\over 2}\,\Gamma^{\mu}\,\partial_{\mu}\,\phi\,+\,
{3\over 8}\,{e^{\phi}\over 2!} \,F_{\mu\nu}^{(2)}\,\Gamma^{\mu\nu}\,\Gamma_{11}\,\,-\,
{1\over 8}\,{e^{\phi}\over 4!} \,F_{\mu\nu\rho\sigma}^{(4)}\,\Gamma^{\mu\nu\rho\sigma}\,
\Big]\,\epsilon\,\,,
\rc\rc
&&\delta\psi_{\mu}\,=\,\Big[\,\nabla_{\mu}\,-\,{e^{\phi}\over 8}\,{1\over 2!} \,
F_{\rho\sigma}^{(2)}\,\Gamma^{\rho\sigma}\,\,\Gamma_{11}\,\Gamma_{\mu}\,-\,
\,{e^{\phi}\over 8}\,{1\over 4!} \,
F_{\mu\nu\rho\sigma}^{(4)}\,\Gamma^{\mu\nu\rho\sigma}\,\Gamma_{\mu}\,
\Big]\,\epsilon\,\,.
\eear
In order to study the supersymmetric metrics of  form (\ref{metric-ansatz}), let us choose the following basis of  frame one-forms:
\bear
&&e^{\mu}\,=\,h^{-{1\over 4}}\,dx^{\mu},\,\,(\mu=0,1,2)\,\,,
\qquad\qquad e^{3}\,=\,h^{{1\over 4}}\,dr\,\,,
\qquad\qquad
e^{4}\,=\,h^{{1\over 4}}\,e^{f}{\cal S}^{\xi}\,\,,\rc\rc
&&e^{5}\,=\,h^{{1\over 4}}\,e^{f}\,{\cal S}^{1}\,\,,\qquad\qquad
e^{6}\,=\,h^{{1\over 4}}\,e^{f}\,{\cal S}^{2}\,\,,\qquad\qquad
e^{7}\,=\,h^{{1\over 4}}\,e^{f}\,{\cal S}^{3}\,\,,\rc\rc
&&e^{8}\,=\,h^{{1\over 4}}\,e^{g}\,E^{1}\,\,,\qquad\qquad
e^{9}\,=\,h^{{1\over 4}}\,e^{g}\,E^{2}\,\,.\qquad\qquad
\label{ten-dim-frame}
\eear

Let us first compute the dilatino variation. One gets:
\bear
&&\delta\lambda\,=\,\Big[\,{1\over 2}\,h^{-{1\over 4}}\,\Gamma_3\,\phi'\,+\,
{3k\over 16}\,e^{\phi}\,h^{-{1\over 2}}\,\Big(\,e^{-2g}\,\Gamma_{89}\,-\,
\eta\,e^{-2f}\,\Gamma_{47}\,-\,\eta\,e^{-2f}\,\Gamma_{56}\,\Big)\,\Gamma_{11}\,+\,\rc\rc
&&\qquad\qquad\qquad\qquad\qquad\qquad
+{e^{\phi}\over 8}\,K\,h^{{1\over 2}}\, \Gamma_{012}\,\Gamma_3\,
\Big]\,\epsilon\,\,.
\label{delta-lambda}
\eear
We will first impose the following projection conditions:
\beq
\Gamma_{47}\,\epsilon\,=\,\Gamma_{56}\,\epsilon\,=\,\Gamma_{89}\,\epsilon\,\,.
\label{Kahler-proj}
\eeq
Then, the vanishing of the dilatino variation, $\delta\lambda\,=\,0$, leads to the following  equation:
\beq
\phi'\,\epsilon\,+\,{3k\over 8}\,e^{\phi}\,h^{-{1\over 4}}\,\big(\,
e^{-2g}-2\eta\,e^{-2f}\,\big)\,\Gamma_{389}\Gamma_{11}\,\epsilon\,-\,
{e^{\phi}\over 4}\,K\,h^{{3\over 4}}\, \Gamma_{012}\,\epsilon\,=\,0\,\,.
\label{dilatino-var}
\eeq
Let us next impose the following projection on $\epsilon$:
\beq
\Gamma_{012}\,\epsilon\,=\,-\,\epsilon\,\,.
\label{D2-proj}
\eeq
Notice that, as $\Gamma_{11}$ is defined as:
\beq
\Gamma_{11}\,=\,\Gamma_{01\cdots 9}\,\,,
\eeq
the two matrices appearing on (\ref{dilatino-var}) are related, namely:
\beq
\Gamma_{389}\,\Gamma_{11}\,=\,\Gamma_{012}\,\Gamma_{4567}\,\,.
\label{Gamma389-11}
\eeq
Since $\Gamma_{4567}\,\epsilon\,=\,-\epsilon$ (see eq. (\ref{Kahler-proj})), one has that the projection (\ref{D2-proj}) implies that:
\beq
\Gamma_{389}\,\Gamma_{11}\,\epsilon\,=\,\epsilon\,\,.
\eeq

Using these projections, the dilatino equation becomes the following first-order differential equation:
\beq
\phi'\,=\,-\,{3k\over 8}\,e^{\phi}\,h^{-{1\over 4}}\,\big(\,
e^{-2g}-2\eta\,e^{-2f}\,\big)\,-\,
{e^{\phi}\over 4}\,K\,h^{{3\over 4}}\,\,.
\label{phiprime}
\eeq

The variation of the components of the gravitino along the Minkowski directions leads to the equations:
\bear
&&h^{{3\over 2}}\,\partial_{x^\mu}\,\epsilon\,-\,{h'\over 8}\,\,\Gamma_{\mu 3}\,\epsilon\,-\,{1\over 8}\,\Big[\,{k\over 2}\,e^{\phi}\,h^{{3\over 4}}\,\Big(\,
e^{-2g}-2\eta\,e^{-2f}\,\Big)\,-\,e^{\phi}\,K\,h^{{7\over 4}}\,\Big]\,
\Gamma_{\mu 3}\,\Gamma_{012}\,\epsilon\,=\,0\,\,.\,\,\qquad\qquad
\label{gravitino-Minkowski}
\eear
When the projection (\ref{D2-proj}) is imposed, (\ref{gravitino-Minkowski}) can be solved by means of a spinor which does not depend on the  cartesian coordinates $x^{\mu}$. Indeed, if  $\partial_{x^\mu}\,\epsilon=0$ eq. (\ref{gravitino-Minkowski}) leads to the following differential equation for $h$:
\beq
h'\,=\,\,{k\over 2}\,e^{\phi}\, h^{{3\over 4}}\, \big(\,e^{-2g}\,-\,2\eta\,e^{-2f}\,\big)\,-\,
\,e^{\phi}\,K\,h^{{7\over 4}}\,\,.
\label{hprime}
\eeq
Let us now consider the equation obtained from the  SUSY variation of the component of the gravitino along the direction $4$.  After using the projections (\ref{Kahler-proj}) and imposing that the spinor does not depend on the internal coordinates, one arrives at the following equation:
\beq
(h'+4hf')\,\epsilon\,+\,4h\,e^{-2f+g}\,\Gamma_{3458}\,\epsilon\,-\,
{1\over 2}\,h^{{3\over 4}}\,\,e^{\phi}\,\Big(\,k\,e^{-2g}\,+2hK\,\Big)\,
\Gamma_{012}\,\epsilon\,=\,0\,\,.
\label{gravitino-4}
\eeq
In order to solve this equation,  let us impose a new projection:
\beq
\Gamma_{3458}\,\epsilon\,=\,-\,\epsilon\,\,.
\label{3458-projection}
\eeq
Using this projection, together with the one in (\ref{D2-proj}), leads to the differential equation:
\beq
h'+4hf'\,=\,-{k\over 2}\,h^{{3\over 4}}\, e^{\phi}\,e^{-2g}\,-\,
e^{\phi}\,K\,h^{{7\over 4}}\,+\,4 \,h\,e^{-2f+g}\,\,.
\label{hprime-fprime}
\eeq
By combining eqs.  (\ref{hprime}) and (\ref{hprime-fprime}) one can easily prove that:
\beq
f'\,=\,{k\over 4} \,h^{-{1\over 4}}\,e^{\phi}\,\big[\,\eta\,e^{-2f}\,-\,e^{-2g}\,\big]
\,+\,\,e^{-2f+g}\,\,.
\label{fprime}
\eeq
Let us next consider the equation obtained from $\delta\psi_8=0$. Again, after imposing  (\ref{Kahler-proj})  and the independence of the spinor on the internal coordinates,  one arrives at:
\bear
&&\big(\,h'+4h\,g'\,\big)\,\epsilon\,+\,4h\,\big(\,e^{-g}\,-\,e^{-2f+g}\,\big)\,
\Gamma_{3458}\,\epsilon\,+\rc\rc
&&\qquad\qquad\qquad\qquad+\,
{1\over 2}\,h^{{3\over 4}}\,e^{\phi}\,
\big(\,2k\eta\,e^{-2f}\,+\,k\,e^{-2g}\,-\,2Kh\,\big)\,
\Gamma_{012}\,\epsilon\,=\,0\,\,.
\label{gravitino-8}
\eear
By using again the projections (\ref{D2-proj}) and (\ref{3458-projection}), we get:
\beq
h'+4h\,g'\,=\,\,{k\over 2}\,e^{\phi}\,h^{{3\over 4}}\, \,\big(\,
e^{-2g}\,+\,2\,\eta\,e^{-2f}\,\big)\,-\,\,e^{\phi}\,K\,h^{{7\over 4}}\,+\,
4 h\,\big(\,e^{-g}\,-\,e^{-2f+g}\,\big)\,\,.\qquad
\eeq
Eliminating $h'$ from (\ref{hprime}) we arrive at:
\beq
g'\,=\,{k\over 2}\,e^{\phi}\,h^{-{1\over 4}}\, \eta\,e^{-2f}\,+\,
e^{-g}\,-\,e^{-2f+g}\,\,.
\label{gprime}
\eeq
Let us finally analyze the  supersymmetry variation of the gravitino component along the radial direction. After imposing (\ref{Kahler-proj}) one arrives at the following equation:
\beq
\partial_r\,\epsilon\,=\,-{1\over 6}\,e^{\phi}\,h^{{3\over 4}}\,K\,\Gamma_{012}\,\epsilon\,\,.
\label{radial-eq}
\eeq
This equation can be easily integrated. First of all we impose (\ref{D2-proj}). Secondly, as shown in section \ref{partial-integration},  from the equations derived above the function $K$ can be written in terms of $\phi$ and $h$ as in (\ref{K-phi-h}) and one can show that:
\beq
e^{\phi}\,h^{{3\over 4}}\,K\,=\,-{d\over dr}\,\log\Big(\,e^{\phi}\,h^{{3\over 4}}\,\Big)\,\,.
\label{K-phi-h-new}
\eeq
Therefore, the Killing spinor equation (\ref{radial-eq})  can be integrated as:
\beq
\epsilon\,=\,e^{-{\phi\over 6}}\,\,h^{-{1\over 8}}\,\,\epsilon_0\,\,,
\label{radial-normalization-epsilon}
\eeq
where $\epsilon_0$ is a constant spinor satisfying the same projections as $\epsilon$.

Eqs. (\ref{phiprime}), (\ref{hprime}), (\ref{fprime}) and (\ref{gprime}) constitute the system of first-order BPS equations  (\ref{BPS-flavored}). They have been obtained by imposing the projections (\ref{full-projections}) and ensure the preservation of two supercharges, both in the unflavored and flavored theories. As we will show in the next subsection, the actual number of supersymmetries is increased for certain particular solutions of the BPS equations due to the fact that some of the projections which are imposed in the generic case are not needed  in these special solutions. In particular, for the case of $AdS$ solutions of sections \ref{AdS-unflavored} and \ref{AdS-flavored}, the projection (\ref{D2-proj}) is not needed and there are four Killing spinors (as it corresponds to ${\cal N}=1$ superconformal supersymmetry in three dimensions). Moreover, for the unflavored ABJM solution one can solve the BPS equations without imposing any of the projections written in (\ref{full-projections}) and, after a detailed study, one can show that there are 24 Killing spinors, as it corresponds to ${\cal N}=6$ in 3d.

\subsection{SUSY for the Anti-de Sitter solutions}

Let us consider the particular solution of the BPS equations which leads to the $AdS_4$ metric. Since the dilaton $\phi$ is constant in this case, it follows from (\ref{phiprime}) that the following relation holds:
\beq
2\eta\,e^{-2f}\,-\,e^{-2g}\,=\,{2Kh\over 3k}\,\,.
\label{f-g-rel-AdS}
\eeq
Actually, by using (\ref{f-g-rel-AdS}) and (\ref{Gamma389-11}) one can show that (\ref{dilatino-var}) is satisfied by imposing only the projections (\ref{Kahler-proj}),  without requiring the condition (\ref{D2-proj}). Moreover, by plugging (\ref{f-g-rel-AdS}) into (\ref{gravitino-Minkowski}) one gets:
\beq
\partial_{x^{\mu}}\,\epsilon\,=\,{1\over 8}\,h^{-{3\over 2}}\,h'\,\,\Gamma_{\mu 3}\,\epsilon\,-\,{1\over 6}\,h^{{1\over 4}}\,e^{\phi}\,K\,\Gamma_{\mu 3}\,\Gamma_{012}\,\epsilon\,\,.
\label{AdS-spinors-partialx}
\eeq
Furthermore, when $\phi$ is constant (\ref{K-phi-h-new}) can be used to relate $K$ to $h'$. This relation can be written as:
\beq
e^{\phi}\,K\,=\,-{3\over 4}\,\, \,h^{-{7\over 4}}\,h'\,\,.
\label{K-phi-hprime}
\eeq
By eliminating $K$ on the right-hand side of (\ref{AdS-spinors-partialx}), one gets:
\beq
\partial_{x^{\mu}}\,\epsilon\,=\,{1\over 8}\,h^{-{3\over 2}}\,h'\,\,\Gamma_{\mu 3}\,
\big(\,1\,+\,\Gamma_{012}\,\big)\,\epsilon\,\,.
\label{AdS-spinors-partialx-2}
\eeq
Since for these solutions $h=L^4/r^4$, where $L$ is the $AdS_4$ radius, we can rewrite (\ref{AdS-spinors-partialx-2}) as:
\beq
\partial_{x^{\mu}}\,\epsilon\,=\,-{r\over 2L^2}\,\Gamma_{\mu 3}\,
\big(\,1\,+\,\Gamma_{012}\,\big)\,\epsilon\,\,.
\label{AdS-spinors-partialx-3}
\eeq
We can now combine this equation with (\ref{radial-eq}) to obtain the dependence of the Killing spinors on the $AdS_4$ coordinates. Indeed, by using (\ref{K-phi-hprime}) in (\ref{radial-eq}) it is straightforward to prove that:
\beq
\partial_{r}\,\epsilon\,=\,-{1\over 2r}\,\,\Gamma_{012}\,\epsilon\,\,.
\label{AdS-spinors-partialr}
\eeq
It is now easy to integrate (\ref{AdS-spinors-partialx-3}) and (\ref{AdS-spinors-partialr}) following \cite{LPT}. One gets:
\beq
\epsilon\,=\,r^{-{\Gamma_{012}\over 2}}\,\,\Big(\,1\,+\,
{1\over 2L^2}\,x^{\mu}\,\Gamma_3\,\Gamma_{\mu}\,\big(\,1\,+\,\Gamma_{012}\,\big)
\,\Big)\,\epsilon_0\,\,,
\label{AdS-spinors}
\eeq
where $\epsilon_0$ is a constant spinor satisfying the projection conditions (\ref{Kahler-proj}). Notice that the spinors $\epsilon$  in (\ref{AdS-spinors}) with 
$\Gamma_{012}\epsilon_0=-\epsilon_0$ satisfy (\ref{D2-proj}) and are independent of the cartesian coordinates.  On the contrary, if we choose  an $\epsilon_0$ such that 
$\Gamma_{012}\epsilon_0=\epsilon_0$, the resulting Killing spinors $\epsilon$ do depend on the cartesian coordinates and do not have a well-defined eigenvalue of 
$\Gamma_{012}$. Moreover, since for these $AdS$ solutions 
$h'+4hf'=h'+4hg'=0$, one can easily verify that the equations obtained from the variation of the gravitino along the internal directions (\ie\ eqs. (\ref{gravitino-4}) and (\ref{gravitino-8})) are satisfied if the following projection:
\beq
\Gamma_{012}\,\Gamma_{3458}\,\epsilon\,=\,\epsilon\,\,,
\label{012-3458-projection}
\eeq
is imposed on $\epsilon$. Notice that the matrix on the left-hand side of (\ref{012-3458-projection}) commutes with the one multiplying the constant spinor $\epsilon_0$ in (\ref{AdS-spinors}). Thus, $\epsilon_0$ must also satisfy (\ref{012-3458-projection}) and these $AdS$ backgrounds preserve four supercharges, as claimed. 

Interestingly,  the BPS equations for the $AdS$ solutions can be recast as the ones corresponding to a compactification with fluxes in an internal manifold with an $SU(3)$-structure (see \cite{Grana:2005jc} for a review). To verify this fact, let us define the fundamental two-form ${\cal J}$ as:
\beq
{\cal J}\,=\,h^{-{1\over 2}}\,e^{-2f}\,\,\Big(\,
e^4\wedge e^7\,+\,e^5\wedge e^6\,+\,e^8\wedge e^9\,\Big)\,\,,
\eeq
where the one-forms $e^{4},\cdots e^{9}$ are the ones written in (\ref{ten-dim-frame}). 
Moreover, let $\Omega_{hol}$ be the holomorphic three-form defined as:
\beq
i\Omega_{hol}\,=\,h^{-{3\over 2}}\,e^{-3f}\,\,
\big(\,e^4+ie^7\,\big)\wedge\big(\,e^5+ie^6\,\big)\wedge \big(\,e^8+ie^9\,\big)\,\,.
\eeq
One can check that these forms satisfy:
\bear
&&d\,{\cal J}\,=\,{3\over 2}\,{\cal W}_1\,{\rm Im}\,\big(\Omega_{hol}\big)\,\,,\rc\rc
&&d\,\Omega_{hol}\,=\,-{\cal W}_1\,{\cal J}\wedge {\cal J}\,-\,{\cal W}_2\,\wedge {\cal J}\,\,,
\eear
where ${\cal W}_1$ and ${\cal W}_2$ are the so-called torsion classes which, for our solutions with $e^{g}=e^{f}/\sqrt{q}$,  are given by:
\bear
&&{\cal W}_1\,=\,{2\over 3}\,{q+1\over \sqrt{q}}\,\,,\rc\rc
&&h^{{1\over 2}}\,e^{2f}\,
{\cal W}_2\,=\,{2\over 3}\,{2-q\over \sqrt{q}}\,\,
\Big(\,e^4\wedge e^7\,+\,e^5\wedge e^6\,-2\,e^8\wedge e^9\,\Big)\,\,.
\label{torsion-classes}
\eear
Notice that, in terms of the one-forms in (\ref{ten-dim-frame}), our ansatz  (\ref{F2-flavored}) for $F_2$ can be written as:
\beq
h^{{1\over 2}}\,e^{2f}\,F_2\,=\,-{k\over 2}\,\,
\Big[\,\eta\,\big(e^4\wedge e^7\,+\,e^5\wedge e^6\big)\,-\,
\sqrt{q}\,\,e^8\wedge e^9\,\Big]\,\,.
\eeq
Then, one can check that, if  the squashing factors $q$ and $\eta$ are related as in (\ref{q-eq-flav}), the two-form $F_2$ is also given in terms of the torsion classes and of the fundamental form by:
\beq
F_2\,=\,-{\sqrt{q}\over 3-q}\,\eta\,k\,
\Big[\,{1\over 4}\,{\cal W}_1\,{\cal J}\,-\,{}^{*_6}\big(\,{\cal W}_2\wedge {\cal J}\,\big)
\,\Big]\,\,,
\eeq
where the ${\cal W}_i$ are given in (\ref{torsion-classes}) and ${}^{*_6}$ denotes the Hodge dual with respect to the six-dimensional internal metric 
$h^{-{1/2}}\,e^{2f}\,\big(\,(e^4)^2\,+\,\cdots +\,(e^9)^2\,\big)$.

\subsection{SUSY for the unflavored ABJM solution}

For the ABJM unflavored solution the squashing factors $q$ and $\eta$ are equal to one. Moreover, it follows from (\ref{fg-ABJM}) and (\ref{hK-ABJM}) that $K\,h\,e^{2g}\,=\,3k/2$ in this case and one can easily verify that the projections (\ref{Kahler-proj}) are not needed to solve the dilatino equation $\delta\lambda=0$. Indeed, it is straightforward to show from (\ref{delta-lambda}) that  the equation for the SUSY variation of the dilatino leads to:
\beq
\epsilon\,=\,\big(\,\Gamma_{4756}-\Gamma_{5689}\,-\,\Gamma_{4789}\,\big)\,\epsilon\,\,.
\label{projection-ABJM}
\eeq
In order to study the solutions of this equation let us work on a representation of the Dirac algebra in which the spinors are characterized by their $\pm 1$ eigenvalue of the following complete commuting set of matrices: 
$\{\Gamma_{01}\,,\,i\Gamma_{23}\,,\,i\Gamma_{47}\,,\,i\Gamma_{56}\,,\,
i\Gamma_{89}\}$. In this representation the matrices on the right-hand side of (\ref{projection-ABJM}) will also act diagonally. Let us parametrize their eigenvalues as:
\beq
\Gamma_{4756}\,\epsilon\,=\, s_1\,\epsilon\,\,,\qquad
\Gamma_{5689}\,\epsilon\,=\, s_2\,\epsilon\,\,,\qquad
\Gamma_{4789}\,\epsilon\,=\, -s_1\,s_2\,\epsilon\,\,.
\eeq
One immediately shows that (\ref{projection-ABJM}) is equivalent to the following condition on $s_1$ and $s_2$:
\beq
s_1-s_2+s_1\,s_2\,=\,1\,\,.
\label{s1-s2-cond}
\eeq
Since only three of the four possible values of $(s_1,s_2)$  satisfy (\ref{s1-s2-cond}), the projection (\ref{projection-ABJM}) preserves 3/4 of the supercharges, \ie\ 24 of them. This is, indeed, the amount of supersymmetry of an ${\cal N}=6$ supersymmetric theory in 3d. Moreover, one can show that the remaining equations for $\epsilon$ can also be solved without imposing any additional projection.

\vskip 1cm
\renewcommand{\theequation}{\rm{C}.\arabic{equation}}
\setcounter{equation}{0}

\section{A consistent truncation}
\label{squashed-unflavored}

Let us see that there exists a consistent truncation of the reduced unflavored BPS system  (\ref{Sigma-Delta-sys}) that allows to find some non-trivial solutions. In these truncations  $\Sigma$ and $\Delta$ are related as:
\beq
e^{\Sigma}\,=\,A\,e^{-\Delta}\,\,,
\label{truncation}
\eeq
with $A$ being a constant. To have a natural interpretation of this truncation, let us 
look at the metric obtained after uplifting to eleven dimensions. 
This metric is given by (\ref{uplifted-metric}). 
For our explicit ansatz (\ref{metric-ansatz})-(\ref{7metric-ansatz}) the uplifted  eleven-dimensional metric has the form:
\bear
&&ds_{11}^2\,=\,e^{-{2\phi\over 3}}\,h^{-{1\over 2}}\,dx^2_{1,2}\,+\,
e^{-{2\phi\over 3}}\,h^{{1\over 2}}\,dr^2\,+\,
e^{-{2\phi\over 3}}\,h^{{1\over 2}}\,e^{2f}\,ds^2_{{\mathbb S}^4}\,+\,\rc\rc
&&\qquad\qquad\qquad
+e^{-{2\phi\over 3}}\,h^{{1\over 2}}\,e^{2g}\,
\Big[\,(E^1)^2\,+\,(E^2)^2\Big]\,+\,e^{{4\phi\over 3}}\,\,(dx_{11}-A_1)^2\,\,.
\label{11d-metric-explicit}
\eear
Notice that the relative coefficient between the $U(1)$ fiber and the $E^1E^2$ parts of $ds_{11}^2$ (the last two terms in (\ref{11d-metric-explicit})) is given by
$e^{2\phi}\,h^{-{1\over 2}}\,e^{-2g}\,=\,e^{2\Sigma+2\Delta}$, which is constant if the truncation condition (\ref{truncation}) holds. Therefore, when (\ref{truncation}) is satisfied these two parts of the metric can give rise to the metric of a three sphere.

Substituting the relation (\ref{truncation}) on the right-hand side of the system (\ref{Sigma-Delta-sys}), we get the following two equations:
\bear
&&\dot\Sigma\,=\,\Big(\,{3kA\over 4}\,-\,1\,\Big)\,e^{-\Delta}\,-\,{kA\over 4}\,e^{\Delta}
\,\,,\rc\rc
&&\dot\Delta\,=\,
\Big(\,2\,-\,{kA\over 4}\,\Big)\,e^{-\Delta}\,-\,\Big(\,1\,+\,{kA\over 4}\,\Big)\,e^{\Delta}\,\,.
\label{dot-Sigma-Delta-truncated}
\eear
The truncation condition (\ref{truncation}) implies that $\dot\Sigma=-\dot \Delta$ which in turn leads to:
\beq
\Big(\,{kA\over 2}\,+\,1\,\Big)\,\Big(\,e^{-\Delta}\,-\,e^{\Delta}\,\Big)\,=\,0\,\,,
\label{constraint-truncation}
\eeq
There are two possible ways to solve (\ref{constraint-truncation}).  The first one is by imposing that $e^{\Delta}=e^{-\Delta}$, which would lead to  $e^{2\Delta}=1$ or $\Delta=0$. This solution with constant squashing corresponds to the unflavored ABJM solution described in section \ref{ABJM-unflavored}. Indeed, from (\ref{truncation}) it follows that $\Sigma$ is constant and, by taking $\Delta=\dot\Delta=\dot\Sigma=0$ in (\ref{dot-Sigma-Delta-truncated}) one discovers that $A=2/k=e^{\Sigma}$ for this solution. Notice that this implies that $k$ must be positive in this case.

The other possibility to solve (\ref{constraint-truncation}) consists in taking $kA=-2$, namely:
\beq
A\,=\,-{2\over k}\,\,,
\eeq
which implies that
\beq
e^{\Sigma}\,=\,-{2\over k}\,e^{-\Delta}\,\,.
\label{Sigma-Delta-trunc}
\eeq
Notice that $k$ must be negative in this case. In this solution the squashing can vary with the radial coordinate.  To continue with the analysis of this case it is much more convenient to come back to the original system (\ref{3-system}) in terms of the radial variable $r$ and the functions $\Lambda$, $f$ and $g$. Actually, from the definitions of $\Sigma$ and $\Delta$, the truncation  (\ref{Sigma-Delta-trunc})  is equivalent to:
\beq
e^{\Lambda}\,=\,-{2\over k}\,e^{g}\,\,.
\label{Lambda-g-trunc}
\eeq
After using (\ref{Lambda-g-trunc}), the equations for $f$ and $g$ in (\ref{3-system}) become:
\bear
&&f'\,=\,{1\over 2}\,\,e^{g-2f}\,+\,{1\over 2}\,e^{-g}\,\,,\rc\rc
&&g'\,=\,-2\,e^{g-2f}\,+\,e^{-g}\,\,.
\label{truncated-BPS}
\eear
It follows by combining these equations that:
\beq
g'\,-\,2f'\,=\,-3\,e^{g-2f}\,\,,
\eeq
which can be immediately integrated, namely:
\beq
e^{2f-g}\,=\,3r+c\,\,,
\eeq 
with $c$ being a constant. Let us use this result in the equation for $g'$ in 
(\ref{truncated-BPS}). If we define:
\beq
y\equiv e^{g}\,\,,
\eeq
then, the equation for $y$ is:
\beq
y'\,+\,{2\over 3r+c}\,y\,=\,1\,\,.
\label{y-ODE}
\eeq
By using the method of variation of constants, we get the general solution of (\ref{y-ODE}), namely:
\beq
y\,=\,{3\over 5}\,\big(r+{c\over 3}\big)\,+\,{3\over 5}\,{a\over \Big(r+{c\over 3}
\Big)^{{2\over 3}}}\,\,,
\eeq
with $a$ being a constant. By defining a new constant $b$ as:
\beq
b\,\equiv\,{3^{{5\over 3}}\,a\over  5}\,\,,
\eeq
we get the following solution for $f$ and $g$:
\bear
&& e^{2f}\,=\,{(3r+c)^2\over 5}\,\,\Big[1\,+\,{b\over (3r+c)^{{5\over 3}}}\,\Big]\,\,,\rc\rc
&& e^{2g}\,=\,{(3r+c)^2\over 25}\,\,\Big[1\,+\,{b\over (3r+c)^{{5\over 3}}}\,\Big]^2\,\,.
\eear
Let us rewrite this solution in terms of the new radial variable $\rho$ defined as:
\beq
3\rho=3r+c\,\,.
\eeq
Introducing a new constant $\mu$, related to $b$ and $a$ as:
\beq
\mu^{{5\over 3}}\,=\,-{b\over 3^{{5\over 3}}}\,=\,-{a\over 5}\,\,,
\eeq
the functions $f$ and $g$ can be written as:
\beq
 e^{2f}\,=\,{9\over 5}\,\,\rho^2\,\Big[\,1\,-\,\Big({\mu\over \rho}\Big)^{{5\over 3}}\,\Big]\,\,,
 \qquad\qquad
  e^{2g}\,=\,{9\over 25}\,\,\rho^2\,\Big[\,1\,-\,\Big({\mu\over \rho}\Big)^{{5\over 3}}\,\Big]^2
  \,\,.
 \eeq
The squashing corresponding to this solution is:
\beq
e^{2\Delta}\,=\,{5\over 1\,-\,\Big({\mu\over \rho}\Big)^{{5\over 3}}}\,\,,
\label{running-Delta}
\eeq
which varies with the radial coordinate except when the constant $\mu$ is chosen to vanish. In this last case one gets the gravity dual of the ${\cal N}=1$ 
model \cite{Ooguri:2008dk} with squashed ${\mathbb C}{\mathbb P}^3$ and $SO(5)\times U(1)$ global symmetry (see below).  Notice that, in this variable $\rho$, the function $\Lambda$ is given by:
\beq
e^{\Lambda}\,=\,{6\over 5 |k|}\,\,\rho\,\,\Big[\,1\,-\,
\Big({\mu\over \rho}\Big)^{{5\over 3}}\,\Big]\,\,,
\eeq
where we have taken into account that $k$ should be negative for this solution  and that $|k|=-k$. Let us now compute the warp factor $h$ from the integral (\ref{warp-integral}). Let us express the result in terms of the quantity $\bar R$ defined as:
\beq
\bar R^6\,=\,{2^5\,5^5\over 3^8}\,\, |k|\,\beta\,\,.
\label{barR}
\eeq
By using the relation $\beta=3\pi^2 N$, one can express $\bar R$ in terms of the rank $N$ of the gauge group as:
\beq
\bar R^6\,=\,{2^5\,5^5\over 3^7}\,\pi^2\,N\,|k|\,\,.
\eeq
From (\ref{warp-integral}) one can now compute the warp factor $h$. One gets:
\beq
h(\rho)\,=\,-{27\over 400}\,\,{\bar R^6\over k^2}\,\,
{1\over \rho \,\Big[\,1\,-\,\big({\mu\over \rho}\big)^{{5\over 3}}\,\Big]}\,\,
\int_{\rho_0}^{\rho}\,\,
{d\xi\over \xi^4\,\Big[\,1\,-\,\big({\mu\over \xi}\big)^{{5\over 3}}\,\Big]}\,\,,
\label{integral-wrap}
\eeq
where $\rho_0$ is a constant that should be determined. The dilaton $\phi$ for this solution can be obtained simply from the relation $e^{\phi}\,=\,e^{\Lambda}\,h^{{1\over 4}}$, which follows from the definition of $\Lambda$ in (\ref{Lambda-def}). Similarly, the function $K$ which parametrizes the RR four-form $F_4$ can be obtained from (\ref{K-beta}).

\subsection{$AdS$ solution with squashing}

Let us consider now the squashed solution obtained above in the case in which the constant $\mu$  vanishes. It follows from (\ref{running-Delta}) that $e^{2\Delta}=5$ and that the functions $f$ and $g$ are given by:
\beq
e^{2f}\,=\,{9\over 5}\,\rho^2\,\,,\qquad\qquad
e^{2g}\,=\,{9\over 25}\,\rho^2\,\,.
\eeq
The integral (\ref{integral-wrap}) giving the warp factor can be straightforwardly computed. If one chooses $\rho_0=-\infty$, one ends up with the result:
\beq
h(\rho)\,=\,{9\over 400}\,{\bar R^6\over |k|^2} \,\,{1\over \rho^4}\,\,,
\eeq
where $\bar R$ has been defined in (\ref{barR}). Moreover,
the dilaton is constant with this election of $\rho_0$ and given by:
\beq
e^{2\phi}\,=\,{27\over 125}\,\,{\bar R^3\over |k|^3}\,\,.
\eeq
Let us now write the metric corresponding to this solution. By rescaling the Minkowski coordinates as $x^{\mu}=\lambda \bar x^{\mu}$ one can show that the coordinates $x^{\mu}$ and $r$ parametrize and $AdS_4$ space. The value of $\lambda$ that one has to choose is:
\beq
\lambda\,=\,{3\over 20}\,\,{\bar R^3\over |k|}\,\,.
\eeq  

In order to have an interpretation of the parameter $\bar  R$, let us look at our  solution uplifted to M-theory. By using eq. (\ref{11d-metric-explicit})  and redefining the coordinate $x_{11}$ as in (\ref{psi-x11}), we get:
\beq
ds_{11}^2\,=\,{\bar R^2\over 4}\,ds^2_{AdS_4}\,+\,
\bar R^2\,\Big[\,{9\over 20}\,ds^2_{{\mathbb S}^4}\,+\,{9\over 100}\,\,\big[\,(E^1)^2\,+\,(E^2)^2\,+\,(E^3)^2\,
\big]\,\,\Big]\,\,,
\label{squashed-in-11d}
\eeq
where the one-form $E^3$ has been defined in (\ref{E3}). The metric (\ref{squashed-in-11d})  corresponds to a space which is the product of $AdS_4$ with radius $\bar R/2$ and the quotient $\tilde {\mathbb S}^7/Z_k$, with $\tilde {\mathbb S}^7$ being the squashed seven-sphere with metric
\beq
ds^2_{\tilde {\mathbb S}^7/Z_k}\,=\,{1\over 4}\,\,
\Big[\,{9\over 5}\,\,ds^2_{{\mathbb S}^4}\,+\,{9\over 25}\,
\sum_i\,\big(\,\tilde\omega^i+A^i\,\big)^2\,\Big]\,\,,
\eeq
where the $\tilde\omega^i$ are the one-forms defined in (\ref{tilde-omega-i}).  The resulting ten-dimensional metric takes the form:
\beq
ds^2_{10}\,=\,\bar L^2\,ds^2_{AdS_4}\,+\,{36\over 25}\,\,
\bar L^2\,ds^2_{\overline{{\mathbb C}\mathbb{P}^3}}\,\,,
\eeq
where $ds^2_{\overline{{\mathbb C}\mathbb{P}^3}}$ is the metric of a squashed ${\mathbb C}\mathbb{P}^3$, given by:
\beq
ds^2_{\overline{{\mathbb C}\mathbb{P}^3}}\,=\,{5\over 4}\,ds^2_{{\mathbb S}^4}\,+\,{1\over 4}\,
\big[\,(E^1)^2\,+\,(E^2)^2\,\big]\,\,,
\eeq
and the Anti-de-Sitter radius $\bar L$ is given by:
\beq
\bar L^4\,=\,{9\over 400}\,{\bar R^6\over |k|^2}\,=\,{250\over 243}\,\,\pi^2\,\,
{N\over |k|}\,\,.
\eeq
The RR four-form $F_4$ for this solution is given by:
\beq
F_4\,=\,{5\over 2}\,|k|\,\bar L^2\,\,\,\Omega_{AdS_4}\,=\,
{25\,\pi\over 18}\,\,\sqrt{{10\over 3}}\,\,\Big(\,|k|\,N\,\Big)^{{1\over 2}}\,\,
\Omega_{AdS_4}\,\,.
\eeq
As an application of the results found above, 
let us compute the conformal dimension of the gauge dual of a D0-brane.  First of all we notice that  the string coupling constant in this squashed case is given by:
\beq
g_s\,=\,e^{\phi}\,=\,2\sqrt{\pi}\,\,\Big[\,{2\over 15}\,\,{N\over |k|^5}\,\Big]^{{1\over 4}}\,.
\eeq
Using this result we can compute the conformal dimension of the D0-brane by means
of the formula $\Delta_{D0}=\bar L\,m_{D0}\,=\,\bar L/g_s$. One gets:
\beq
\Delta_{D0}={5\over 6}\,|k|\,\,,
\label{Delta-D0-squashed}
\eeq
which suggests that the some of the scalar bifundamental  fields have dimension $5/6$ in this case.

\vskip 1cm
\renewcommand{\theequation}{\rm{D}.\arabic{equation}}
\setcounter{equation}{0}

\section{Kappa symmetry}
\label{Kappa}

Let us consider a flavor D6-brane embedded in the ten-dimensional background (\ref{metric-ansatz}) and let us choose the following set of worldvolume coordinates:
\beq
\zeta^{\alpha}\,=\,(x^{\mu}, r, \xi, \hat \psi, \varphi)\,\,.
\eeq
The embedding is then defined by the equations:
\beq
\hat\theta\,\,,\hat\varphi\,\,=\,{\rm constant}\,\,,\qquad\qquad \theta=\theta(r)
\,\,.
\label{theta-ansatz}
\eeq
We want to determine, by using kappa symmetry,  the function $\theta(r)$ that makes the embedding supersymmetric. First of all we compute the pullback of the frame one-forms:
\bear
&& \hat e^{\mu}\,=\,h^{-{1\over 4}}\,dx^{\mu}\,\,,\qquad\qquad
\hat e^{3}\,=\,h^{{1\over 4}}\,dr\,\,,\qquad\qquad
\hat e^{4}\,=\,{2\over 1+\xi^2}\,\,h^{{1\over 4}}\,e^{f}\,d\xi\,\,,
\qquad\qquad\qquad\qquad\rc\rc
&&\hat e^{5}\,=\,0\,\,,\qquad\qquad
\hat e^{6}\,=\,{\xi\over 1+\xi^2}\,\,h^{{1\over 4}}\,e^{f}\,\sin\theta\,d\hat\psi\,\,,
\qquad\qquad
\hat e^{7}\,=\,-{\xi\over 1+\xi^2}\,\,h^{{1\over 4}}\,e^{f}\,\cos\theta\,d\hat\psi\,\,,\rc\rc
&&\hat e^{8}\,=\,h^{{1\over 4}}\,e^{g}\,\theta'\,dr\,\,,\qquad
\hat e^{9}\,=\,h^{{1\over 4}}\,e^{g}\,\sin\theta\,\Big(\,d\varphi\,-\,{\xi^2\over 1+\xi^2}\,d\hat\psi
\,\Big)\,\,.
\label{pullback-es}
\eear
The induced gamma matrices are:
\bear
&&\gamma_{x^{\mu}}\,=\,h^{-{1\over 4}}\Gamma_{\mu}\,\,,\qquad\qquad
\gamma_{r}\,=\,h^{{1\over 4}}\,\Big(\,\Gamma_3\,+\,e^{g}\,\theta'\,\Gamma_8\,\Big)\,\,,
\qquad\qquad
\gamma_{\xi}\,=\,{2\over 1+\xi^2}\,\,h^{{1\over 4}}\,e^{f}\,\Gamma_4\,\,,
\qquad\qquad
\rc\rc
&&\gamma_{\hat \psi}\,=\,{\xi\over 1+\xi^2}\,h^{{1\over 4}}\,e^{f}\,\sin\theta\,\,
\Big[\,\Gamma_{6}\,-\,\cot\theta\,\Gamma_7\,-\,\xi\,e^{g-f}\,\Gamma_9\,\Big]\,\,,
\qquad\qquad
\gamma_{\varphi}\,=\,h^{{1\over 4}}\,e^{g}\,\sin\theta\,\Gamma_9\,\,.
\eear
Let $\gamma_*$ be the antisymmetrized product of all induced gamma matrices, namely:
\beq
\gamma_{*}\,\equiv\,\gamma_{x^0\,x^1\,x^2\,r\,\xi\,\hat\psi\,\varphi}\,\,.
\eeq
In terms of the flat 10d Dirac matrices $\gamma_*$ is:
\beq
\gamma_*\,=\,{2\xi\over (1+\xi^2)^2}\, h^{{1\over 4}}\,e^{2f+g}\,(\,\sin\theta\,)^2\,\,
\Gamma_{012}\,\,\Big(\,\Gamma_3\,+\,e^g\,\theta'\,\Gamma_8\,\Big)\,
\Gamma_4\,\Big(\,\Gamma_6\,-\,\cot\theta\,\Gamma_7\,\Big)\,\Gamma_9\,\,.
\eeq
Notice that the kappa symmetry matrix $\Gamma_{\kappa}$ is just:
\beq
\Gamma_{\kappa}\,=\,{1\over \sqrt{-\det \hat g}}\,\,\gamma_*\,\,,
\eeq
where $\hat g$ is the induced metric on the worldvolume. With our notations the supersymmetric embeddings are those that satisfy $\Gamma_{\kappa}\,\epsilon\,=\,-\epsilon$. Using the projection $\Gamma_{012}\epsilon\,=\,-\epsilon$
(see eq. (\ref{D2-proj})), one obtains that $\gamma_*$ acts on the Killing spinors as:
\bear
&&\gamma_*\,\epsilon\,=\,-{2\xi\over (1+\xi^2)^2}\, h^{{1\over 4}}\,e^{2f+g}\,
(\,\sin\theta\,)^2\,\,\times
\qquad\qquad\qquad\qquad\qquad\qquad\rc\rc
&&\times\Bigg[\,
\Gamma_{3469}\,\epsilon\,-\,\cot\theta\,\Gamma_{3479}\,\epsilon\,+\,
e^g\,\theta'\,\Gamma_{8469}\,\epsilon\,+\,
e^g\,\theta'\,\cot\theta\,\Gamma_{8479}\,\epsilon\,\Bigg]\,\,.
\eear
From the projections satisfied by the Killing spinors it follows that:
\beq
\Gamma_{3469}\,\epsilon\,=\,-\Gamma_{8479}\,\epsilon\,=\,\epsilon\,\,,
\qquad\qquad
\Gamma_{3479}\,\epsilon\,=\,\Gamma_{8479}\epsilon\,=\,-\Gamma_{38}\,\epsilon\,\,.
\eeq
Therefore:
\beq
\gamma_*\,\epsilon\,=\,-{2\xi\over (1+\xi^2)^2}\, h^{{1\over 4}}\,e^{2f+g}\,
(\,\sin\theta\,)^2\,\,
\Big[\,1\,+\,e^g\,\theta'\,\cot\theta\,+\,\big(\cot\theta\,-\,e^{g}\,\theta'\,\big)\,
\Gamma_{38}\,\Big]\epsilon\,\,.
\eeq
In order to fulfill the condition $\Gamma_{\kappa}\,\epsilon\,=\,-\epsilon$ we should require that the terms not containing the unit matrix vanish. Thus, we are led to the following differential equation for $\theta(r)$:
\beq
{d \theta\over dr}\,=\,e^{-g}\,\cot\theta\,\,.
\label{kappa-ode}
\eeq

The induced metric  for an embedding like the one in our ansatz is given by:
\bear
&&ds^2_{wv}\,=\,h^{-{1\over 2}}\,(dx^{\mu})^2\,+\,
h^{{1\over 2}}\,\big(\,1\,+\,e^{2g}\,(\,\theta'\,)^2\,\big)\,\,(dr)^2\,+\,
h^{{1\over 2}}\,e^{2f}\,{4\over (1+\xi^2)^2}\,\,(d\xi)^2\,+\,\rc\rc
&&\qquad\qquad+
h^{{1\over 2}}\,e^{2f}\,{\xi^2\over (1+\xi^2)^2}\,\,(d\hat\psi)^2\,+\,
h^{{1\over 2}}\,e^{2g}\,\sin^2\theta\,\Big(\,d\varphi\,-\,{\xi^2\over 1+\xi^2}\,
d\hat\psi\,\Big)^2\,\,,
\eear
whose determinant is just:
\beq
\sqrt{-\det \hat g}\,=\,{2\xi\over (1+\xi^2)^2}\,h^{{1\over 4}}\,
e^{2f+g}\,\sin\theta\,\sqrt{1+e^{2g}\,(\,\theta'\,)^2}\,\,.
\label{det-induced-metric}
\eeq
If the BPS equation (\ref{kappa-ode}) holds, the determinant of the induced metric becomes:
\beq
\sqrt{-\det \hat g}_{|\,BPS}\,=\,{2\xi\over (1+\xi^2)^2}\,h^{{1\over 4}}\,
e^{2f+g}\,\,.
\eeq
Therefore, we have:
\beq
\Gamma_{\kappa}\,\epsilon_{|\,BPS}\,=\,-(\,\sin\theta\,)^2\,\,
\Big[\,1\,+\,e^g\,\theta'\,\cot\theta\,\Big]\epsilon_{|\,BPS}\,=\,-\epsilon\,\,,
\eeq
which proves the supersymmetric character of the embeddings satisfying (\ref{kappa-ode}). 

Let us now integrate the BPS equation (\ref{kappa-ode}). The integration of this equation is immediate, namely:
\beq
\log\big(\,\cos\theta\,\big)\,=\,-\int^{r}\,e^{-g(z)}\,dz\,+\,{\rm constant}\,\,.
\eeq
This result can be rewritten as:
\beq
\cos\theta\,=\,C\,\exp\Big[-\int^{r}\,e^{-g(z)}\,dz\Big]\,\,,
\label{general-integral-kappa}
\eeq
where $C$ is a constant. Let us suppose that the function $g$ is given by:
\beq
e^{g}\,=\,{r\over b}\,\,,
\eeq
with $b$ being a  positive constant. This is the form of $e^{g}$ for the different $AdS_4$ solutions (see, for example,  (\ref{f-g-flavored})).  Using this expression of $e^g$  in (\ref{general-integral-kappa}) and defining $r_*$ as $C=r_*^{b}$, one gets:
\beq
\cos\theta\,=\,\Big(\,{r_*\over r}\,\Big)^{b}\,\,.
\eeq
Notice that $r_*$ is the minimal value of the coordinate $r$ (which occurs at $\theta=0$) and that $\theta\to\pi/2$ when $r\to\infty$. Moreover, when $r_*=0$, which corresponds to the massless case,  the angle $\theta$ takes the constant value $\theta=\pi/2$. 

In the unflavored ABJM case, the constant $b$ is equal to one and the embedding is given by:
\beq
\cos\theta\,=\,{r_*\over r}\,\,.
\eeq
Furthermore, in the squashed ABJM model $b=5/3$ and $\theta(r)$ is:
\beq
\cos\theta\,=\,\Big({r_*\over r}\,\Big)^{{5\over 3}}\,\,.
\eeq

\subsection{Equations of motion of the probe}

The dynamics of the  D6-brane probe is governed by the DBI+WZ action. The first of these two terms is given by:
\beq
S_{DBI}\,=\,-T_{D_6}\,\int e^{-\phi}\,\,\sqrt{-\det \hat g}\,\,\,
d^7\zeta\,\,.
\eeq
For our ansatz, the determinant of the induced metric has been computed in (\ref{det-induced-metric}).  Let us integrate over the coordinates $\xi$, $\hat\psi$ and $\varphi$. This integration generates the following constant multiplicative factor:
\beq
\int_0^{\infty}\,{2\xi\over (1+\xi^2)^2}\,\,d\xi\,\int_0^{4\pi}\,d\hat\psi\,\,\int_0^{2\pi}\,d\varphi
\,=\,8\pi^2\,\,.
\eeq
Therefore, the DBI lagrangian density in the remaining coordinates is given by:
\beq
{\cal L}_{DBI}\,=\,-8\pi^2\,T_{D_6}\,e^{-\phi}\,h^{{1\over 4}}\,e^{2f+g}\,\,\sin\theta\,\,
\sqrt{1+e^{2g}\,(\,\theta'\,)^2}\,\,.
\label{DBI-loc}
\eeq
Let us next consider the WZ term of the action, which is given by:
\beq
S_{WZ}\,=\,T_{D_6}\,\int\,\hat C_7\,\,,
\eeq
with $\hat C_7$ being the pullback of the RR seven-form potential of $F_8\equiv -{}^*F_2$ (\ie\ $F_8=dC_7$). It is clear from the calibration condition  (\ref{calibration-conds}) that ${\cal K}$ provides such a potential, namely one can choose $C_7$ as:
\beq
C_7\,=\,e^{-\phi}\,\,{\cal K}\,\,.
\label{C7-explicit}
\eeq
Thus, the WZ action takes the form:
\beq
S_{WZ}\,=\,T_{D_6}\,\int e^{-\phi}\,\,\hat{{\cal K}}\,\,.
\eeq
By looking at the expression of the calibration form ${\cal K}$ in (\ref{cal-K-explicit}) and the pullbacks of the basis one-forms in (\ref{pullback-es}), one easily concludes that only the terms with $e^{3469}$ and $e^{4789}$ contribute. The result for $\hat{{\cal K}}$ is the following:
\beq
\hat{{\cal K}}\,=\,{2\xi\over (1+\xi^2)^2}\,\,h^{{1\over 4}}\,e^{2f+g}\,\,\sin\theta\,\,
\Big[\,\sin\theta\,+\,\cos\theta\,e^{g}\,\theta'\,\Big]\,\,d^7\zeta\,\,.
\eeq
Integrating again over $\xi$, $\hat\psi$ and $\varphi$, we get the following WZ lagrangian density:
\beq
{\cal L}_{WZ}\,=\,8\pi^2\,T_{D_6}\,e^{-\phi}\,h^{{1\over 4}}\,e^{2f+g}\,\,\sin\theta\,\,
\Big[\,\sin\theta\,+\,\cos\theta\,e^{g}\,\theta'\,\Big]\,\,.
\label{WZ-lagrangian-local}
\eeq
Thus, the total lagrangian density is:
\beq
{\cal L}\,=\,- 8\pi^2\,T_{D_6}\,e^{-\phi}\,h^{{1\over 4}}\,e^{2f+g}\,\,\sin\theta\,\,\Big[\,\,
\sqrt{1+e^{2g}\,(\,\theta'\,)^2}\,-\,\Big(\,\sin\theta\,+\,\cos\theta\,e^{g}\,\theta'\,\Big)\,
\Big]\,\,.
\eeq
From this expression  we compute the two derivatives that enter the equations of motion of the probe, namely:
\bear
&&{\partial {\cal L}\over \partial \theta}\,=\,- 8\pi^2\,T_{D_6}\,e^{-\phi}\,h^{{1\over 4}}\,e^{2f+g}\,\Big[
\cos\theta\, \sqrt{1+e^{2g}\,(\,\theta'\,)^2}-2\sin\theta\cos\theta-
(\cos^2\theta\,-\,\sin^2\theta\,)\,e^{g}\,\theta'\Big]\,\,,\rc\rc
&&{\partial {\cal L}\over \partial \theta'}\,=\,- 8\pi^2\,T_{D_6}\,e^{-\phi}\,
h^{{1\over 4}}\,e^{2f+g}\,\,\sin\theta\,\,\Big[\,\,
{e^{2g}\,\theta'\over \sqrt{1+e^{2g}\,(\,\theta'\,)^2}}\,-\,\cos\theta\,e^{g}\,\,\Big]\,\,.
\label{derivatives-Lprobe}
\eear
One can verify straightforwardly that the terms in parentheses on the right-hand side of (\ref{derivatives-Lprobe}) vanish when the BPS equation for the embedding (\ref{kappa-ode}) holds.  This shows that, indeed, the equations of motion of the probe are satisfied for the kappa symmetric configuration. 

\vskip 1cm
\renewcommand{\theequation}{\rm{E}.\arabic{equation}}
\setcounter{equation}{0}

\section{Equations of motion}
\label{eoms}

In this appendix we write the equations of motion for the supergravity plus branes system and we verify that the first-order BPS system  (\ref{BPS-flavored}) implies the second-order equations of motion for the different fields. In what follows we will work in Einstein frame.  The  total action is given by:
\beq
S\,=\,S_{IIA}\,+\,S_{{\rm sources}}\,\,,
\label{total-action}
\eeq
with the type IIA term being given by:
\beq
S_{IIA}\,=\,{1\over 2\kappa_{10}^2}\,\,\Bigg[\,
\int \sqrt{-g}\,\Big(R\,-\,{1\over 2}\,\partial_{\mu}\phi\,\partial^{\mu}\,\phi\,\Big)\,-\,
{1\over 2}\,\int\,\Big[\,e^{{3\phi\over 2}}\,\,{}^*F_2\wedge F_2+
e^{{\phi\over 2}}\,\,{}^*F_4\wedge F_4\Big]\Bigg]\,\,,
\label{IIA-action}
\eeq
and the source contribution is just the sum of the DBI  and WZ action of the D6-branes, which can be written as:
\beq
S_{{\rm sources}}\,=\,-T_{D_6}\,\int\,\Big(\,e^{{3\phi\over 4}}\,{\cal K}\,-\,C_7\,
\Big)\,\wedge\,\Omega\,\,,
\label{source-action}
\eeq
where the calibration form ${\cal  K}$ is given by (\ref{cal-K-explicit}), with the $e^a$'s being the one-forms written in (\ref{ten-dim-frame}) multiplied by $e^{-\phi/4}$, as it corresponds 
to the Einstein frame.

The Maxwell equations for the forms $F_2$ and $F_4$ derived from (\ref{total-action}) are just:
\beq
d\,\Big(\,e^{{3\phi\over 2}}\,\,{}^*F_2\,\Big)\,=\,0\,\,,\qquad\qquad
d\,\Big(\,e^{{\phi\over 2}}\,\,{}^*F_4\,\Big)\,=\,0\,\,,
\eeq
and one can show that they are satisfied for our ansatz as a consequence of the BPS equations (\ref{BPS-flavored}). Similarly, the equation for the dilaton $\phi$  derived from (\ref{IIA-action}) is just:
\beq
d\,{}^*d\phi\,=\,{3\over 4}\,\,e^{{3\phi\over 2}}\,\,{}^*F_2\wedge F_2+{1\over 4}\,
e^{{\phi\over 2}}\,\,{}^*F_4\wedge F_4\,+\,
{3\over 2}\,\kappa_{10}^2\,T_{D_6}\,\,e^{{3\phi\over 4}}\,{\cal K}\wedge\,\Omega\,\,,
\eeq
and it is also fulfilled as a consequence of the first-order equations (\ref{BPS-flavored}). Let us now study Einstein equations, which read:
\bear
&&R_{\mu\nu}\,-\,{1\over 2}\,g_{\mu\nu}\,R\,=\,
{1\over 2}\,\partial_{\mu}\phi\,\partial_{\nu}\phi\,-\,{1\over 4}\,g_{\mu\nu}\,
\partial_{\rho}\phi\,\partial^{\rho}\phi\,+\,
{1\over 4}\,e^{{3\phi\over 2}}\,\Big[\,2F_{\mu\rho}^{(2)}\,F_{\nu}^{(2)\,\,\rho}\,-\,{1\over 2}\,g_{\mu\nu}\,F_2^2\,\Big]\,+\,\rc\rc
&&\qquad\qquad\qquad\qquad
+{1\over 48}\,
e^{{\phi\over 2}}\,\Big[\,4F_{\mu\rho\sigma\lambda}^{(4)}\,F_{\nu}^{(4)\,\,\rho\sigma\lambda}\,-\,
{1\over 2}\,g_{\mu\nu}\,F_4^2\,\Big]\,+\,T_{\mu\nu}^{{\rm sources}}\,\,,
\label{Einstein-eq}
\eear
where $T_{\mu\nu}^{{\rm sources}}$ is just the stress-energy tensor for the flavor branes, defined as:
\beq
T_{\mu\nu}^{{\rm sources}}\,=\,-{2\kappa_{10}^2\over \sqrt{-g}}\,\,
{\delta S_{{\rm sources}}\over \delta g^{\mu\nu}}\,\,.
\label{T-def}
\eeq
In order to write the explicit expression of $T_{\mu\nu}^{{\rm sources}}$ derived from (\ref{T-def}), let us define the following operation for any two $p$-forms $\omega_{(p)}$ and 
$\lambda_{(p)}$:
\beq
	\omega_{p} \lrcorner \lambda_{(p)} = \frac{1}{p!} \omega^{\mu_1 ... \mu_p}
	 \lambda_{\mu_1 ... \mu_p}\,\,.
\eeq
Then, by computing explicitly the derivative of the action (\ref{source-action}) with respect to the metric,  one can check that:
\beq
T_{\mu\nu}^{{\rm sources}}\,=\,\kappa_{10}^2\,T_{D_6}\,\,e^{{3\phi\over 4}}\,
\Big[\,\,g_{\mu\nu}\,
{}^*{\cal K} \lrcorner\,\Omega\,-\,
{1\over 2}\,\Omega_{\mu}^{\,\,\,\rho\sigma}\,\big({}^*{\cal K}\big)_{\nu\rho\sigma}
\Big]\,\,.
\eeq
The different flat components of this tensor, in the basis analogue to (\ref{ten-dim-frame}) in Einstein frame,  are:
\bear
&&T_{00}=-T_{11}=-T_{22}=-T_{33}\,=\,k(\eta-1)\,e^{-2f-g+{3\phi\over 2}}\,\,h^{-{3\over 4}}\,\,,\rc\rc
&&T_{ab}\,=\,-{k\over 2}\,(\eta-1)\,e^{-2f-g+{3\phi\over 2}}\,\,h^{-{3\over 4}}\delta_{ab}\,\,,
\qquad\qquad (a,b=4,\cdots, 9)\,\,.
\eear
By using these values one can verify that the Einstein equations (\ref{Einstein-eq}) are indeed satisfied as a consequence of the first-order system (\ref{BPS-flavored}).



\begin{thebibliography}{99}
 
 
  \bibitem{jm} J.~M.~Maldacena, ``The large $N$ limit of superconformal field
theories and supergravity'', {\it Adv.\ Theor.\ Math.\ Phys.}\  {\bf 
2} (1998) 231,
{\rm hep-th/9711200}.

 
\bibitem{BL}
  J.~Bagger, N.~Lambert,
 ``Modeling Multiple M2's,''
  Phys.\ Rev.\  {\bf D75}, 045020 (2007).
  [hep-th/0611108];
  J.~Bagger, N.~Lambert,
 ``Gauge symmetry and supersymmetry of multiple M2-branes,''
  Phys.\ Rev.\  {\bf D77 } (2008)  065008.
  [arXiv:0711.0955 [hep-th]];
  J.~Bagger, N.~Lambert,
 ``Comments on multiple M2-branes,''
  JHEP {\bf 0802}, 105 (2008).
  [arXiv:0712.3738 [hep-th]].
 
\bibitem{Gustavsson:2007vu}
  A.~Gustavsson,
  ``Algebraic structures on parallel M2-branes,''
  Nucl.\ Phys.\  {\bf B811}, 66-76 (2009).
  [arXiv:0709.1260 [hep-th]].
 
 
 
\bibitem{Aharony:2008ug}
  O.~Aharony, O.~Bergman, D.~L.~Jafferis and J. M. Maldacena,
``N=6 superconformal Chern-Simons-matter theories, M2-branes and their gravity duals,'' JHEP {\bf 0810}, 091 (2008).
  [arXiv:0806.1218 [hep-th]].
  
  
\bibitem{Drukker:2010nc}
  N.~Drukker, M.~Mari\~no, P.~Putrov,
``From weak to strong coupling in ABJM theory,''
  [arXiv:1007.3837 [hep-th]].

\bibitem{Herzog:2010hf}
  C.~P.~Herzog, I.~R.~Klebanov, S.~S.~Pufu, T.~Tesileanu,
 ``Multi-Matrix Models and Tri-Sasaki Einstein Spaces,''
  Phys.\ Rev.\  {\bf D83}, 046001 (2011).
  [arXiv:1011.5487 [hep-th]].

  
  
\bibitem{Aharony:2008gk}
  O.~Aharony, O.~Bergman, D.~L.~Jafferis,
``Fractional M2-branes,''
  JHEP {\bf 0811}, 043 (2008).
  [arXiv:0807.4924 [hep-th]].
  
  
  
  
  
  
  
\bibitem{Gaiotto:2009mv}
  D.~Gaiotto, A.~Tomasiello,
``The gauge dual of Romans mass,''
  JHEP {\bf 1001}, 015 (2010).
  [arXiv:0901.0969 [hep-th];  D.~Gaiotto, A.~Tomasiello,
``Perturbing gauge/gravity duals by a Romans mass,''
  J.\ Phys.\ A {\bf A42}, 465205 (2009).
  [arXiv:0904.3959 [hep-th]].



\bibitem{Fujita:2009kw}
  M.~Fujita, W.~Li, S.~Ryu, T.~Takayanagi,
  ``Fractional Quantum Hall Effect via Holography: Chern-Simons, Edge States, and Hierarchy,''
  JHEP {\bf 0906}, 066 (2009).
  [arXiv:0901.0924 [hep-th]].
  
  
  
  
  
\bibitem{Imamura:2008nn}
  Y.~Imamura, K.~Kimura,
``On the moduli space of elliptic Maxwell-Chern-Simons theories,''
  Prog.\ Theor.\ Phys.\  {\bf 120}, 509-523 (2008).
  [arXiv:0806.3727 [hep-th]].
  
\bibitem{Jafferis:2008qz}
  D.~L.~Jafferis, A.~Tomasiello,
``A Simple class of N=3 gauge/gravity duals,''
  JHEP {\bf 0810}, 101 (2008).
  [arXiv:0808.0864 [hep-th]].
  
\bibitem{Imamura:2008ji}
  Y.~Imamura, S.~Yokoyama,
``N=4 Chern-Simons theories and wrapped M-branes in their gravity duals,''
  Prog.\ Theor.\ Phys.\  {\bf 121}, 915-940 (2009).
  [arXiv:0812.1331 [hep-th]].
  
  
  
\bibitem{Martelli:2008si}
  D.~Martelli, J.~Sparks,
  ``Moduli spaces of Chern-Simons quiver gauge theories and AdS(4)/CFT(3),''
  Phys.\ Rev.\  {\bf D78}, 126005 (2008).
  [arXiv:0808.0912 [hep-th]].
  
  
\bibitem{Hanany:2008cd}
  A.~Hanany, A.~Zaffaroni,
  ``Tilings, Chern-Simons Theories and M2 Branes,''
  JHEP {\bf 0810}, 111 (2008).
  [arXiv:0808.1244 [hep-th]]; A.~Hanany, D.~Vegh, A.~Zaffaroni,
  ``Brane Tilings and M2 Branes,''
  JHEP {\bf 0903}, 012 (2009).
  [arXiv:0809.1440 [hep-th]].

  
  
  
   
  
\bibitem{Ooguri:2008dk}
  H.~Ooguri, C.~-S.~Park,
``Superconformal Chern-Simons Theories and the Squashed Seven Sphere,''
  JHEP {\bf 0811}, 082 (2008).
  [arXiv:0808.0500 [hep-th]].



  
  

\bibitem{Hohenegger:2009as}
  S.~Hohenegger, I.~Kirsch,
  ``A Note on the holography of Chern-Simons matter theories with flavour,''
  JHEP {\bf 0904}, 129 (2009).
  [arXiv:0903.1730 [hep-th]].
  
  
\bibitem{Gaiotto:2009tk}
  D.~Gaiotto, D.~L.~Jafferis,
``Notes on adding D6 branes wrapping RP**3 in AdS(4) x CP**3,''
   [arXiv:0903.2175 [hep-th]].

  
  
  
\bibitem{Fujita:2009xz}
  M.~Fujita, T.~-S.~Tai,
 ``Eschenburg space as gravity dual of flavored N=4 Chern-Simons-matter theory,''
  JHEP {\bf 0909}, 062 (2009).
  [arXiv:0906.0253 [hep-th]].




\bibitem{Jafferis:2009th}
  D.~L.~Jafferis,
  ``Quantum corrections to N=2 Chern-Simons theories with flavor and their AdS(4) duals,''[arXiv:0911.4324 [hep-th]].
  
  
  
\bibitem{Benini:2009qs}
  F.~Benini, C.~Closset, S.~Cremonesi,
 ``Chiral flavors and M2-branes at toric CY4 singularities,''
  JHEP {\bf 1002}, 036 (2010).
  [arXiv:0911.4127 [hep-th]].
  
\bibitem{Benini:2011cm}
  F.~Benini, C.~Closset, S.~Cremonesi,
 ``Quantum moduli space of Chern-Simons quivers, wrapped D6-branes and AdS4/CFT3,'' [arXiv:1105.2299 [hep-th]].


  
  
  
  
  
  
\bibitem{Hikida:2009tp}
  Y.~Hikida, W.~Li, T.~Takayanagi,
``ABJM with Flavors and FQHE,''
  JHEP {\bf 0907}, 065 (2009).
  [arXiv:0903.2194 [hep-th]].

  
  
  
\bibitem{Jensen:2010vx}
  K.~Jensen,
  ``More Holographic Berezinskii-Kosterlitz-Thouless Transitions,''
  Phys.\ Rev.\  {\bf D82}, 046005 (2010).
  [arXiv:1006.3066 [hep-th]].


\bibitem{Ammon:2009wc}
  M.~Ammon, J.~Erdmenger, R.~Meyer {\it et al.},
 ``Adding Flavor to AdS(4)/CFT(3),''
  JHEP {\bf 0911}, 125 (2009).
  [arXiv:0909.3845 [hep-th]].






\bibitem{Gauntlett:1997pk}
  J.~P.~Gauntlett, G.~W.~Gibbons, G.~Papadopoulos, P.~K.~Townsend,
  ``Hyper-Kahler manifolds and multiply intersecting branes,''
  Nucl.\ Phys.\  {\bf B500}, 133-162 (1997).
  [hep-th/9702202].




\bibitem{Bianchi}
  M.~S.~Bianchi, S.~Penati, M.~Siani,
  ``Infrared stability of ABJ-like theories,''
  JHEP {\bf 1001}, 080 (2010).
  [arXiv:0910.5200 [hep-th]]; M.~S.~Bianchi, S.~Penati, M.~Siani,
  ``Infrared Stability of N = 2 Chern-Simons Matter Theories,''
  JHEP {\bf 1005}, 106 (2010).
  [arXiv:0912.4282 [hep-th]].






\bibitem{Bigazzi:2005md}
  F.~Bigazzi, R.~Casero, A.~L.~Cotrone, E.~Kiritsis, A.~Paredes,
 ``Non-critical holography and four-dimensional CFT's with fundamentals,''
  JHEP {\bf 0510}, 012 (2005).
  [hep-th/0505140].




\bibitem{Nunez:2010sf}
  C.~Nunez, A.~Paredes, A.~V.~Ramallo,
``Unquenched flavor in the gauge/gravity correspondence,''
  Adv.\ High Energy Phys.\  {\bf 2010}, 196714 (2010).
  [arXiv:1002.1088 [hep-th]].







\bibitem{Tomasiello:2007eq}
  A.~Tomasiello,
  ``New string vacua from twistor spaces,''
  Phys.\ Rev.\  {\bf D78}, 046007 (2008).
  [arXiv:0712.1396 [hep-th]].





\bibitem{Aldazabal:2007sn}
  G.~Aldazabal, A.~Font,
``A Second look at N=1 supersymmetric AdS(4) vacua of type IIA supergravity,''
  JHEP {\bf 0802}, 086 (2008).
  [arXiv:0712.1021 [hep-th]].





\bibitem{Callan:1988wz}
  C.~G.~Callan, Jr., C.~Lovelace, C.~R.~Nappi, S.~A.~Yost,
  ``Loop Corrections to Superstring Equations of Motion,''
  Nucl.\ Phys.\  {\bf B308}, 221 (1988).
  
\bibitem{Bigazzi:2008zt}
  F.~Bigazzi, A.~L.~Cotrone, A.~Paredes,
  ``Klebanov-Witten theory with massive dynamical flavors,''
  JHEP {\bf 0809 } (2008)  048.
  [arXiv:0807.0298 [hep-th]].



\bibitem{HoyosBadajoz:2008fw}
  C.~Hoyos-Badajoz, C.~Nunez, I.~Papadimitriou,
  ``Comments on the String dual to N=1 SQCD,''
  Phys.\ Rev.\  {\bf D78}, 086005 (2008).
  [arXiv:0807.3039 [hep-th]].


\bibitem{Emparan:1999pm}
  R.~Emparan, C.~V.~Johnson, R.~C.~Myers,
`Surface terms as counterterms in the AdS / CFT correspondence,''
  Phys.\ Rev.\  {\bf D60}, 104001 (1999).
  [hep-th/9903238].





\bibitem{Lee:2006ys}
  K.~-M.~Lee, H.~-U.~Yee,
``New AdS(4) x X(7) Geometries with CN=6 in M Theory,''
  JHEP {\bf 0703}, 012 (2007).
  [hep-th/0605214].





\bibitem{Santamaria:2010dm}
  R.~C.~Santamaria, M.~Marino, P.~Putrov,
``Unquenched flavor and tropical geometry in strongly coupled Chern-Simons-matter theories,''  [arXiv:1011.6281 [hep-th]].






\bibitem{Wilson}
  J.~M.~Maldacena,
``Wilson loops in large N field theories,''
  Phys.\ Rev.\ Lett.\  {\bf 80}, 4859-4862 (1998).
  [hep-th/9803002];
  S.~-J.~Rey, J.~-T.~Yee,
  ``Macroscopic strings as heavy quarks in large N gauge theory and anti-de Sitter supergravity,''
  Eur.\ Phys.\ J.\  {\bf C22}, 379-394 (2001).
  [hep-th/9803001].


\bibitem{Drukker:2008zx}
  N.~Drukker, J.~Plefka, D.~Young,
  ``Wilson loops in 3-dimensional N=6 supersymmetric Chern-Simons Theory and their string theory duals,''
  JHEP {\bf 0811}, 019 (2008).
  [arXiv:0809.2787 [hep-th]].


\bibitem{Gubser:2002tv}
  S.~S.~Gubser, I.~R.~Klebanov, A.~M.~Polyakov,
  ``A Semiclassical limit of the gauge / string correspondence,''
  Nucl.\ Phys.\  {\bf B636}, 99-114 (2002).
  [hep-th/0204051].
  



\bibitem{CNP}
  R.~Casero, C.~Nunez, A.~Paredes,
  ``Towards the string dual of N=1 SQCD-like theories,''
  Phys.\ Rev.\  {\bf D73}, 086005 (2006).
  [hep-th/0602027];
   R.~Casero, C.~Nunez, A.~Paredes,
  ``Elaborations on the String Dual to N=1 SQCD,''
  Phys.\ Rev.\  {\bf D77}, 046003 (2008).
  [arXiv:0709.3421 [hep-th]].






\bibitem{conifold}
  F.~Benini, F.~Canoura, S.~Cremonesi, C.~Nunez, A.~V.~Ramallo,
  ``Unquenched flavors in the Klebanov-Witten model,''
  JHEP {\bf 0702}, 090 (2007).
  [hep-th/0612118];
 F.~Benini, F.~Canoura, S.~Cremonesi, C.~Nunez, A.~V.~Ramallo,
  ``Backreacting flavors in the Klebanov-Strassler background,''
  JHEP {\bf 0709 } (2007)  109.
  [arXiv:0706.1238 [hep-th]].
  

\bibitem{Ahn:2008ua}
  C.~Ahn,
  ``Squashing Gravity Dual of N=6 Superconformal Chern-Simons Gauge Theory,''
  Class.\ Quant.\ Grav.\  {\bf 26}, 105001 (2009).
  [arXiv:0809.3684 [hep-th]].



\bibitem{Bigazzi:2009gu}
  F.~Bigazzi, A.~L.~Cotrone, A.~Paredes, A.~V.~Ramallo,
  ``Screening effects on meson masses from holography,''
  JHEP {\bf 0905}, 034 (2009).
  [arXiv:0903.4747 [hep-th]].
  
  
 
\bibitem{D3-D7}
  F.~Bigazzi, A.~L.~Cotrone, J.~Mas, A.~Paredes, A.~V.~Ramallo, J.~Tarrio,
 ``D3-D7 Quark-Gluon Plasmas,''
  JHEP {\bf 0911}, 117 (2009).
  [arXiv:0909.2865 [hep-th]];
   F.~Bigazzi, A.~L.~Cotrone, J.~Tarrio,
  ``Hydrodynamics of fundamental matter,''
  JHEP {\bf 1002}, 083 (2010).
  [arXiv:0912.3256 [hep-th]];
   F.~Bigazzi, A.~L.~Cotrone, J.~Mas, D.~Mayerson, J.~Tarrio,
  ``D3-D7 Quark-Gluon Plasmas at Finite Baryon Density,''
  JHEP {\bf 1104}, 060 (2011).
  [arXiv:1101.3560 [hep-th]].
  
  
  
\bibitem{Aharony:2009fc}
  O.~Aharony, A.~Hashimoto, S.~Hirano, P.~Ouyang,
  ``D-brane Charges in Gravitational Duals of 2+1 Dimensional Gauge Theories and Duality Cascades,''
  JHEP {\bf 1001 } (2010)  072.
  [arXiv:0906.2390 [hep-th]].
  

\bibitem{Martucci:2005rb}
  L.~Martucci, J.~Rosseel, D.~Van den Bleeken and A.~Van Proeyen,
``Dirac actions for D-branes on backgrounds with fluxes,''
  Class.\ Quant.\ Grav.\  {\bf 22}, 2745 (2005)
  [arXiv:hep-th/0504041].



\bibitem{LPT}H. Lu, C. N. Pope and P. K. Townsend, 
``Domain walls from Anti-de-Sitter spacetime", 
{\sl \pl} {\bf B391} (1997) 39, {\rm hep-th/9607164}.



\bibitem{Grana:2005jc}
  M.~Grana,
 ``Flux compactifications in string theory: A Comprehensive review,''
  Phys.\ Rept.\  {\bf 423}, 91-158 (2006).
  [hep-th/0509003].





  
  


    
    
    
    
  
\end{thebibliography}
\end{document}